\documentclass[prd,aps,superscriptaddress,eqsecnum,nofootinbib,showpacs,preprintnumbers]{revtex4-2}

\usepackage{leftindex}
\usepackage{tensor}
\usepackage{bbding}
\usepackage{slashed}
\usepackage{array}
\usepackage{graphicx}
\usepackage{amsmath}
\usepackage{amssymb}
\usepackage{subcaption} 
\usepackage{amsfonts}
\usepackage{indentfirst}
\usepackage{color}
\usepackage{braket}
\usepackage{nameref} 
\usepackage[absolute,overlay]{textpos}
\usepackage[hidelinks]{hyperref}

\begin{document}

\preprint{\leftline{KCL-PH-TH/2025-{\bf 20}}}

\title{On internal mechanical properties  of Electroweak Magnetic Monopoles and their effects on stability}

\author{K. Farakos}
	\affiliation{Physics Division, School of Applied Mathematical and Physical Sciences,
		National Technical University of Athens,  Zografou Campus,
		Athens 15780, Greece.}

 \author{G. Koutsoumbas}    
\affiliation{Physics Division, School of Applied Mathematical and Physical Sciences,
		National Technical University of Athens,  Zografou Campus,
		Athens 15780, Greece.}

\author{Nick E. Mavromatos}
	\affiliation{Physics Division, School of Applied Mathematical and Physical Sciences,
		National Technical University of Athens,  Zografou Campus,
		Athens 15780, Greece.}
	\affiliation{Theoretical Particle Physics and Cosmology Group, Physics Department, King's College London, Strand, London WC2R 2LS, UK.}

\author{Alexandros Zarafonitis }
\affiliation{Physics Division, School of Applied Mathematical and Physical Sciences,
		National Technical University of Athens,  Zografou Campus,
		Athens 15780, Greece.}

%\date{}
%\noaffiliation

\begin{abstract}
    By considering properties of the energy-momentum tensor of the electroweak magnetic monopole and its Born-Infeld extension, we attempt to make comments on the stability of these configurations. 
    Specifically, we
    perform a study of the behaviour of the so-called internal force and pressure of these extended field-theoretic solitonic objects, which are derived from the energy-momentum tensor. Our method is slightly different from the so-called Laue's criterion for stability of nuclear matter, a local form of which had been proposed and applied in the earlier literature to the `t Hooft-Polyakov (HP) magnetic monopole, and found to be violated.
        By applying our method first to HP monopole, 
    we also observe that, despite its topological stability, the total (finite) internal force (which has only radial components) is directed inwards, towards the centre of the monopole, which would imply instability. 
    Thus this mechanical criterion for stability is arguably violated in the case of the HP monopole, as is the local version of Laue's criterion. The criterion is satisfied for the short-range part of the energy momentum tensor, in which the long-range part, due to the massless photon of the U(1) subgroup, is subtracted. 
\color{black} This makes the HP monopole mechanically stable by our criterion, which is also confirmed due to its proven topological stability. \color{black}
    Par contrast, the total internal force of the Cho-Maison (CM) electroweak monopole has both radial and angular components, which diverge at the origin, leading to rotational instabilities, \color{black} violating the short-range Laue's criterion for stability. \color{black} Finally, by studying extensions of the CM, in which the latter is embedded in theories with non-minimal couplings of the hypercharge and Higgs sectors, as well as higher-derivative electromagnetic interactions of Born-Infeld type, we find that the total force, integrated over space, is finite, but, the Born-Infeld case, it has also angular components. The latter feature is interpreted as indicating that, \color{black} unlike the rest of the CM extensions, \color{black} the Born-Infeld-CM monopole might be subject to rotations upon the action of perturbations, but \color{black} this does not necessarily imply mechanical instabilities of the configuration. \color{black} For such unstable composite monopoles, one expects a decay after production into charged constituent $W^\pm$ bosons, which are in principle detectable at colliders.
\end{abstract}

\maketitle
\tableofcontents

\section{Introduction}

The concept of the magnetic pole (magnetic monopole), whose existence would “symmetrize” Maxwell’s equations of electromagnetism,
had been conjectured as early as 1894 by Pierre Curie~\cite{Curie}, and the motion of a charged particle in the presence of such isolated magnetic poles had been studied mathematically for the first time by H. Poincar\'e~\cite{Poincare:1896}, 
in an attempt to explain the experiments by Birkeland~\cite{Birkeland:1896,Birkeland:1896B}, indicating the focusing of the cathodic beams in a Crookes tube in the presence of a magnet (the experiment is explained of course by the nature of the cathodic beam being electrons, but this was not known in 1896). Subsequently, J.J. Thomson~\cite{Thomson:1904} had demonstrated an effect analogous to the Dirac quantization condition ({\it cf.} \eqref{diraccond}) for the product of magnetic and electric charge, by considering the quantization of the angular momentum of an electron in the system used by Poincar\'e to explain the Birkeland experiment.

Nonetheless, it was Dirac~\cite{Dirac:1931,Dirac:1948um,Dirac:1978ie} who has been the first to formulate rigorously the concept of a magnetic pole in a quantum field-theoretic way. Dirac had used the concept of the monopole as a source of a singular at the origin magnetic field
\begin{align}\label{mmf}
    {\vec B}_{{\rm mono}} = g_m \frac{{\vec r}}{r^2} = \vec{\nabla} \times {\vec A}(r)  \,,
\end{align}
where $g_m$ is the magnetic charge.
Dirac observed that the solution for the electromagnetic potential $\vec A(r)$ in \eqref{mmf} is proportional to a singular gauge transformation along the direction of the celebrated (infinite in length) Dirac string that accompanies the isolated magnetic pole. 
By considering an electron looping around far away from the monopole centre $r \to 0$, Dirac noticed that the single-valuedness of the 
electron wavefunction, and hence the non-observability of the string, requires the quantization condition
\begin{align}\label{diraccond}
    g_m \, q_e = \frac{n}{2}\, \hbar \, c \, , \quad n \in \mathbb Z\,,
\end{align}
where $q_e$ is the electric charge, which in the case of the electron equals $e$. 
Dirac used the concept of the magntic monopole as a source for the magnetic field \eqref{mmf} without further specifying whether is a composite or elementary object. In principle, if exists, it can be a new elementary particle.

Dirac's suggestion had inspired `t Hooft and Polyakov, independently~\cite{tHooft:1974kcl,Polyakov:1974ek}, to consider several years later, the  magnetic monopole as a solitonic (extended-object with finite energy) solution of a gauge field theoretic system with spontaneous gauge symmetry breaking. This monopole solution is composite though of gauge bosons and Higgs fields. An important ingredient was the presence of the spontaneous gauge symmetry,  which provides a Higgs field configuration such that, the latter is vanishing at the centre of the monopole ($r \to 0$), and acquires a constant value at infinity ($r \to \infty$), far away from the monopole centre. Roughly speaking, inside the monopole core the gauge symmetry is unbroken, while (far) outside the monopole core the symmetry breaks. The model of `t Hooft was the SU(2) Georgi-Glashow field theory~\cite{Georgi:1974sy}, while Polyakov dealt with phenomenologically realistic grand unified theories (GUT), such as those based on, e.g., the SU(5) gauge group. An important feature of these solutions was that they were based on simply connected gauge groups G broken down to a subgroup H, which had crucial non-trivial homotopy properties that guarantee the topological stability of the monopole. 

For example~\cite{Daniel:1979yz}, in the case of the SU(5) which breaks down to H$_{\rm E}$=SU(3)$_{\rm c} \otimes $U$_{\rm em}$(1) (with c indicating the colour and em the (compact) electromagnetic groups, respectively) one has:
\begin{align}
    \pi_2(\rm SU(5)/H_{\rm E}) = \pi_1 (H_{\rm E}) = \{ x^n, \, \, n \in \mathbb Z \}\,,
\end{align}
where $\pi_1 (\rm H_{E})$ is the first homotopy group of the unbroken group, and 
$x$ denotes the homotopy class of closed curves in H$_{\rm E}$ which start and end at the unit element (the group H$_{\rm E}$ is the set of pairs ($g, h$) 
identified with $(ga, h a^{-1})$, 
with $g \in \rm SU_c(3)$, and $h \in {\rm U}_{\rm em}(1)$. the unit element is given by $(1,1)= (a^2, \, a)= (a, \, a^2)$). The homotopy group $\pi_1(\rm H_E)$ is isomorphic to $\pi_1{\rm U_{em}(1)}$ of the compact electromagnetic U(1) group, which is generated by the set of integers. Thus, the topological magnetic charge of the SU(5) monopole is characterised by an integer, which implies its absolute stability for topological reasons.

In a similar fashion, the SU(2) monopole of `t Hooft is also topologically stable for similar reasons, given that 
\begin{align}\label{su2hom}
    \pi_2(\rm SU(2)) = \mathbb Z\,,
\end{align}
since the triplet of the scalar fields $\phi_a$, $a=1,2,3$, that characterises the Georgi-Glashow model in the broken phase has the topology of a sphere $S_2$, $\sum_{a=1}^3 \phi_a^2 = v^2$ (with $v$ the vacuum expectation value). The homotopy \eqref{su2hom} expresses the number of times a sphere $S^2$ wraps around the field-theoretic sphere $S_2$ of the configuration of $\phi_1$, defined by the gauge group SU(2).  

Since the Georgi-Glashow model is not realised in Nature, it was believed in view of the above considerations that one is left with the GUT monopoles, whose masses is close to the GUT scale. If this scale is close to inflation, such monopoles have been diluted by inflation, and hence their densities today are negligible. Nonetheless cosmic searches for them have taken place. For a relatively recent review on magnetic monopoles and their searches we refer the reader to \cite{Mavromatos:2020gwk}. 

The gauge group SU(2) $\otimes $ U$_{\rm Y}$(1) of standard model (SM) of particle physics, not embedded in any GUT group, is not simply connected due to the non-compact hypercharge-U(1) group factor, and 
thus one cannot apply the above-described homotopy arguments for the existence of topologically stable monopole configurations. Hence many believed that there are no magnetic-monopole configurations in the SM. 

However, Cho-Maison (CM)~\cite{Cho:1996qd} have constructed a monopole solution within the SM, which is a hybrid between the HP and Dirac monopole, in the sense that it is characterised by a Dirac string. They argued that there is an alternative way to guarantee a topological stability, by looking at the homotopy not of the gauge group itslef, but of the Higgs sector. Viewing the Higgs configuration as a CP$^1$ field, which is isomoprhic to the Riemann sphere, they have arrived at the conclusion that their monopole configuration satisfies the homotopy:
\begin{align}
\pi_2 (\rm CP^1) = \mathbb Z\,,
\end{align}
which has been argued in \cite{Cho:1996qd} to provide the required topological stability of the monopole. 
However, this argument relies on a specific representation of the Higgs doublet, and does not appear to be as robust as the homotopy analysis of the HP monopole. 

The hope of CM was that this monopole would have mass of order at most a few TeV, and thus observable at current or future colliders. They termed it electroweak monopole. 
However, the mass of the original CM monopole turns out to be formally infinite, at least classically, which makes it problematic both from a physical point of view, but also from a formal point of view, since a soliton (which the CM monopole is supposed to be) should have, by definition, a finite energy. This infinity is due to the point-like U(1) at the origin, but there is the possibility that gravity can alleviate such singularities, as it can provide proper cutoffs via the creation of event horizons that shield the U(1) charge~\cite{Gervalle:2022vxs}. This has not been demonstrated explicitly though, as yet.

Attempts to produce finite energy CM monopole configurations in flat spacetimes do exist in the literature. They range from {\it ad hoc} phenomenological extensions of the SM, in which quantum corrections (arising from integration of heavy new particles that might exist in such extensions) produce modified susceptibilities of the hypercharge U(1) kinetic terms (and thus electromagnetic terms)~\cite{Cho:2013vba,Ellis:2016glu}, to embedding the SM in string-inspired Born-Infeld (BI) higher-derivative corrections of, say, the hypercharge sector~\cite{Arunasalam:2017eyu,Mavromatos:2018kcd}. Such higher-derivative corrections, which imply the corresponding corrections to the electromagnetic  sector, result in alleviation of the infinities at the origin $r \to 0$ that characterise the toral energy of the CM monopole within the traditional SM, and lead to estimates of its mass to be at least of order $\mathcal O(10)$~TeV, which make it of interest for the production at the next generation colliders (since the monopoles are produced in pairs with their antiparticles, the minimum mass of interest to their production at the LHC is about 5 TeV in order of magnitude. For the monopole of \cite{Mavromatos:2018kcd}, a mass of at least 14 TeV emerges from calculations). 
Moreover, since BI theories involve light-by-light scattering, and the latter has been observed in current LHC experiments~\cite{dEnterria:2013zqi,ATLAS:2017fur,CMS:2018erd}, one can infer lower bounds for the BI parameter~\cite{Ellis:2017edi,Ellis:2022uxv}, which are such that the resulting mass of the CM-like monopole is way above the limits of LHC, hence more suitable for future colliders. 

Leaving aside the possibility of direct observation of CM-like monopoles, the important question on their stability remains, given that the CP$^1$ homotopy argument is not beyond doubt, as we have already mentioned, due to its dependence on the Higgs configuration. 
A linear stability of the CM monopole (against small perturbations) has been demonstrated in \cite{Gervalle:2022npx}, but this does not prove stability beyond the linear approximation. It is therefore important to try and discuss different ways of understanding the stability of the CM-like monopole configurations. 

This is the point of the current work. In fact, in this article we shall be concerned with the development of novel criteria of stability of monopole configurations by looking at properties of the respective energy-momentum tensors, in particular those associated with the internal pressure (radial and polar components), which determine mechanical properties of the monopoles~\cite{Panteleeva:2023aiz}. In addition, such  studies can also give information about local stability of a solitonic system (with a finite energy)~\cite{Perevalova:2016dln},
given that~\cite{Polyakov:2002yz} 
the spatial components of the energy momentum tensor $T_{ij}$ are associated with the distribution of shear forces $s(r)$ and the elastic pressure $p(r)$ in the system. For a spherically symmetric system, 
$T_{ij}$ is expressed as:
\begin{align}\label{localT}
    T_{ij} (\vec r) = \big(\frac{x_i \, x_j}{r^2} - \frac{1}{3} \, \delta_{ij} \Big) \, s(r) + \delta_{ij}\, p(r)\,, \quad i,j=1,2,3\,.
\end{align}
If a composite system, such as a monopole, is characterised by a core, in the interior of which (roughly) the gauge symmetry is unbroken,
while in its exterior the symmetry is spontaneously broken, then an obvious criterion is a balance between the exterior and interior forces. In terms of the pressure $p(r)$, this criterion can be simply expressed as~\cite{Perevalova:2016dln} 
\begin{align}\label{lauecr}
    \int_0^\infty dr r^2 \, p(r) = 0\,,
\end{align}
which is Laue criterion for stability~\cite{Laue:1911lrk}.
This criterion is necessary for stability but not sufficient, given that it is also satisfied by unstable systems. 

Because of this, the authors of \cite{Perevalova:2016dln} attempted to find a stronger {\it local} criterion for stability, in contrast to the global Laue criterion \eqref{lauecr}.
In ~\cite{Perevalova:2016dln} it was argued that if $s(r) >0$, then the normal force is always positive (directed outwards from the centre of the soliton $ r \to 0$). While if $s(r) < 0$,  then the normal force on a solitonic system is negative ({\it i.e.} directed inwards, towards the centre $r \to 0$) and the system will collapse, thus unstable under internal pressure.
We define the normal component of the total (radial) force exerted by the system on an infinitesimal area $dA$, which we denote as $p_{\rm total}(r)$, via the relation:
\begin{align}\label{Fptot}
  dF^i(r) \equiv p_{\rm total}(r) \, dA^i = T^{ij}\, dA_j \stackrel{\eqref{localT}}{=} \Big(\frac{2}{3}\, s(r) + p(r) \Big)\, dA^i \,, \, \, i, j=1,2,3\,.
\end{align}
The positivity of the normal component of the force, 
that is the case in which the normal force points to the outward direction (away from the solitonic object's centre $r \to 0$),  
would avoid collapse of the system, and thus serves 
as a necessary requirement for stability, which, on account of \eqref{Fptot}, implies 
\begin{align}\label{strongercr}
    \frac{2}{3}\, s(r) + p(r) > 0 \,,
\end{align}
Although this criterion, as Laue's one \eqref{lauecr}, is necessary but not sufficient, nonetheless, as a local condition is stronger than \eqref{lauecr}. In \cite{Panteleeva:2023aiz}, this criterion was examined for the case of the HP magnetic monopole, and it was found to be violated.  

In our approach we shall use a somewhat modified mechanical criterion for stability, which avoids the decomposition \eqref{localT} into shear forces and pressure. Specifically, we shall concentrate on the details of  the total internal force of the monopole, defined in terms of pressure components, and in particular 
its strength as a function of the distance from the monopole centre. 
By applying our methods to the case of the HP monopole, which is known to be topologically stable, we shall demonstrate 
the violation of the mechanical criterion that a negative (inwards pointing) radial ({\it i.e.} normal component of the) force indicates instability of the system, as the latter tends to collapse. 
Indeed, in the HP case, we do find that the internal force has only a radial component, pointing inwards, with a magnitude diminishing with an increasing distance from the monopole centre. 
The force is finite throughout, and although it has a non-zero value near the monopole's centre, this is not sufficient to induce a collapse of the monopole configuration.
As suggested in \cite{Panteleeva:2023aiz} the failure of the mechanical criteria for stability might be associated with the existence of long range (gauge, electromagnetic) forces in the HP monopole, due to the massless photon of the compact U(1) $\subset$ SU(2) gauge subgroup. 
The magnitude of the force becomes smaller and smaller as one approaches the Bogomolny-Prasad-Sommerfeld (BPS) limit~\cite{Bogomolny:1975de,Prasad:1975kr}, at which the internal force vanishes, and the situation resembles that of isotropic matter. 

When we apply this analysis in the CM monopole, which we know is characterised by a divergent energy, due to the  electromagnetic U(1) gauge group, 
we do find that the total force 
has non-trivial angular components as well, whose magnitude diverges as we approach the monopole centre $r \to 0$. Due to the divergent non-trivial angular components of the force in this case, one might conclude that the latter imply angular instability of the CM monopole, given that they will induce infinitely fast rotations as one approaches the monopole centre $r \to 0$.
Par contrast, 
when this study of the mechanical properties is  applied to the case of the extensions of the CM into higher-derivative electromagnetic sectors, sch as BI hypercharge extensions of the SM, which might be inspired from string theory, 
we do find that, although the stress-energy tensor components diverge at the origin $r \to 0$ of the monopole, 
nonetheless, the global physical observables given by the total internal force and pressure, integrated over space, are finite. However, the fact that the internal force has, as in the CM case, non-trivial angular components, which however - par contrast to the CM monopole- are finite at the monopole centre in this case, there is 
the possibility that the monopole configuration might be subject to rotation upon perturbations which could be due to either quantum effects of the configuration, or to the spacetime background itself. However, this cannot be interpreted as indicating an instability of the BI electroweak monopole, and \color{black} a further analysis of time-dependent perturbations, beyond linear stability, is required to settle unambiguously this issue in this case.\color{black}

The structure of the article is the following: in the next section \ref{EMT}, we discuss the mechanical properties that stem from the energy-momentum tensor which we shall use in our stability analysis for the monopole configurations. In section \ref{Energy Conditions} we give the energy conditions (in Minkowski spacetime), the validity of which will be examined for each of the monopole solutions duiscussed in the current article. In 
section \ref{HPReview}, we apply our stability criteria based on the aforementioned mechanical properties to the HP monopole, which we know is topologically stable. We explain how our criteria can be adapted so as to reflect this stability. We also verify that all energy conditions are satisfied in the case. 
In section \ref{Electroweak Monopole Analysis}, we  extend our analysis to incorporate first the initial CM magnetic monopole (section \ref{sec:CMclassicalmodel}), which, 
at least classically and in flat spacetime, is characterised by an infinite energy, and, thus, is considered as an unphysical solution, and then the finite-energy extensions of the CM monopole, which are described in sections \ref{sec:cmfinite} and \ref{BI}. The model in section \ref{sec:cmfinite}, is a class of phenomenological extensions of the CM monopole 
with non-minimal couplings between the Higgs and hypercharge sectors, 
in which the coefficient of the hypercharge gauge-boson kinetic term in the pertinent Lagrangian density  is a function of the Higgs field. Such a model is constrained phenomenologically, by requiring agreement of the predictions of the model with LHC data. On the other hand, the model in section \ref{BI} has a more microscopic origin in string theory, and involves a non-linear Born-Infeld-type extension of the hypercharge sector of the CM monopole Lagrangian. All such solutions are characterised by a topological in nature magnetic charge, as is the case of the HP monopole, which we construct in section \ref{sec:topmagcharge}. In section \ref{NumericalResults} we study numerically some properties of the CM solution and its finite-energy extensions, which we shall use in the following sections \ref{EWpressureAnalysis} and \ref{sec:intforce}, where we study these models from the point of view of the mechanical stability criteria we developed for the HP monopole in section \ref{EMT}. Our study indicates  that the mechanical properties of the original CM monopole
are ill-defined in the sense that the associated internal force diverge at the monopole centre. This implies angular instabilities, given that the force has, in addition to the radial component, also angular ones. Par contrast, for the 
Born-Infeld CM extension,  our study shows that the global (total) internal force, integrated over space, has non trivial angular components, which are finite at the monopole origin. This indicates the possibility that perturbations induce rotations of the monopole, but does not necessarily imply instabilities. 
In section \ref{sec:extnBI} we discuss also some further extensions of the hypercharge sector of the electroweak model, and found that there are violations of the Laue local stability criterion for such models, in a similar fashion as the non-minimal Higgs-hypercharge sector couplings, discussed in section \ref{sec:nmcoupl}.
In subsection \ref{encm} we discuss the validity of the energy conditions for the CM electroweak monopole and its variants, and find that all models satisfy the weak and dominant energy conditions. Only the strong energy condition is found not to be satisfied by the string-inspired (hypercharge sector) Born-infeld extension of the CM. 
Finally, section \ref{concl} contains our conclusions. Some technical aspects of our approach, associated with the numerical methods used in solving various field equations and other systems of differential equations, are briefly discussed in 
a short Appendix.

\section{Internal Force Field from the Energy Momentum Tensor}\label{EMT}

This section is devoted to a study of some basic properties of the energy-momentum tensor (EMT) of systems with spherical symmetry, stemming from its conservation. In particular, we shall be interested in diagonal EMTs, expressed in spherical coordinates and use their local covariant conservation equations to obtain some useful conditions, which we shall make use of in our subsequent discussion on the stability of the magnetic-monopole solutions of the field equations of appropriate systems. In this respect, we are also going to discuss below how an internal force field of a particular field-theoretic system can be described in terms of EMT elements.
\subsection{The form of the Energy-Momentum tensor for spherically-symmetric systems}
For a (3+1)-dimensional field-theoretic model  described by an action $S$ the energy momentum tensor (in Minkowski spacetime, we are going to consider throughout this work) is given by:\footnote{Throughout this work we use natural units 
$\hbar = c =1$. The Minkowski metric signature convention is:
$\eta_{\mu\nu}=diag(1,-1,-1,-1)$. 
For the convenience of the reader we note that, in spherical coordinates $(t,r,\theta,\psi)$, where $\theta \in [0, \pi]$ denotes the polar and $\psi \in (0, 2\pi]$ the azimuthal angles, respectively, the Minkowski metric is expressed as:
$g_{\mu\nu}= diag(1,-1,-r^2,-r^2\sin^2(\theta))$.
In this coordinate system, the Minkowski metric has non-zero Christoffel symbol components $\Gamma^\mu_{\nu \alpha}= \Gamma^\mu_{\alpha \nu}$:
\begin{equation}\label{Minkcsc}
\Gamma^{r}_{\mu a}=
\begin{pmatrix}
 0 & 0 & 0 & 0 \\
 0 & 0 & 0 & 0 \\
 0 & 0 & -r & 0 \\
 0 & 0 & 0 & -r \sin^2(\theta)  \\
\end{pmatrix}_{\mu a} \,,  \qquad \Gamma^{\theta}_{\mu \alpha}=\begin{pmatrix}
 0 & 0 & 0 & 0 \\
 0 & 0 & \frac{1}{r} & 0 \\
 0 & \frac{1}{r} & 0 & 0 \\
 0 & 0 & 0 & -\rm sin (\theta ) \rm cos (\theta) \\
\end{pmatrix}_{\mu \alpha}\,.
\end{equation}
}
\begin{equation}\label{EMT_General}
    T^{\mu\nu}= T^{\nu\mu} = \frac{2}{\sqrt{-g}}\frac{\delta S}{\delta g_{\mu\nu}}\,,
\end{equation}
and at the end of the computation we take the metric to be that of the flat Minkowski spacetime. 
In this paper we are going to work with energy momentum tensors of the following diagonal form in spherical coordinates  $(t,r,\theta,\psi)$, which characterises all types of spherical monopole configurations we shall consider in this work:
\begin{equation}
    T^{\mu \nu} =diag(T^{t t},T^{r r} ,T^{\theta \theta} ,T^{\psi \psi})
\end{equation}
Diagonal spatial elements $T^{i i}$  (no summation over $i=t, r, \theta,\psi$) give us force along direction $i$ over spatial surface of constant $i$. This is what we call normal stress, or simply pressure. On the other hand diagonal time element $T^{tt}$ give us the matter density of particular system, or in other words the Hamiltonian density. 

In \cite{Panteleeva:2023aiz}, \cite{Polyakov:2002yz} the EMT for static spherical symmetric solitonic field configurations  has been introduced for spin-0 and
spin-1/2 targets, in Cartesian coordinates $x^i=\{x,y,z\}$, where it assumes the form \eqref{localT}. In  spherical polar coordinates $x^{\prime i}=\{r,\theta,\psi\}$, 
the EMT is diagonal, even in the presence of shear forces:
\begin{equation}\label{EnergyMomentumTensorSphericalCoordJulia}
    T'_{ij}=\begin{pmatrix}
        \frac{2}{3}s(r)+p(r) & 0 & 0 \\
        0 & r^2[p(r)-\frac{1}{3}s(r)] & 0 \\
        0 & 0 & r^2\sin^2(\theta)[p(r)-\frac{1}{3}s(r)]  
    \end{pmatrix}_{ij}\,.
\end{equation}

In our analysis in this work, we are not going to apply such a decomposition for the EMT, but instead we are going to calculate directly its components from the action of each monopole system we shall consider using the definition \eqref{EMT_General}.   
We note, nonetheless, that such a definition also leads to a diagonal form of the Energy Momentum Tensor for all the types of magnetic-monopole actions we shall study here.

\subsection{Equilibrium Conditions }
Local covariant conservation of the EMT is described by the set of equations, 
\begin{equation}\label{EMTcons}
    0 = \nabla_{\mu} \, T^{\mu\nu} = 
    \partial_{\mu}T^{\mu\nu} + \Gamma^{\mu}_{\mu \alpha} \, T^{\nu \alpha} + \Gamma^{\nu}_{\mu \alpha} \, T^{\mu \alpha}  = 0 \,,
 \end{equation}  
 where $\Gamma^{\nu}_{\mu \alpha}= \Gamma^{\nu}_{\alpha \mu}$ are the Christoffel symbols
\eqref{Minkcsc}, corresponding to  the Minkowski metric when expressed in spherical polar coordinates. It will be 
convenient for our analysis below, to rewrite slightly \eqref{EMTcons}, by multiplying \eqref{EMTcons} with the non-vanishing $\sqrt{-g} \ne 0$, and using that
$\nabla_\mu (\sqrt{-g}) = \partial_\mu (\sqrt{-g}) - \sqrt{-g}\, \Gamma^\alpha_{\alpha \mu}$, and the covariant constancy of the metric tensor $\nabla_\mu g_{\alpha \nu}=0$. This leads to:
    \begin{align}\label{Energy Momentum Conservation}
    0 = \partial_{\mu}(\sqrt{-g}T^{\mu\nu})+\sqrt{-g}\, \Gamma^{\nu}_{\mu \alpha} T^{\mu \alpha}\,,
\end{align}
which we shall use below.

Following standard treatments, we define the components of the pressure $\mathcal P$,  as:\footnote{Although, as already mentioned, we are not going to use the decomposition 
$(\ref{EnergyMomentumTensorSphericalCoordJulia})$, nonetheless we mention for completeness that, should one compare \eqref{pressurecomps} with that equation,  they would obtain
 the following expressions of the pressure components in terms of $s(r)$ and $p(r)$:
\begin{equation}
    \mathcal{P}_R (r)=\frac{2}{3}s(r)+p(r)\,, \quad 
    \mathcal{P}_{\Theta}(r)=\frac{1}{r^2}\Big(p(r)-\frac{s(r)}{3}\Big) \,.
\end{equation}
In view of \eqref{strongercr}, the first of these equations translated to the positivity of the radial pressure as a criterion for stability.}
\begin{align}\label{pressurecomps}
\mathcal{P}_R =T^{r r}\,, \quad  \mathcal{P}_{\Theta}=T^{\theta \theta}\,, \quad \mathcal{P}_{\Psi} =T^{\psi \psi}\,. 
\end{align}
We now observe that, for $\nu=\theta$, the conservation Eq.~$(\ref{Energy Momentum Conservation})$ yields:
\begin{equation}\label{EQcond2}
    \mathcal{P}_{\Theta}(r)=\sin^2(\theta)\mathcal{P}_{\Psi}(r,\theta) \,,
\end{equation}
 while, for $\nu=r$, it  implies:
\begin{equation}\label{EQcond1}
    \frac{d \mathcal{P}_R (r)}{d r}+\frac{2}{r} (\mathcal{P}_{R}(r) -r^2\mathcal{P}_{\Theta}(r))=0\,.
\end{equation}
Then by integrating equation $(\ref{EQcond1})$ and assuming that $\mathcal{P}_R (r) \xrightarrow{ r \rightarrow \infty} 0 $, we obtain:
\begin{equation}\label{Pressure Equilibruim}
    \mathcal P_R (r)=\Sigma(r)\,, 
\end{equation}
where,
\begin{equation}
    \Sigma(r)=-\frac{2}{r^2}\int^{\infty}_r  dr'r^{'3}\mathcal{P}_{\Theta}(r')
\end{equation}
Equation $(\ref{Pressure Equilibruim})$ describes  the condition for local pressure equilibrium. It is useful to note that $(\ref{Pressure Equilibruim})$ provide us with the following useful identity:
\begin{equation}\label{UsefulInd}
    \int^{\infty}_0dr' r^{'3}\mathcal{P}_{\Theta}(r')=-\lim_{r \to 0 }\left[\frac{r^2}{2}\mathcal{P}_R (r)\right]
\end{equation}
During our analysis below we are going to modify the equilibrium condition $(\ref{EQcond1})$ by introducing external forces associated with am external pressure $\mathcal{P}_{ext}(r)$, which we are going to specify for each field-theoretic system under study. Specifically, $(\ref{EQcond1})$ would be modified as follows:
\begin{equation}\label{Eq_Ext}
    \frac{d \mathcal{P}_R(r)}{d r}+\frac{2}{r} (\mathcal{P}_{R}(r) -r^2\mathcal{P}_{\Theta}(r))=\frac{\mathcal{P}_{ext}(r) }{r}
\end{equation}
Then by integrating $(\ref{Eq_Ext})$ we obtain:
\begin{equation}
    \mathcal{P}_R(r)+P_{ext}(r)= \Sigma(r)\,, 
\end{equation}
where,
\begin{equation}\label{ExtPressure}
    P_{ext}(r)=\frac{1}{r^2}\int^{\infty}_{r}r'\mathcal{P}_{ext}(r')dr'
\end{equation}
Such equations describe local pressure equilibrium in the presence of the external pressure $P_{ext}(r)$. 
\subsection{Internal Force Field and Stability Criteria}\label{InternalForceGeneralStructure}
The spatial elements of the EMT \eqref{pressurecomps} give rise to infinitesimal force field components $dF^i$ via the differential relation:
\begin{equation}\label{Force_Element}
    dF^i=T^{ij}dA_j\,,
\end{equation}
where repeated indices $i,j=r, \theta , \psi$ denote summation, as usual, and  $dF^i$ denotes an infinitesimal force, pointing outwards from the infinitesimal area element $dA_i$. 

In order to obtain the expression of $dA_i$ in spherical polar coordinates, we are going to consider the infinitesimal vector $d\vec{r}$ in this coordinate system:
\begin{equation}
d\vec{r}=dr\hat{r}+rd\theta \hat{\theta} +r\sin(\theta)d\psi \hat{\psi}
\end{equation}
The radial surface element $d\mathcal{\vec{A}}_R$, pertaining to 
the infinitesimal region from $\theta$ to $\theta+d\theta$ and $\psi$ to $\psi+d\psi$, for constant $r$, is given by:
\begin{equation}
    d\mathcal{\vec{A}}_R=\left(\frac{\partial \vec{r}}{\partial\theta}\times \frac{\partial \vec{r}}{\partial\psi} \right)\, d\theta \, d \psi  = r^2\sin(\theta)\, d\theta \, d\psi\, \hat{r}
\end{equation}
On the other hand, the polar surface element $d\mathcal{\vec{A}}_{\Theta}$,  
corresponding to the infinitesimal region 
from $r$ to $r+dr$ and $\psi$ to $\psi+d\psi$, for constant $\theta$, is given by:
\begin{equation}
    d\mathcal{\vec{A}}_{\Theta}=\left(\frac{\partial \vec{r}}{\partial \psi}\times \frac{\partial \vec{r}}{\partial r} \right)\, dr \, d \psi =r\sin(\theta)\, dr\, d\psi\, \hat{\theta}
\end{equation}
Finally, the azimuthal component of the surface element, $d\mathcal{\vec{A}}_{\Psi}$, pertaining to the region from $r$ to $r+dr$ and $\theta$ to $\theta+d\theta$, for constant $\psi$ reads:
\begin{equation}
    d\mathcal{\vec{A}}_{\Psi}=\left(\frac{\partial \vec{r}}{\partial r}\times \frac{\partial \vec{r}}{\partial\theta} \right)\, dr \, d \theta = r\, dr\, d\theta\, \hat{\psi} 
\end{equation}
Then, the covariant components $dA_j$ in the spherical polar coordinate system are given by:
\begin{equation}
    dA_r = d\mathcal{A}_R\,, \quad dA_{\theta} = r\, d\mathcal{A}_{\Theta}\, \quad dA_{\psi}=r\, \sin(\theta)d\mathcal{A}_{\Psi}\,,
\end{equation}
while the contravariant  components of the infinitesimal force element $dF^i$ \eqref{Force_Element} read:
\begin{equation}
    dF^r=d\mathcal{F}_R \quad dF^{\theta}=\frac{1}{r}d\mathcal{F}_{\Theta} \quad dF^{\psi}=\frac{1}{r\sin(\theta)}d\mathcal{F}_\Psi  
\end{equation}

In terms of the internal pressure components \eqref{pressurecomps}, we then obtain the components of the total force, by integrating $(\ref{Force_Element})$ appropriately:
\begin{equation}\label{pressurecomps2}
    \mathcal{F}_R (r) = 4\pi r^2 \mathcal{P}_R (r), \qquad  \mathcal{F}_\Theta (\theta) = 2\pi \sin(\theta) \int^{\infty}_0drr^3 \mathcal{P}_{\Theta} (r), \qquad \mathcal{F}_{\Psi} = \int^{\pi}_0 d\theta \sin^2(\theta)\int^{\infty}_0 d r r^3 \mathcal{P}_{\Psi}(r,\theta)   
\end{equation}
We can apply  $ (\ref{EQcond2})$ and $ (\ref{UsefulInd})$, to obtain more practical expressions for $\mathcal{F}_{\Theta}(\theta)$ and $\mathcal{F}_{\Psi}$, which we shall make use of in our subsequent analysis in this paper:
\begin{equation}\label{Polar Force}
    \mathcal{F}_{\theta}(\theta)=-\pi \sin(\theta) \lim_{r \to 0}[r^2\mathcal{P}_R (r)]  \,,
\end{equation}
and
\begin{equation}\label{Azimuthal Force}
    \mathcal{F}_{\Psi} =\pi \int^{\infty}_0 dx'x^{'3}\mathcal{P}_{\Theta}(x')=-\frac{\pi}{2}\lim_{r \to 0}[r^2 \mathcal{P}_R (r)] \,.
\end{equation}
The reader should note that the force is 
independent of the azimuthal angle. 
Note that, if $\mathcal{P}_R (r)$ has a finite value at $r=0$, then both $\mathcal{F}_{\Theta}(\theta)$ and $\mathcal{F}_{\Psi}$ are zero. If this is the case, then the  internal force field is radial. 

In general, the internal force field is given in spherical coordinates by the following expression:
\begin{equation}
    \vec{\mathcal{F}}_{int}(r,\theta)=\mathcal{F}_R(r)\hat{r}+\mathcal{F}_{\Theta}(\theta)\hat{\theta}+\mathcal{F}_{\Psi}\hat{\psi}
\end{equation}
As a final note, we mention that, in the presence of an external pressure $(\ref{ExtPressure})$, the total radial force becomes:
\begin{equation}
    \mathcal{F}_{R\, total}(r)=4\pi r^2[\mathcal{P}_R(r)+P_{ext}(r)]
\end{equation}
Below we shall examine the behaviour of the various components of the total internal force  in several magnetic-monopole solutions, placing the emphasis on their stability. 
In addition to a potential {\it violation} of the equilibrium condition \eqref{Pressure Equilibruim}, another straightforward criterion for mechanical instability of the monopole solutions, which we examine here, is that the total {\it radial} force points towards the centre of the configuration, which would indicate that the system has a tendency of collapsing. The reader should notice that this is equivalent to the criterion  \eqref{strongercr}, in view of \eqref{pressurecomps} and \eqref{pressurecomps2}. Moreover, if the {\it angular components} of the pressure {\it diverge} as we approach the monopole centre, this would indicate rotational instabilities, as the system near the origin would spin with an absurdly fast angular velocity.
We shall also look for less straightforward criteria of (in)stability, for instance, following the analysis in \cite{Panteleeva:2023aiz}, we shall examine the behaviour of the (components of the) pressure 
after subtraction of the long-range force component, 
corresponding to the magnetic U(1) gauge subgroup, which characterises the magnetic monopole solutions. 

Below we shall study these criteria for a variety of monopole solutions available in the current literature, specifically the `t Hooft-Polyakov (HP) monopole, the Cho-Maison (CM) monopole and its (Abelian U(1)) Born-Infeld extension, inspired from string theory. 
We commence our analysis from the case of the HP magnetic monopole. 
\section{Energy Conditions}\label{Energy Conditions}
Energy conditions in general relativity is simply the generalization of the statement that the energy density of a region of space cannot be negative. There conditions are: \\ \\
\textbf{(1) Weak energy conditions (WEC):} For every timelike vector field $X$ holds:
\begin{equation}
    T_{\mu\nu}X^{\mu}X^{\nu}\geq 0
\end{equation}
This ensures that the energy for an observe moving along $X$, is positive. The same statement can be extended to null vectors. \\ \\
\textbf{(2) Strong Energy conditions (SEC):} For every timelike vector field $X$ holds:
\begin{equation}
    (T_{\mu\nu}-\frac{T}{2}g_{\mu\nu}) X^{\mu}X^{\nu} \geq 0
\end{equation}
This ensures that timelike curves converge, which a statement that gravity is an attractive force. The same can said for null vectors.\\ \\

\textbf{(3) Dominant Energy conditions (DEC):} Define for every timelike vector field $X$ the current:
\begin{equation}
    J^{\mu}=-T^{\mu\nu}X_{\nu}
\end{equation}
Then $J_{\mu}J^{\mu} \geq 0$, which states that energy can not travel faster than light. \\ \\ 

Note that condition $T_{\mu\nu}X^{\mu}X^{\nu} \geq 0$ for every null vector is implied by both weak and strong conditions. 
\\ \\
In our case $T^{\mu\nu}=diag(\mathcal{H},\mathcal{P}_R,\mathcal{P}_{\Theta},\mathcal{P}_{\Psi})$ and we can pick general timelike vector $X_{\mu}=(1,0,0,0)$ to obtain from WEC, $\mathcal{H} \geq 0$.
As for the null vectors we can pick $X_{\mu}=(1,1,0,0)$ and obtain from WEC, $\mathcal{H}+\mathcal{P}_R \geq 0$. Thus we conclude:
\begin{equation}
    WEC=
    \begin{cases}
  \mathcal{H}\geq 0 \,, \quad  {\rm timelike-vectors} \\
  \mathcal{H}+\mathcal{P}_R \geq 0\,, \quad {\rm null-vectors}
\end{cases}
\end{equation}
$SEC$ for timelike vector $X_{\mu}=(1,0,0,0)$ provide us with $\mathcal{H}+\mathcal{P}_R+2r^2\mathcal{P}_{\Theta}\geq 0 $. As null vector $X_{\mu}=(1,1,0,0)$ we get $\mathcal{H}+\mathcal{P}_R\geq 0$. Thus we conclude:
\begin{equation}
    SEC=
    \begin{cases}
  \mathcal{H}+\mathcal{P}_R+2r^2\mathcal{P}_{\Theta}\geq 0 \,, \quad  {\rm timelike-vectors} \\
  \mathcal{H}+\mathcal{P}_R \geq 0 \,,\quad {\rm null-vectors}
\end{cases}
\end{equation}
\\ 
DEC for timelike vector $X_{\mu}=(1,0,0,0)$ provide us with $(\mathcal{H})^2\geq 0$ and for null vector $X_{\mu} =(1,1,0,0)$ we get $(\mathcal{H})^2-(\mathcal{P}_R)^2 \geq 0$. Thus we conclude:
\begin{equation}
    DEC=
    \begin{cases}
  (\mathcal{H})^2\geq 0 \,, \quad    {\rm timelike-vectors} \\
  (\mathcal{H})^2-(\mathcal{P}_R)^2 \geq 0\,,  \quad {\rm null-vectors}
\end{cases}
\end{equation}
In this article we shall examine the validity of these energy conditions for the various magnetic monopole solutions studied, in parallel to their mechanical stability, so as to form a more complete picture about their properties. 

\section{`t Hooft-Polyakov Monopole}\label{HPReview}

The HP monopole~\cite{tHooft:1974kcl,Polyakov:1974ek} is a stable  monopole configuration solution in theoretical physics. 
As we have discussed in the introduction, its stability is due to its non-trivial homotopy properties. 
In this section, as a calibration of our stability criteria, we are going to review such a solution in order to establish a basic understanding of the mechanical properties of a well defined monopole solution. In the next subsection we give a brief review of the mathematical details of the solution. 

\subsection{Brief review of the formalism}

The HP magnetic monopole, in its $SU(2)$ version,\footnote{The HP monopole also, and most importantly, characterises phenomenologically realistic grand unified theories, which we shall not examine in our article, which concentrates on the so-called electroweak monopoles, that is monopoles associated with standard-model group and its Abelian Born-Infeld extensions.} is a solution of the field equations of the $SU(2)$ Georgi-Glashow model, and 
 a well-established theoretical topological solution in field theories \cite{tHooft:1974kcl}, \cite{Shnir:2005vvi}, \cite{Ryder:1985wq}. It consists of a non-Abelian $SU(2)$ field $\vec{A}_{\mu}$ and a Higgs triplet $\vec{\phi}$, with a Lagrangian density given by:
\begin{equation}
    \mathcal{L}=-\frac{1}{4}\vec{F}_{\mu\nu}\cdot \vec{F}^{\mu\nu}+\frac{1}{2}(D_{\mu}\vec{\phi})\cdot( D^{\mu}\vec{\phi})-\frac{\lambda}{4}\left(\vec{\phi}\cdot \vec{\phi}-\frac{\mu^2}{\lambda}\right)^2\,,
\end{equation}
where the covariant derivative is given by:
\begin{equation}\label{covgauge}
    D_{\mu}\vec{\phi}= \partial_{\mu}\vec{\phi} - e \vec{A}_{\mu}\times \vec{\phi} \,.
\end{equation}
 Above, $A^a_\mu$, where  the Latin index $a=1,2,3$ is a gauge group index, denote the SU(2) gauge bosons, and 
we used the compact  notation for the exterior product $\times$ among SU(2) vectors:  $\vec A_\mu \times \vec \phi \rightarrow  \epsilon^{abc}A_\mu^b \, \phi^c$\,,  $a,b,c =1,2,3$, where the totally antisymmetric Levi-Civita symbol $\epsilon^{abc}$ denotes the structure constants of the su(2) algebra (we follow the summation convention for the group (Latin) indices, as was the case for the spacetime (Greek) world indices, $\mu, \nu$, that is, a repeated index implies summation). 

The non-Abelian $SU(2)$ tensor is given by:
\begin{equation}\label{naFmn}
    \vec{F}_{\mu\nu}=\partial_{\mu}\vec{A}_{\nu}-\partial_{\nu}\vec{A}_{\mu}-e\vec{A}_{\mu}\times \vec{A}_{\nu}
\end{equation}
After spontaneous symmetry breaking ($SSB$) of the gauge group $SU(2)\rightarrow U(1)$, the physical spectrum of the model consists of two massive $W$-bosons, one massive Higgs and a massless photon. The respective masses are given by:
\begin{equation}
    M_W =\frac{\mu}{\sqrt{\lambda}}\,e=u\,e \qquad M_H = \sqrt{2\lambda}\,u\,,
\end{equation}
where $u$ is the vacuum expectation value (vev) of the scalar (Higgs) triplet, 
that is, in the $SSB$ phase one has 
\begin{align}\label{spherephi}
\phi^a \phi^a = u^2\,.
\end{align}
The Higgs triplet has a topology of a sphere $S^2$, and, since the vacuum manifold is $\mathcal{M}= SU(2)/U(1)\thicksim S^2$, the homotopy group of this particular set-up is $\pi_2 (\mathcal{M})=\mathbb{Z}$. Thus, as we have  mentioned in  the introduction, such a model is well suited for magnetic monopole configurations. The solution is static (time independent) and given by the following set of expressions~\cite{tHooft:1974kcl}:
\begin{equation}\label{Sol1HiggsΗοoft}
    \vec{\phi}= \frac{H(r)}{er}\hat{r}\,,
\end{equation}
and
\begin{equation}\label{Sol2GaugeHooft}
 \vec{A}_{\mu}=\frac{1-f(r)}{e}\hat{r}\times\partial_{\mu}\hat{r}\,,
\end{equation}
with boundary conditions: 
\begin{equation}\label{BoundaryCondHooft}
    f(0)=1 \quad f(\infty)=0 \qquad \qquad H(0)=0 \quad \frac{H(r)}{r}\xrightarrow{r \rightarrow \infty} M_W 
\end{equation}
These conditions are essential in 
guaranteeing that  the solution corresponds to the minimum of the energy functional of the model at the boundary and is well defined at the origin $r=0$.  
\par The EMT of this particular model, obtained from \eqref{EMT_General},  reads:
\begin{equation}\label{Energy-MomentumTensorHooft}
    T^{\mu}_{\; \;\nu} =-\vec{F}^{\mu \sigma}\cdot \vec{F}_{\nu\sigma}+D^{\mu}\vec{\phi} \cdot D_{\nu}\vec{\phi} -\delta^{\mu}_{\nu}\mathcal{L}\,,
\end{equation}
where the dot notation denotes inner product among SU(2) vectors $\vec A \cdot \vec B = A^a B^a$, $a=1,2,3$. 
From the purely temporal ($tt$) components of the EMT we construct the Hamiltonian density $\mathcal H$ of the (static) monopole configuration:
\begin{align}\label{Hamiltonian Hooft}
(e^2r^2)\, T^{t t} =(e^2r^2)\, \mathcal{H} =-(e^2 r^2)\, \mathcal{L} &= \frac{(1-f^2)^2}{2r^2} +(f')^2+\frac{1}{2 r^2} (H'r-H)^2+\frac{H^2f^2}{r^2} \nonumber \\
&+\lambda\frac{H^4}{e^2r^2}-\frac{\mu^2}{2}H^2+\frac{\mu^4 e^2}{4\lambda} r^2 
\end{align}
The magnetic charge is given by~\cite{tHooft:1974kcl,Polyakov:1974ek}:
\begin{equation}
    q_m =\int dS^i \epsilon^{ijk} \frac{1}{u}\vec{F}_{jk}\cdot \vec{\phi}=\frac{1}{e}\int dS^i \epsilon^{ijk}(\partial_j \hat{r} \times \partial_k \hat{r}) \cdot \hat{r} =\frac{4\pi}{e} \,,
\end{equation}
where the integral $\frac{1}{4\pi}\int dS^i \epsilon^{ijk}(\partial_j \hat{r} \times \partial_k \hat{r}) \cdot \hat{r} =1 $ is the Brower degree (winding number) of the map $\hat{r}: S^2 \rightarrow S^2$, which justifies formally the {\it topological nature}  of the magnetic charge in this case.   

The energy functional, on the other hand, is given by the spatial integral of $\mathcal H$:
\begin{align}\label{enerfunctional}
    E= \int d^3x \,\mathcal{H}=\frac{4\pi}{e^2} \int^{\infty}_0 dr \left[\frac{(1-f^2)^2}{2r^2}+(f')^2+\frac{(H'r-H)^2}{2r^2}+(\frac{Hf}{r})^2
    +\frac{\gamma}{4} \frac{H^4}{r^2}-\frac{\mu^2}{2}H^2 +r^2\frac{\mu^4 }{4\gamma} \right]\,,
\end{align}
where we have let $\gamma = \lambda/e^2$.
In view of the static nature of the configuration, we may apply the Euler-Lagrange equations to the energy functional and obtain the following system of differential equations:
\begin{equation}\label{HooftProfileDiffScalar}
    H'' =\frac{2f^2H}{r^2}+\gamma H[\frac{H^2}{r^2}-(M_W)^2]
\end{equation}
\begin{equation}\label{HooftProfileDiffGauge}
    f'' = \frac{H^2f}{r^2}+\frac{f(f^2-1)}{r^2}  
\end{equation}
Near $r=0$ we let:
\begin{equation}
f(r)=1+\Delta_1 (r) \qquad |\Delta_1|<<1 
\end{equation}
Thus, Eqs.~$(\ref{HooftProfileDiffScalar})$ and $(\ref{HooftProfileDiffGauge})$ become:
\begin{equation}\label{HooftScalarSmallx}
    H''+ (\gamma M^2_W -\frac{2}{r^2})H =0 
\end{equation}
\begin{equation}\label{HooftGaugeSmallx}
    \Delta''_1 -2\frac{\Delta_1 }{r^2}=H^2
\end{equation}
The non-homogeneous term $H^2$ provide us with a very small correction to $\Delta_1$, so we can safely ignore it. Then these equations are solved by: 
\begin{equation}
    H(r)\xrightarrow{ r \rightarrow 0} c_1  r^2
\end{equation}
\begin{equation}
f(r)\xrightarrow{r \rightarrow 0}1-c_2 r^2
\end{equation}
Furthermore, as $r\rightarrow \infty$  we let:
\begin{equation}
    \frac{H(r)}{r}=M_W+\Delta_2(r) \qquad |\Delta_2|<<1\,,
\end{equation}
and $f(r)$ approach zero. Then $(\ref{HooftProfileDiffScalar})$ and $(\ref{HooftProfileDiffScalar})$ become:
\begin{equation}
    \Delta^{''}_2+\frac{2}{r}\Delta^{'}_2-2\gamma M^2_W \Delta_2 =0 \,.
\end{equation}
\begin{equation}
    f^{''}-M^2_W f = 0
\end{equation}
By solving these equations we obtain the asymptotic behaviors of $H(r)$ and $f(r)$ as $r\rightarrow \infty$:
\begin{equation}
    \frac{H(r)}{r} \xrightarrow{ r\rightarrow \infty} M_W -c_4 \frac{\exp(-\sqrt{2\gamma }M_W r)}{r}
\end{equation}
\begin{equation}
    f(r) \xrightarrow{r \rightarrow \infty} c_3 \, \exp(-M_W r)
\end{equation}
In table $\ref{AsymptoticConstantsHooft}$ below we give the values of the free constant parameters $c_i$, $i=1, \dots, 4$, of the model for some values of $\gamma$.
\begin{table}[h]
\caption{The constants $c_1$ $c_2$, $c_3$ and $c_4$ for various values of $\gamma$.}
\centering
  \begin{tabular}{|| c || c | c | c |  c |}
  \hline
    $\gamma$ & $c_1$ & $c_2$ & $c_3$ & $c_4$ \\
     \hline\hline
 1  & 0.87 & 0.39 & 2.90 & 2.00\\ 
 \hline
 0.5 & 0.73  & 0.32 & 3.00 & 1.80\\
 \hline
 0.1 & 0.54 & 0.26 & 3.95 & 1.20 \\
 \hline
 0.0001 & 0.39 & 0.19 & 8.50 & 0.62 \\
 \hline
  \end{tabular}
\label{AsymptoticConstantsHooft}
\end{table}

\par
\begin{figure}[ht]
    \centering
\includegraphics[width=0.55\textwidth]{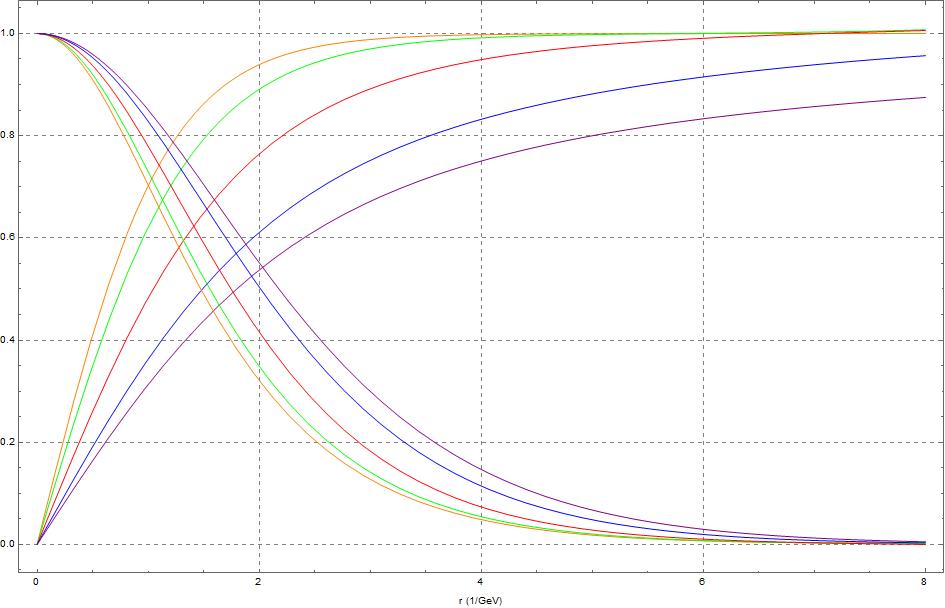}
    \caption{Solutions of $(\ref{HooftProfileDiffScalar})$ and $(\ref{HooftProfileDiffGauge})$ for various $\gamma$ values with $\mu=\lambda$ and $M_W =1~{\rm GeV}$ fixed, in the HP magnetic monopole. $f(r)$ solutions approach zero asymptotically and $H(r)/r$ solutions approach asymptotically $H(r)/r \xrightarrow{r \rightarrow \infty} 1$. Orange, green, red, blue and purple lines correspond to $\gamma$ $1,0.5,0.1,0.0001$ and $0$ respectively. }\label{HooftSolutions}
\end{figure}

\begin{figure}[ht]
    \centering
\includegraphics[width=0.55\textwidth]{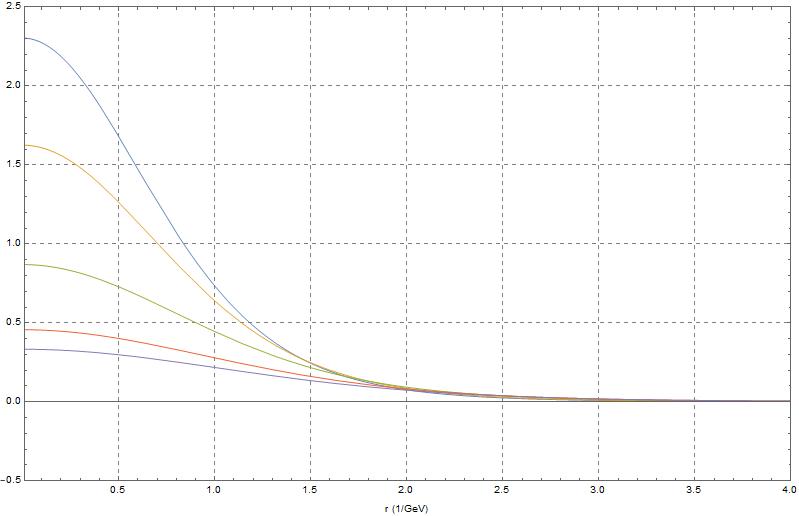}
    \caption{The HP-monopole Hamiltonian density $\mathcal{H}$  according to $(\ref{Hamiltonian Hooft})$ for various $\lambda_0$ values, with $\mu=\lambda$ and $M_W=1$ $\rm GeV$ fixed. Blue, orange, green, red and purple lines correspond to $\gamma = 1,0.5,0.1,0.0001,0$ respectively.}
\label{HamiltionianHooft}
\end{figure}

 This particular system has an analytic solution in the limit $(\lambda,\mu )\rightarrow (0,0)$, while keeping $M_W$ fixed:
\begin{align}\label{bpslimit}
    f(r)= \frac{M_W r}{\sinh(M_W r)}\,, \qquad 
    H(r)=(M_W r)\, \coth(M_W r)-1\,.
\end{align}
This continuous limit is known~\cite{Shnir:2005vvi} as the BPS limit~\cite{Bogomolny:1975de,Prasad:1975kr} and provides a lower bound of the monopole mass. In figure $\ref{HooftSolutions}$ we exhibit some solutions, approaching the BPS limit $(\lambda,\mu)\rightarrow (0,0)$ with $M_W = 1~{\rm GeV}$ fixed. This is done for $\lambda = \frac{\mu}{\rm GeV}$ for each solution. 
We observe that, the closer we are to the BPS limit, the slower the profile functions $H(r)/r$ and $f(r)$ approach their asymptotic behavior. In figure \ref{HamiltionianHooft} we demonstrate the behavior of the Hamiltonian density $(\ref{Hamiltonian Hooft})$ for various values of $\gamma$, as we approach BPS limit. We observe that, as $(\lambda,\mu
)\rightarrow (0,0)$, the Hamiltonian density becomes smaller and smaller.

\par We are now well equipped to study the mechanical properties of the HP monopole configuration via the behaviour of the energy momentum tensor, giving emphasis on the stability
of the solution, following the discussion in section \ref{InternalForceGeneralStructure}. 

\subsection{Internal Pressure Analysis}\label{InternalPressureAnalysisHooft}

To this end, we first calculate in this subsection the diagonal spatial elements of $(\ref{Energy-MomentumTensorHooft})$, which correspond to normal stress of monopole configuration, also known as pressure. The calculation is performed in spherical coordinates, in which the radial pressure is given by the following expression:
\begin{equation}\label{Radial Pressure Hooft}
    e^2\mathcal{P}_R (r)= e^2 T^{r r}  = \frac{1}{2r^4}(H'r-H)^2-\frac{H^2f^2}{r^4}-\frac{\gamma}{4}\left(\frac{H^2}{r^2}-M^2_W\right)^2-\frac{(1-f^2)^2}{2r^4}+\left(\frac{f'}{r}\right)^2
\end{equation}
Taking into account the results of the previous subsection, we observe that, in the limit $r\rightarrow 0$, the radial pressure asymptotes a constant finite value
\begin{align}
    e^2\mathcal{P}_R (r) &\xrightarrow{ r  \rightarrow 0} \left[-\frac{M^4_W \gamma}{4} + \frac{c^2_1}{2}+2(c_2)^2\right]+\left[2(c_2)^3+\frac{2(c_1)^2 M^2_W \gamma}{2}\right]r^2
    \nonumber \\ 
&\stackrel{ r \to 0}{\simeq} \left[-\frac{M^4_W \gamma}{4} + \frac{c^2_1}{2}+2(c_2)^2\right]
    \,,
\end{align}
whilst, as $r \rightarrow \infty$, the radial pressure asymptotes to:
\begin{equation}
    e^2\mathcal{P}_R (r) \xrightarrow{ r\rightarrow \infty}-\frac{1}{2r^4}\,.
\end{equation}
On the other hand, the polar $T^{\theta \theta}$  component of the pressure reads:
\begin{equation}\label{Polar Pressure Hooft}
e^2\mathcal{P}_{\Theta} (r)=e^2T^{\theta \theta}=-\frac{1}{2r^6}(H'r-H)^2-\frac{\gamma}{4r^2}\left(\frac{H^2}{r^2}-M^2_W\right)^2+\frac{(1-f^2)^2}{2r^6}\,,
\end{equation}
while the reader should recall from \eqref{EQcond2} that the azimuthal components (related to $T^{\psi \psi}$) are expressed in terms of $\mathcal P_\Theta$ as $\mathcal{P}_{\Theta}(r)=\sin^2(\theta)\mathcal{P}_{\Psi}(r,\theta)$, which implies that for fixed $\theta \ne 0$, the asymptotic behaviour (as $r \to 0, \, \rm{or} \,  \infty$) of the azimuthal pressure is similar to that of the polar pressure. 

In the limit $r\rightarrow 0$, the polar (and azimuthal) pressure behaves as:
\begin{equation}
    e^2 \mathcal{P}_{\Theta}(r) \xrightarrow{r \rightarrow 0} -\frac{1}{r^2}[\frac{(c_1)^2}{2}+\frac{M^4_W}{4}-2(c_2)^2]+[2(c_1)^2 c_2 +2(c_2)^3 +\frac{(c_1)^2}{2}M^2_W \gamma ]\,,
\end{equation}
whilst, as $ r\rightarrow \infty$, it  behaves as:
\begin{equation}
    e^2 \mathcal{P}_{\Theta}(r) \xrightarrow{ r\rightarrow \infty} \frac{1}{2r^6}\,.
\end{equation}

\begin{figure}[ht]
    \centering
    % Left figure
    \begin{subfigure}{0.35\textwidth}
        \centering
        \includegraphics[width=\linewidth]{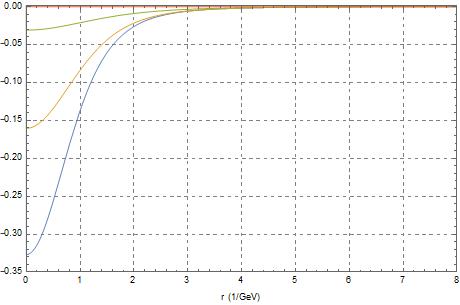} 
        \caption{Radial Pressure $(\ref{Radial Pressure Hooft})$ }
\end{subfigure}
\hspace{0.1\textwidth} 
\begin{subfigure}{0.35\textwidth}
    \centering
       \includegraphics[width=\linewidth]{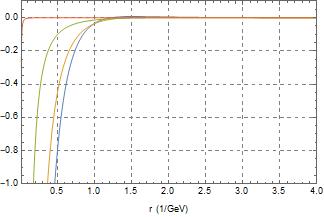} % replace with your figure file
        \caption {Polar Pressure $(\ref{Polar Pressure Hooft})$}
        \label{Polar Pressure}
    \end{subfigure}
\caption{Radial and Polar Pressure for various $\gamma$ values with $\mu (\frac{1}{\rm GeV})=\lambda$ and $M_W =1~\rm GeV$ fixed, in the HP magnetic monopole. Blue, orange, green, red and purple lines correspond to $\gamma = 1,0.5,0.1$ and $0.0001$  respectively. At the BPS limit both radial and polar pressure vanish.}\label{Polar and Radial Pressure Hooft}
\end{figure}
It is important to recall that in our case the non-diagonal elements of the EMT are zero, and therefore this tensor has the following structure in spherical coordinates $(t,r,\theta,\psi)$:
\begin{equation}
    T^{\mu \nu} = diag(\mathcal{H},\mathcal{P}_R,\mathcal{P}_{\Theta},\mathcal{P}_{\Psi} )\,
\end{equation}
whilst in the BPS limit, since Eqs.~$(\ref{Radial Pressure Hooft})$ and $(\ref{Polar Pressure Hooft})$ vanish, the EMT becomes in spherical coordinates:
\begin{equation}
    T^{\mu\nu} = diag(\mathcal{H},0,0,0)
\end{equation}
In figure $\ref{Polar and Radial Pressure Hooft}$, we present both radial and polar pressure for various values of $\gamma$ as we approach the BPS limit. The observed pattern is that, as we approach such a limit, both radial and polar pressure become smaller and smaller and vanish at $(\lambda,\mu)=(0,0)$. This suggests that the mechanical properties of the HP monopole solution are mainly dominated by the scalar potential $V(\phi)=\frac{\lambda}{4}(\vec{\phi}\cdot \vec{\phi}-\frac{\mu^2}{\lambda})^2$, and at the BPS limit the HP monopole configuration behaves as isotropic matter \cite{Panteleeva:2023aiz}. 

In the next subsection we proceed to discuss the mechanical stability of the configuration, according to the mechanical stability criteria, based on the behaviour of the total internal force, as discussed in section \ref{InternalForceGeneralStructure}.

\subsection{Internal Force Field}
\begin{figure}[ht]
    \centering
        \centering
        \includegraphics[width=0.4\textwidth]{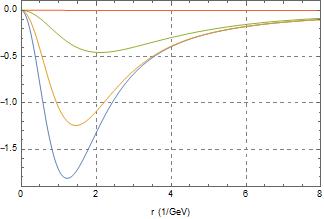} 
    \caption{Radial Force $(\ref{RadialForceHooft})$ for various $\gamma$ values with $\mu (\frac{1}{\rm GeV})=\lambda$ and $M_W =1~\rm GeV$ fixed, in the HP magnetic monopole. Blue, orange, green and red lines correspond to $\gamma = 1,0.5,0.1$ and $0.0001$   respectively.}\label{Polar and Radial Force Hooft}
\end{figure}

\begin{figure}[ht]
    \centering
    % Left figure
    \begin{subfigure}{0.3\textwidth}
        \centering
        \includegraphics[width=\linewidth]{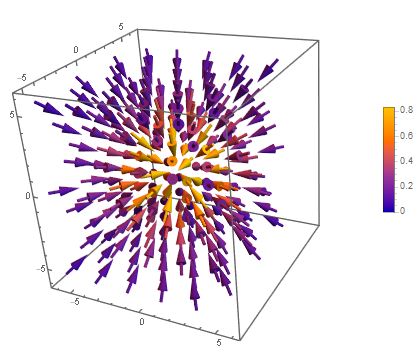} 
        \caption{Internal Force field for $\gamma = 1$ }
    \end{subfigure}
    \hspace{0.1\textwidth} 
    \begin{subfigure}{0.3\textwidth}
        \centering
        \includegraphics[width=\linewidth]{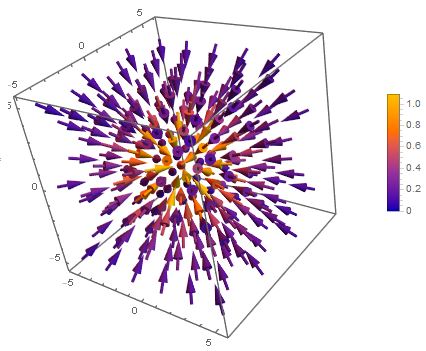} % replace with your figure file
        \caption {Internal Force field for $\gamma =0.5$}
        \label{Force Field Hooft 05}
    \end{subfigure}
    \caption{Internal Force Fields for $\gamma =1$ and $\gamma =0.5$, in the HP magnetic monopole.}\label{Force Hooft Field Hooft l=1,05}\label{FirstPairForceFieldHooft}
\end{figure}

\begin{figure}[ht]
    \centering
    % Left figure
    \begin{subfigure}{0.3\textwidth}
        \centering
        \includegraphics[width=\linewidth]{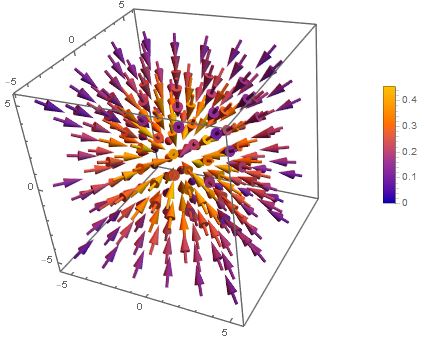} 
        \caption{Internal Force field for $\gamma = 0.1$ }
    \end{subfigure}
    \hspace{0.1\textwidth} 
    \begin{subfigure}{0.3\textwidth}
        \centering
        \includegraphics[width=\linewidth]{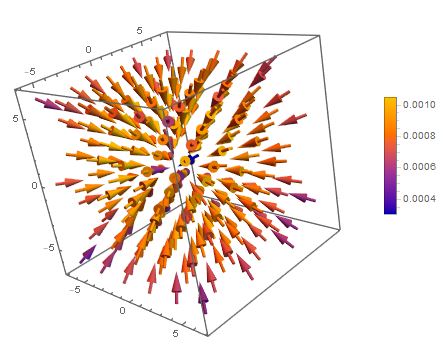} % replace with your figure file
        \caption {Internal Force field for $\gamma =0.0001$.}
        \label{Force Field Hooft 0001}
    \end{subfigure}
    \caption{Internal Force Fields for $\gamma=0.1$ and $\gamma=0.0001$, in the HP magnetic monopole.}\label{Force Hooft Field Hooft l=0.1,0.0001}\label{SecondPairForceFieldHooft}
\end{figure}

To this end, we commence our discussion by examining the polar and azimuthal components of the total forces, which, as discussed in section \ref{InternalForceGeneralStructure}, are given by:
\begin{equation}
    \mathcal{F}_{\theta}(\theta)=-\pi\sin(\theta)\lim_{r \to 0}[r^2 \mathcal{P}_R (r)]=0\,,
\end{equation}
and 
\begin{equation}
    \mathcal{F}_{\Psi}=\pi \int^{\infty}_0 dx'x^{'3}\mathcal{P}_{\Theta}(x') = -\frac{\pi}{2}\lim_{r \to 0}[r^2 \mathcal{P}_R (r)]=0\,.
\end{equation}
Such forces are zero, since $ e^2\mathcal{P}_R(r)$ has a finite value at the origin as we saw in section \ref{InternalPressureAnalysisHooft}.
Therefore, the internal force field is given by:
\begin{equation}\label{RadialForceHooft}
     e^2 \vec{\mathcal{F}}_{int}(r) = 4\pi r^2 e^2\mathcal{P}_R (r) \hat{r}
\end{equation}
In figure $\ref{Polar and Radial Force Hooft}$ we show the behaviour of $e^2F_R(r)$ for various values of $\gamma$.
 We observe that the radial force component becomes less and less negative as we approach the BPS limit, and vanishes for $\gamma =0$. In figures $\ref{FirstPairForceFieldHooft}$ and $\ref{SecondPairForceFieldHooft}$, we showcase the three-dimensional internal force of the monopole configuration.
\par Unfortunately, as we observe, the force is directed inwards, towards the centre of the monopole $r \to 0$, and therefore we are facing a similar situation as in \cite{Panteleeva:2023aiz}, in which the stability criteria are violated, which was to be expected by the equivalence between our criterion of the total force and the criterion  \eqref{strongercr}, in view of \eqref{pressurecomps} and \eqref{pressurecomps2}, as discussed in section \ref{InternalForceGeneralStructure}.
Nonetheless, the internal force field of the HP monopole configuration is radial and has a magnitude that becomes smaller and smaller as we approach the BPS limit, where the monopole configuration behaves as an isotropic matter, as mentioned above. 

\subsection{Energy Momentum Tensor Decomposition into short and long-range parts}\label{slEMTparts}
As discussed in ref.~\cite{Panteleeva:2023aiz} ({\it cf.} sections III D,E thereof) when one subtracts appropriately the long-range part of the EMT due to the massless photon of the U(1) $\subset $ SU(2), the mechanical stability criteria  can be satisfied for the remaining short-range part. A similar conclusion is expected to hold in our case as well. 
\par To this end, we introduce the smooth definition of the electromagnetic field tensor used in \cite{Panteleeva:2023aiz} (in our convention for the gauge covariant derivative \eqref{covgauge}):\footnote{There is no unique way to define the Abelian field strength of the Georgi-Glashow model~\cite{Georgi:1974sy} (or in fact any other unfied theory) used by `t Hooft~\cite{tHooft:1974kcl} (Polyakov~\cite{Polyakov:1974ek}) in the solution of the field-theoretic magnetic monopole. The only requirement is that this definition is SU(2) gauge invariant, and in the unitary gauge coincides with 
${\mathcal F}_{\mu\nu} = \partial_\mu A_\nu^3 - \partial_\nu A_\mu^3 $.
`t Hooft defined the Abelian field as 
\begin{align}\label{thfmn}
\mathcal{F}_{\mu\nu}= \hat{\phi}\cdot \vec{F}_{\mu\nu}+\frac{c_1}{e}\hat{\phi}\cdot(D_{\mu}\hat{\phi} \times D_{\nu}\hat{\phi})\,,
\end{align}
where $c_1$ cannot be fixed by the above requirements. In \cite{tHooft:1974kcl},
$\hat \phi^a \equiv \phi^a/|\vec \phi|$, $a=1,2,3$, and 
the choice $c_1=1$ has been made in order for the magnetic charge density to coincide with the topological charge density. For our purposes, the disadvantage of `t Hooft's definition of the magnetic charge density is its singular nature at the monopole's centre ($r \to 0$). Par contrast, other definitions of the Abelian field strength \eqref{thfmn} exist, which are smooth at the monopole centre, and have been proposed by
Faddeev~\cite{Fadeev:1975}, with $c_1=0$ and 
$\hat \phi^a \equiv \phi^a/u$, $a=1,2,3$,
and by Boulware {\it et al.}~\cite{Boulware:1976tv}, who used $\hat \phi^a \equiv \phi^a/|\vec \phi|$, $a=1,2,3$. In our approach, we follow the smooth definition \eqref{pantfmn} of \cite{Panteleeva:2023aiz}, which at long distances, where $SSB$ occurs \eqref{spherephi}, coincides with `t Hooft's definition \eqref{thfmn}.}
\begin{equation}\label{pantfmn}
    \mathcal{F}_{\mu\nu}= \hat{\phi}\cdot \vec{F}_{\mu\nu}+\frac{1}{e}\hat{\phi}\cdot(D_{\mu}\hat{\phi} \times D_{\nu}\hat{\phi})
\end{equation}
where $\hat{\phi}=\vec{\phi}/u$. Then the associated (traceless) energy momentum tensor of this long-range electromagnetic field in a background metric $g_{\mu\nu}$ is given by:
\begin{equation}\label{ememt}
  \mathcal{T}_{\mu\nu}=\mathcal{F}_{\mu a} \,\mathcal{F}^a_{\; \; \nu} + \frac{1}{4}g_{\mu\nu}\, \mathcal{F}^{ab}\mathcal{F}_{ab}\,.
\end{equation}
The spatial components of the magnetic field intensity in Cartesian coordinates is defined as  $\mathcal B^i = - \frac{1}{2} \, \epsilon^{ijk} \, \mathcal F_{jk}\,, \, i,j,k=1,2,3$  
For the magnetic monopole solution there is only a radial component of the magnetic field, $\mathcal B_r$, and  no electric field. In spherical polar coordinates, of interest to us here, 
$$ r=\sqrt{-x_ix^i}\,, \quad \theta = {\rm arccos}\Big(\frac{x^3}{r}\Big)\,, \quad  \psi = {\rm sgn}(x^2) \, {\rm arccos}\Big(\frac{x^1}{\sqrt{(x^1)^2 + (x^2)^2}}\Big)\,,$$
the electromagnetic stress tensor reads
\begin{equation}
   \mathcal{F}_{r\theta}=\mathcal{F}_{r \psi}=0 \,,
\end{equation}
and
\begin{equation}
   e \mathcal{F}_{\theta \psi}=-\sin(\theta)Q(r)\,,
\end{equation}
where the quantity
\begin{equation}
    Q(r)=\frac{Hf^2}{M_W r}(1-\frac{H^2}{M^2_W r^2})-\frac{H}{M_W r}
\end{equation}
is associated with the magnetic charge across the magnetic monopole configuration. Indeed, consider the magnetic field $\mathcal{B}^i=-\frac{1}{2}\epsilon^{ijk}\mathcal{F}_{jk}$:
\begin{equation}
    e\mathcal{B}^r = \frac{Q(r)}{r^2}
\end{equation}
Such formula give us the expected magnetic charge:
\begin{equation}
    q_m = \oint dS_r \mathcal{B}^r = \frac{4\pi}{e}
\end{equation}
Moreover, it is useful to note that the magnetic charge density $ \vec{\nabla} \cdot \vec{\mathcal{B}}=4\pi \rho_M (r) $ is given by:
\begin{equation}
     e \rho_{M} (r)=\frac{1}{4\pi r^2}\frac{d Q}{dr}\,,
\end{equation}
and the radial and polar components of EM tensor are given by:
\begin{equation}\label{LongRangeRadialPressureHooft}
    e^2\mathcal{P}^{LR}_R (r)= e^2 \mathcal{T}^{rr}=-\frac{Q^2(r)}{2r^4}\,,
\end{equation}
and 
\begin{equation}\label{LongRangePolarHooft}
    e^2\mathcal{P}^{LR}_{\Theta} (r)= e^2 \mathcal{T}^{\theta \theta}= \frac{Q^2(r)}{2r^6}\,.
\end{equation}
We subtract the terms \eqref{ememt} from the total stress energy tensor of the SU(2) magnetic monopole $T^{\mu \nu}$, \eqref{Energy-MomentumTensorHooft},
to obtain the short-range (SR) contributions:
\begin{align}\label{sremt}
T^{\rm SR \mu \nu} \equiv T^{\mu\nu} - \mathcal{T}^{\mu\nu}\,.
\end{align}
We obtain then the short-range radial pressure:
\begin{equation}\label{SRRadial Pressure Hooft}    
   e^2\mathcal{P}^{SR}_R (r)=e^2 T^{SR rr} = \frac{1}{2r^4}(H'r-H)^2-\frac{H^2f^2}{r^4}-\frac{\gamma}{4}\left(\frac{H^2}{r^2}-M^2_W\right)^2-\frac{(1-f^2)^2}{2r^4}+ \left(\frac{f'}{r}\right)^2 +\frac{Q^2(r)}{2r^4}\,,
\end{equation}
and 
\begin{equation}
e^2\mathcal{P}^{SR}_{\Theta}(r)=-\frac{1}{2r^6}(H'r-H)^2-\frac{\gamma}{4r^2}\left(\frac{H^2}{r^2}-M^2_W\right)^2+\frac{(1-f^2)^2}{2r^6}-\frac{Q^2(r)}{2r^6}\,.
\end{equation}
From this we shall now evaluate the internal force and pressure, following our previous analysis, and examine the mechanical stability criteria for the short-range contributions, in similar spirit to the study in \cite{Panteleeva:2023aiz}.

Defining the short-range force components as
\begin{equation}\label{srforce}
    dF^{SRi}=T^{SR ij}dA_{j}
\end{equation}
we obtain for the short-range radial force:
\begin{align}\label{radialDR}
    e^2 \mathcal{F}^{SR}_R(r) = 4\pi r^2 e^2\mathcal{P}^{SR}_R (r)\,.
\end{align}
\par 
Separation of the EMT into short and long range parts suggests that the equilibrium condition must be modified as:
\begin{equation}\label{SRequilibruimEqHooft}
    e^2\frac{d \mathcal{P}^{SR}_R}{dr}+\frac{2}{r}(e^2\mathcal{P}^{SR}_R-r^2 e^2\mathcal{P}^{SR}_{\Theta})=e^2\frac{\mathcal{P}_{ext}(r)}{r} \,,
\end{equation}
and 
\begin{equation}\label{LRequilibruimEqHooft}
    e^2\frac{d \mathcal{P}^{LR}_R}{dr}+\frac{2}{r}(e^2\mathcal{P}^{LR}_R-r^2 e^2\mathcal{P}^{LR}_{\Theta})=-e^2\frac{\mathcal{P}_{ext}(r)}{r} \,,
\end{equation}
where we have introduced the external pressure:
\begin{equation}
    e^2\mathcal{P}_{ext}(r) =\frac{4\pi \rho_M (r) Q(r)}{r}\,.
\end{equation}
This is calculated via $(\ref{LRequilibruimEqHooft})$, by using $(\ref{LongRangeRadialPressureHooft})$ and $(\ref{LongRangePolarHooft})$. We integrate out equation $(\ref{SRequilibruimEqHooft})$, $(\ref{LRequilibruimEqHooft})$ and we obtain:
\begin{equation}
    e^2 \mathcal{P}^{SR}_R (r)+e^2P_{ext}(r) = e^2\Sigma^{SR}(r)\,,
\qquad 
    e^2 \mathcal{P}^{LR}_R (r)-e^2P_{ext}(r) = e^2\Sigma^{LR}(r)\,,
\end{equation}
where, 
{\small \begin{equation}
e^2P_{ext}(r) = \frac{1}{r^2}\int^{\infty}_r dr'r'e^2\mathcal{P}_{ext}(r')\,,
\quad
    \Sigma^{SR}(r)=-\frac{2}{r^2}\int^{\infty}_r  dr'r^{'3}\mathcal{P}^{SR}_{\Theta}(r')\,, \quad 
    \Sigma^{LR}(r)=-\frac{2}{r^2}\int^{\infty}_r  dr'r^{'3}\mathcal{P}^{LR}_{\Theta}(r')\,.
\end{equation}}
\par Such a pressure gives rise to a Coulomb force associated with the interaction of a magnetically charged sphere $Q(r)$ acting on the magnetic charge density $\rho_M (r)$. The short-range equation $(\ref{SRequilibruimEqHooft})$ describes the balance between the ``short-range stress”, which pulls the monopole inwards, towards its center, and the
repulsive magnetic ``Coulomb force”, which pushes the monopole outwards. On the other hand, the long-range equation $(\ref{LRequilibruimEqHooft})$ describes magnetostatic equilibrium between the ``Coulomb stress” pushing the monopole
outwards and the magnetic ``Coulomb force” pulling the monopole inwards, towards the center.
\par
\begin{figure}[ht]
    \centering
        \centering
        \includegraphics[width=0.5\textwidth]{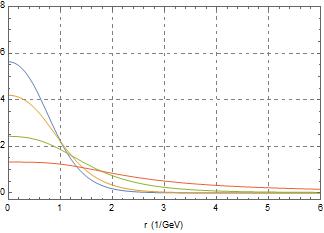} 
    \caption{Total short range radial force $(\ref{Total Radial Force})$ for various $\gamma$ values with $\mu (\frac{1}{\rm GeV})=\lambda$ and $M_W =1~\rm GeV$ fixed, in the HP magnetic monopole. Blue, orange, green and red lines correspond to $\gamma = 1,0.5,0.1$ and $0.0001$   respectively.}\label{TotalRadialForceGraph}
\end{figure}

 Then the total radial force is given by:
\begin{equation}\label{Total Radial Force}
    e^2\vec{\mathcal{F}}^{SR}_{total}(r)=e^2\vec{\mathcal{F}}^{SR}_R(r)+e^2\vec{\mathcal{F}}_{ext}(r)=4\pi r^2( e^2 \mathcal{P}^{SR}_R (r)+e^2P_{ext}(r))\hat{r}
\end{equation}
where,
\begin{equation}
    \vec{\mathcal{F}}_{ext}(r)=4\pi r^2 P_{ext} (r)\hat{r}\,.
\end{equation}
In figure $\ref{TotalRadialForceGraph}$, we showcase the total short-range radial force $(\ref{Total Radial Force})$.
Notice that such a force field is positive at each point, therefore local stability condition $F^{SR}_{total}(r)\geq 0$ is fulfilled. Because of its local character such a condition is a strong one, and suggests that the HP monopole configuration is mechanically stable.
\par Finally, the reader should note that the EMT decomposition to short and long range contributions gives rise to 
short-range polar and azimuthal forces, respectively, 
\begin{equation}\label{ShortRangePolarForceHooft}
    \mathcal{F}^{SR}_{\Theta}(\theta)=2\pi \sin(\theta)\int^{\infty}_0dx'x^{'3}\mathcal{P}^{SR}_{\Theta}(x')\,,
\end{equation}
and
\begin{figure}[ht]
    \centering
        \centering
        \includegraphics[width=0.5\textwidth]{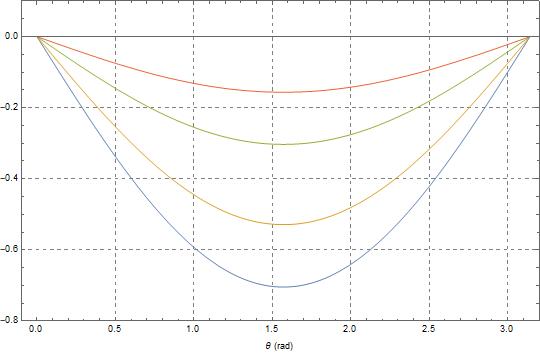} 
    \caption{Short range polar force $(\ref{ShortRangePolarForceHooft})$ for various $\gamma$ values with $\mu (\frac{1}{\rm GeV})=\lambda$ and $M_W =1~\rm GeV$ fixed, in the HP magnetic monopole. Blue, orange, green and red lines correspond to $\gamma = 1,0.5,0.1$ and $0.0001$   respectively.}\label{ShortRangePolarForceGraph}
\end{figure}
\begin{equation}\label{ShortRangeAzimuthalForce}
 \mathcal{F}^{SR}_{\Psi}=\pi \int^{\infty}_0dx'x^{'3}\mathcal{P}^{SR}_{\Theta}(x')\,.
\end{equation}
Moreover, we present the values of $\mathcal{F}^{SR}_{\Psi}$ in table \ref{TableSRazimuthalForce}.
\begin{table}[ht]
\caption{Short Range Azimuthal Force $(\ref{ShortRangeAzimuthalForce})$ for various values of $\gamma$.}
\centering
  \begin{tabular}{||c|c||}
  \hline
    $\gamma$  & $e^2\mathcal{F}^{SR}_{\Psi}$ \\
    \hline
    1                     & -1.41 \\
    \hline
    0.5                     & -1.06 \\
    \hline
    0.1                     & -0.61\\
    \hline
    0.0001                     & -0.31\\
    \hline
  \end{tabular}
  \label{TableSRazimuthalForce}
\end{table}
We observe that both short range polar and azimuthal force components are well behaved for the HP monopole.   

After this study we proceed now to examine the aforementioned mechanical-stability criteria for the case of the CM electroweak monopole.
\subsection{Energy Conditions}
In section \ref{Energy Conditions} we have introduced the weak, strong and dominant energy conditions. These are given by:
\begin{equation}
    WEC=
    \begin{cases}
  \mathcal{H}\geq 0 \,,\quad   {\rm timelike-vectors} \\
  \mathcal{H}+\mathcal{P}_R \geq 0\,, \quad {\rm null-vectors}
\end{cases}
\end{equation}
\begin{equation}
    SEC=
    \begin{cases}
  \mathcal{H}+\mathcal{P}_R+2r^2\mathcal{P}_{\Theta}\geq 0\,, \quad   {\rm timelike-vectors} \\
  \mathcal{H}+\mathcal{P}_R \geq 0\,, \quad {\rm null-vectors}
\end{cases}
\end{equation}
\begin{equation}
    DEC=
    \begin{cases}
  (\mathcal{H})^2\geq 0\,, \quad  {\rm  timelike-vectors }\\
  (\mathcal{H})^2-(\mathcal{P}_R)^2 \geq 0\,, \quad {\rm null-vectors}
\end{cases}
\end{equation}
By using numerical results from this section, we find that all energy conditions are satisfied by the monopole configuration in $SU(2)$ Georgi-Glashow model. 
\section{The Cho-Maison  Electroweak Monopole and its extensions}\label{Electroweak Monopole Analysis}

In this section we first review the monopole solutions in the electroweak model of Cho and Maison~\cite{Cho:1996qd}, and its variants, which lead to finite energy~\cite{Cho:2013vba,Ellis:2016glu}. After that, we are going to study the model's mechanical properties and compare them with those of the well established HP monopole, discussed in the previous section.

\subsection{Monopole configurations in the electroweak model}
\label{sec:CMclassicalmodel}
\subsubsection{The Model}
The CM monopole is a solution of the Electroweak model without fermionic matter (lepton and quark) fields. The model is based on the $SU_L(2)\times U_Y(1)$ gauge symmetry, with 
a field content consisting of 
 the $SU_L(2)$ non-Abelian gauge field $\vec{A}_{\mu}$, the $U_Y(1)$ Abelian (hypercharge) gauge field $B_{\mu}$ and a Higgs complex doublet field $\phi$. The corresponding Lagrangian density reads:
\begin{equation}\label{lagew}
    \mathcal{L}=-\frac{1}{4}\vec{F}^{\mu\nu} \cdot \vec{F}_{\mu\nu}-\frac{1}{4}G_{\mu\nu}G^{\mu\nu}+D_{\mu}\phi^{\dagger}D^{\mu}\phi-\frac{\lambda}{2}\left(\phi^{\dagger}\phi -\frac{\mu^2}{\lambda}\right)^2\,,
\end{equation}
where
\begin{equation}
    G_{\mu\nu}=\partial_{\mu}B_{\nu}-\partial_{\nu}B_{\mu}\,,
\end{equation}
is the Abelian $U_Y(1)$ hypercharge ``Maxwell-like'' tensor, 
while the quantity 
\begin{equation}
\vec{F}_{\mu\nu}=\partial_{\mu}\vec{A}_{\nu}-\partial_{\nu}\vec{A}_{\mu}+g\vec{A}_{\mu}\times\vec{A}_{\nu}\,.
\end{equation}
denotes the non-Abelian $SU_L(2)$ field strength. The covariant derivative of the Higgs field $\phi$ is given is by:
\begin{equation}
    D_{\mu}\phi= \partial_{\mu}\phi-\frac{ig}{2}(\vec{\sigma} \cdot \vec{A}_{\mu})\phi-\frac{ig'}{2}B_{\mu}\phi\,.
\end{equation}
As well known, after spontaneous symmetry breaking $(SSB)$ $SU_L(2) \times U_Y(1)\rightarrow U_{em}(1)$, where $U_{em}(1)$
is the Abelian subgroup of Electromagnetism, 
the mass spectrum of this theory consists of four massive bosons $W^{\pm}$, $Z$, $H$ (Higgs) and one massless boson $\gamma$ (photon). The masses are:
\begin{equation}
    M_W = \frac{gu}{2} \qquad M_Z = \frac{\sqrt{g^2+g^{'2}}}{2}u \qquad M_H = \sqrt{2\lambda}u \,,
\end{equation}
where we have denoted the vacuum expectation value of the scalar doublet field $\phi$ as 
\begin{align}\label{phivev}
\langle \phi \rangle =  
  \begin{pmatrix}
  0 \\ u
  \end{pmatrix}\,,
\end{align}
with 
$u=|\mu|/\sqrt{\lambda}$, with $\mu \in \mathbb R$, $\lambda > 0$. 

\subsubsection{Topological Argument and Cho-Maison Magnetic-monopole Solution}
\par A field theory of a particular symmetry  group $G$ subject to $SSB$, has a vacuum manifold described by $\mathcal{M}=G/h $, where $h$ is the unbroken subgroup of $G$. 
According to \cite{Kibble:1976sj}, in order for a field theory to have a monopole configuration solution, the second homotopy group $\pi_2 (\mathcal{M})$ must be non-trivial. As already mentioned, this can be traced back to the fact that a non-trivial $\pi_2 (\mathcal{M})$ provide us with a Brouwer degree of the map $\hat{\phi}:$ $S^2 \rightarrow S^2$, which is associated with a topological magnetic current  \cite{Shnir:2005vvi}, \cite{Ryder:1985wq}, 
which was the case of the HP monopole~\cite{tHooft:1974kcl,Polyakov:1974ek}.

In the case of the electroweak model we have $\mathcal{M}=\frac{SU_L(2) \times U_Y(1)}{U_{em}(1)}\thicksim S^3$. That $\mathcal{M}\thicksim S^3$, is also understood from the fact that $\phi^{\dagger}\phi=\sum\limits^4_{i=1}|\phi_i|^2$, which is the equation of a three sphere $S^3$. Such a theory has a trivial homotopy group  $\pi_2 (\mathcal{M}) = \mathcal{I}$. 
\par From the above argumentation it follows that the topology of the Electroweak model does not allow for monopole configurations. 
Nonetheless, according to Cho and Maison~\cite{Cho:1996qd} this is not necessarily the case, because there is a {\it hidden} non-trivial topology in the Higgs sector. Indeed, these authors argued that the construction of the following map from $\phi$ to $\hat{\phi}$:
\begin{equation}
    \hat{\phi} = \xi^{\dagger} \vec{\sigma} \xi\,,
\end{equation}
where $\vec{\sigma} = (\sigma^1,\sigma^2,\sigma^3)$ are the Pauli matrices, $\xi = \phi/ |\phi|$ and $|\phi|=\sqrt{\phi^{\dagger}\phi}$, so that $\hat{\phi}$ transforms as a triplet under $G$, and therefore it has the topology of a sphere $S^2$, allows one to define a non trivial homotopy group, given that, as we shall discuss below ({\it cf.} Eq.~\eqref{charts}), the scalar field $\hat \phi$ is now viewed~\cite{Cho:1996qd} as 
a $CP^1$ field, and it is well known that~\cite{Eguchi:1980jx} 
\begin{align}\label{cp1homotopy}
    \pi_2(CP^1) = \mathbb Z\,.
\end{align} 
This supports the idea of the existence of monopole solutions in the Electroweak theory. 
Further support to this is provided by the fact that the monopole configuration, as a soliton solution, must have a boundary which corresponds to the minimum of the energy functional  \cite{Ryder:1985wq}. This holds when:
\begin{equation}\label{kineticHigssBound}
D_{\mu}\hat{\phi} = \partial_{\mu}\hat{\phi}+g\vec{A}_{\mu} \times \hat{\phi}  \xrightarrow{x \rightarrow \infty} 0\,.
\end{equation}
The reader should note that $D_{\mu}\hat{\phi}$ does not appear in the kinetic term of the electroweak Lagrangian \eqref{lagew}. However $D_{\mu}\xi$ does, and therefore $(\ref{kineticHigssBound})$ can be seen as a direct consequence of $D_{\mu}\xi$ being zero at the boundary. Also, it should be remarked that the structure of the covariant derivative in $(\ref{kineticHigssBound})$ originates form the fact $\hat{\phi}$ transforms as a triplet under the Lie gauge group $G$. 
\par 
Following Nambu~\cite{Nambu:1977ag},
we first proceed to the solution  of $(\ref{kineticHigssBound})$,
\begin{equation}
\vec{A}_{\mu}\xrightarrow{x \rightarrow \infty } -\frac{1}{g}\hat{\phi}\times \partial_{\mu}\hat{\phi}\,,
\end{equation}
which, for finite distances, leads to the Cho-Maison monopole solution~\cite{Cho:1996qd}, which, in spherical polar coordinates $(r,\theta,\psi)$, is expressed as:
\begin{equation}\label{Non Abelian Configuration}
    \vec{A}_{\mu}= \frac{1}{g}(f(r)-1)\hat{\phi}\times \partial_{\mu}\hat{\phi}\,,
\end{equation}
and
\begin{equation}\label{Abelian Configuration}
    B_{\mu}= -\frac{1}{g'}(1-cos(\theta))\partial_{\mu}\psi \,,
\end{equation}
with the scalar doublet in the so-called {\it radial gauge} being given by~\cite{Cho:1996qd} 
\begin{equation}\label{Scalar Configuration}
    \phi = \frac{1}{\sqrt{2}}\rho(r) \xi \,,
\qquad 
    \xi =i \begin{pmatrix}
       \sin(\theta /2)\,
       e^{-i\psi} \\
       -\cos(\theta /2)\,.
    \end{pmatrix}\,,
\end{equation}
and
\begin{equation}\label{RadialGauge}
    \hat{\phi} = \xi^{\dagger} \vec{\sigma} \xi = - \hat{r}\,.
\end{equation}
\par 
The Cho-Maison solution has been derived in a particular gauge, which is called radial gauge, since the scalar  $\hat{\phi}$ lies along the radial direction. Notice should be taken of the fact that this monopole solution has, apart from the npn-Abelian $SU_L(2)$, an additional Abelian gauge structure, which stems from the hypercharge boson field $B_{\mu}$ \eqref{Abelian Configuration}. Thus it is a hybrid between the 't Hooft-Polyakov and Dirac monopoles. Furthermore, the profile functions $f(r)$ and $\rho(r)$ must satisfy the boundary conditions:
\begin{equation}\label{BoundaryCondition}
    f(0)=1\,, \quad f(\infty)=0 \,,\qquad \qquad \rho(0)=0\,, \quad \rho(\infty)= \rho_0 =\sqrt{2}\,u\,,
\end{equation}
since the configuration must satisfy $(\ref{kineticHigssBound})$ at spatial infinity and be well defined at the origin. 
\par The Abelian structure is essential for the solution. To understand this, let one consider the analysis of the Abelian monopole field given by Wu-Yang \cite{Wu:1975es}. In this well-known description of a monopole configuration, one splits the space of the spherically symmetric magnetic field into two regions, call them the North (N) and South (S) hemisphere, for concreteness. At each region the Abelian gauge field is well defined:
\begin{align}\label{Bpsi}
B_{\psi} =
\begin{cases}
B^N_{\psi} = -\frac{1}{g'}\frac{1-\cos\theta}{r\, \sin\theta} \hat{\psi} & \text{,} \quad  0 \leq \theta \leq\frac{\pi}{2}+\frac{\epsilon}{2}   \\
  B^S_{\psi} = -\frac{1}{g'}\frac{1+\cos\theta}{r\, \sin\theta} \hat{\psi} & \text{,} \quad -\frac{\epsilon}{2}+\frac{\pi}{2} \leq \theta \leq \pi\,,
\end{cases}
\end{align}
Where $\epsilon \rightarrow 0^+$.
These two configurations of the hypercharge field $B_\mu$ are related via a $U_Y(1)$ transformation:
\begin{equation}
    B^N_{\psi}=B^S_{\psi}-\frac{i}{g'}U^{-1}\vec{\nabla}U \qquad U= e^{2i\psi}
\end{equation}     
The scalar doublet $\phi$ also admits two different descriptions at the north and south hemispheres, respectively:
\begin{equation}
    \phi^N =\frac{\rho}{\sqrt{2}}i\begin{pmatrix}
        \sin(\theta/2)e^{-i\psi} \\
        -\cos(\theta/2)
    \end{pmatrix},\quad 0 \leq \theta \leq\frac{\pi}{2}+\frac{\epsilon}{2}\,, 
\end{equation}    
and
\begin{equation}    
    \phi^S = \frac{\rho}{\sqrt{2}}i\begin{pmatrix}
        \sin(\theta/2)e^{i\psi} \\
        -\cos(\theta/2)e^{2i\psi}
    \end{pmatrix}, \quad -\frac{\epsilon}{2}+\frac{\pi}{2} \leq \theta \leq \pi \,.
\end{equation}
This particular structure of the scalar doublet allows one to define two charts for the complex plane $\mathbb{C}$:
\begin{align}\label{charts}
    z_1 &= e^{\dagger}_1  [\frac{1}{i\frac{\rho}{\sqrt{2}}\cos(\theta/2)} \phi^N   ]=\tan\frac{\theta}{2}e^{-i\psi}=\frac{x-iy}{1-z} \,, \quad 
{\rm and} \nonumber \\
    z_2 & = e^{\dagger}_2  [\frac{1}{i\frac{\rho}{\sqrt{2}}\, \sin(\theta/2)e^{2i\psi} }\phi^S ] = \cot\frac{\theta}{2}e^{i\psi}=\frac{x+iy}{1+z} \,, \nonumber \\
&{\rm where} \quad 
    e_1 =\begin{pmatrix}
        1 \\
        0
    \end{pmatrix} \,, \qquad   e_2 =\begin{pmatrix}
        0 \\
        -1
    \end{pmatrix}\,.
\end{align}
These are in fact stereographic projections. The inverse maps of $z_1(x,y,z)$ and $z_2(x,y,z)$ map the whole complex plane $\mathbb{C}$ to the unit sphere $S^2$. From this we can therefore see that the Higgs doublet has a $CP^1$ structure, a fact which is possible from the introduction of the Abelian monopole field $B_{\mu}$. By viewing the Higgs doublet as a $CP^1$ is what allow us to have a topology $CP^1 \thicksim S^2$, that leads to the non-trivial second homotopy group \eqref{cp1homotopy}, as mentioned previously.
\\

\subsubsection{Energy Functional and Equations of Motion}
\par The EMT of such a model can be calculated via
\eqref{EMT_General}:
\begin{equation}\label{Energy-Momentum-Tensor-Cho}
    T^{\mu \nu}= -g^{\mu\nu}\mathcal{L} +2D^{\mu}\phi^{\dagger}D^{\nu}\phi -\vec{F}^{\mu \sigma}\cdot \vec{F}^{\nu}_{ \; \; \sigma } -G^{\mu\sigma} G^{\nu}_{\; \; \sigma }
\end{equation}
\par By substituting the field configurations $(\ref{Non Abelian Configuration})$, $(\ref{Abelian Configuration})$, $(\ref{Scalar Configuration})$ on the purely temporal ($tt$) component of $(\ref{Energy-Momentum-Tensor-Cho})$, we obtain the Hamiltonian density of the configuration in the radial gauge:
\begin{equation}
    \mathcal{H}=T^{t t}=-\mathcal{L} =\frac{1}{2(g')^2r^4}+\frac{(f')^2}{g^2r^2}+\frac{(\rho')^2}{2} +\frac{(f^2-1)^2}{2g^2r^4}+\frac{f^2\rho^2}{4r^2}+\frac{\lambda}{2}\left(\frac{\rho^2}{2}-u^2\right)^2\,,
\end{equation}
implying that the energy functional is given then by:
\begin{align}\label{EnergyFuncChoMaison}   
    E &= \frac{4\pi}{g^2}M_W\int^{\infty}_0 dx \left[\frac{g^2}{2(g')^2x^2}\right]+\frac{4\pi}{g^2}M_W\int^{\infty}_0 dx [(f')^2+4x^2(H')^2
    \nonumber \\
    &+\frac{(f^2-1)^2}{2x^2}+ 2f^2H^2+2\gamma x^2(H^2-1)^2]  \,,
\end{align}
where $x=M_W r$, $H(x)=\rho(x)/\rho_0$ and $\gamma =4\lambda/g^2$.
Since the configuration is static, we apply, as in the case of the HP monopole discussed in section \ref{HPReview}, the Euler-Lagrange equations on the expression \eqref{EnergyFuncChoMaison}
for the energy functional, and obtain the following system of differential equations:
\begin{equation}\label{EquationScalarProfile}
    f''=\frac{f(f^2-1)}{x^2}+2fH^2\,,
\end{equation}
and 
\begin{equation}\label{EquationGaugeProfile}
    H''=-\frac{2}{x}H'+\frac{f^2H}{2x^2}+\gamma H(H^2-1)\,.
\end{equation}
Near $x=0$ we write the gauge profile function $f(x)$ as:
\begin{equation}
    f(x)=1+\Delta_1 (x) \quad |\Delta_1|<<1\,.
\end{equation}
The differential equations $(\ref{EquationScalarProfile})$ and $(\ref{EquationGaugeProfile})$ become:
\begin{equation}\label{DiffScalarProfilesSmallX}
    H^{''}+\frac{2}{x}H^{'} +(\gamma -\frac{1}{2x^2})H=0\,,
\end{equation}
and
\begin{equation}\label{DiffGaugeProfilesSmallX}
    \Delta^{''}_1-\frac{2 \Delta_1}{x^2} = 2H^2\,,
\end{equation}
respectively. Eq.~$(\ref{DiffGaugeProfilesSmallX})$ is solved by means of Bessel function of the first kind, $J_{\sqrt{3}/2}(\sqrt{\gamma}x)$. On taking $H^2 <<1$, then, we obtain from  $(\ref{DiffGaugeProfilesSmallX})$ that $\Delta_1 \thicksim x^2 $. Considering $H^2$ as an inhomogeneous part of $(\ref{DiffGaugeProfilesSmallX})$, one obtains a small correction to $\Delta_1 \thicksim x^2 $. In particular, for $\gamma = 1.21382$ (which corresponds to the standard model bare parameters $\lambda=0.129$ and $g=0.652$~\cite{ParticleDataGroup:2024cfk})
we obtain:
\begin{equation}
    \Delta_1 (x) = -c_2 x^2 +x^{\sqrt{3}}\, [0.1309\, x +\mathcal{O}(x^3)]\,.
\end{equation}
For small $x$ such extra correction terms have a minor contribution, for instance, for $x= 0.01 $ one obtains a correction of order $10^{-7}$. Ignoring such inhomogeneous contributions we have the following asymptotic behaviour as $x \to 0$
\begin{equation}\label{EWHiggsProfileNearZero}
    H(x) \xrightarrow{ x \rightarrow 0} \frac{c_1}{\sqrt{x\sqrt{\gamma}}}J_{\frac{\sqrt{3}}{2}}(x\sqrt{\gamma})\,,
\end{equation}
and \begin{equation}\label{EWGaugeProfileNearZero}
    f(x) \xrightarrow{x \rightarrow 0}1- c_2 x^2\,.
\end{equation}
The first derivative of the Higgs profile function near $x=0$ can be calculated from $(\ref{EWHiggsProfileNearZero})$, and reads: 
\begin{equation}
    H'(x) \xrightarrow{x \rightarrow 0} \frac{c_1 \sqrt{\gamma} \left(J_{-1+\frac{\sqrt{3}}{2}}\left(\sqrt{\gamma} x\right)-J_{1+\frac{\sqrt{3}}{2}}\left(\sqrt{\gamma} x\right)\right)}{2 \sqrt{\sqrt{\gamma} x}}-\frac{c_1 \sqrt{\gamma}
   J_{\frac{\sqrt{3}}{2}}\left(\sqrt{\gamma} x\right)}{2 \left(\sqrt{\gamma} x\right)^{3/2}}
\end{equation}
For the standard model value~\cite{ParticleDataGroup:2024cfk} $\gamma=1.21382$, we obtain:
\begin{equation}
     H'(x) \xrightarrow{x \rightarrow 0} x^{\frac{\sqrt{3}}{2}} \left(-\frac{0.554889}{x^{3/2}}+\frac{1.83129}{x}\right)\,.
\end{equation}
Thus, $H'(x)$ is singular at $x=0$. As for the behavior at $x\rightarrow \infty$ we 
write
\begin{equation}
    H(x) = 1+\Delta_2(x) \,,\qquad |\Delta_2|<<1\,,
\end{equation} 
while $f(x)$ approaches zero in this limit. Then $(\ref{EquationScalarProfile})$
and $(\ref{EquationGaugeProfile})$ become, repsectively:
\begin{equation}
    \Delta^{''}_2 + \frac{2}{x}\Delta^{'}_2-2\gamma \Delta_2 =0\,, \qquad 
    f^{''}-2f =0\,,
\end{equation}
which can be solved, leading to the following asymptotic behavior, as $x\rightarrow \infty$:
\begin{equation}
H(x) \xrightarrow{ x\rightarrow \infty}1-c_4 \, \frac{\exp(-\sqrt{2\gamma}x)}{x}\,, \qquad 
    f(x) \xrightarrow{ x \rightarrow \infty} \, c_3 \, \exp[-\sqrt{2}x]
    \,.
\end{equation}
In table \ref{AsymptoticConstantsEW} we give the values of the free constant parameters $c_i$, $i=1, \dots, 4$, for $\gamma=1.21382$. We observe that the Abelian gauge field contribution produces a singular term in $(\ref{EnergyFuncChoMaison})$ proportional to $1/x^2$. Therefore, the energy of the configuration is singular and thus non-physical, as it cannot be produced in physical situations, {\it e.g.} through collisions of particles of finite energy.

\begin{table}
\caption{Constants $c_1,$ $c_2,$ $c_3$ and $c_4$ for the standard model value $\gamma=1.21382$.}
 \begin{center}
\begin{tabular}{|| c || c | c | c |  c |} 
 \hline
 $\gamma$ & $c_1$ & $c_2$ & $c_3$ & $c_4$   \\ [0.5ex] 
 \hline\hline
 1.21382  & 1.85576 & 1.50000 & 2.41000 & 0.12000  \\ 
 \hline
\end{tabular}
\end{center}
\label{AsymptoticConstantsEW}
\end{table}

Variants of the CM model that lead to {\it finite energy} have been considered in 
\cite{Cho:2013vba,Ellis:2016glu}, in a phenomenological approach, in which one considers extensions of the electroweak Lagrangian \eqref{lagew} involving non-minimal couplings between the Higgs and hypercharge sectors, specifically a Higgs-field-dependent  coefficient of the kinetic term  of the hypercharge-gauge field.
The respective models are described briefly in subsection \ref{sec:cmfinite}. 
On the other hand, there are also extensions of the CM model which have a semi-microscopic origin, {\it e.g.} in string theory~\cite{Arunasalam:2017eyu,Mavromatos:2018kcd,Ellis:2022uxv}. The latter entail Born-Infeld-type extensions of the hypercharge sector of the CM Lagrangian \eqref{lagew}, and are discussed in subsection \ref{BI}.\footnote{This is a simplified extension of the string-inspired Born-Infeld version of the CM monopole, given that in principle, in the context of string theory, higher-derivative corrections in the non-Abelian weak sector should be considered. However, this goes beyond to the scope of the current article.} 

\subsection{Finite-energy CM monopole with non-minimal couplings between the Higgs snd hypercharge sectors}\label{sec:cmfinite}

The first type of finite-energy extensions of the CM model, have been proposed initially in \cite{Cho:2013vba}, and subsequently from a more realistic point of view, in the sense of examining conditions under which these extensions are consistent with current(LHC)  phenomenology, in \cite{Ellis:2016glu}. The pertinent modifications consist of non-linear couplings between the Higgs and hypercharge sectors, and specifically, a non-trivial Higgs function that multiplies the hypercharge kinetic term in the extended electroweak model \eqref{lagew}:
\begin{equation}\label{lagew2}
    \mathcal{L}=-\frac{1}{4}\vec{F}^{\mu\nu} \cdot \vec{F}_{\mu\nu}-\frac{1}{4} \, h\Big(\frac{\phi^\dagger \phi}{u^2}\Big)\, G_{\mu\nu}G^{\mu\nu}+D_{\mu}\phi^{\dagger}D^{\mu}\phi-\frac{\lambda}{2}\left(\phi^{\dagger}\phi -\frac{\mu^2}{\lambda}\right)^2\,.
\end{equation}
In \cite{Cho:2013vba}, Cho, Kim and Yoon (CKY), 
using the scalar field representation \eqref{Scalar Configuration} in the radial gauge, proposed the simplest possible function $h(\phi^\dagger \phi)$ that guaranteed a finite energy functional for the monopole, which had the form:
\begin{align}\label{cky}
  h\Big(\frac{\phi^\dagger\phi}{u^2}\Big)_{\rm CKY} =  \Big(\frac{\rho}{\rho_0}\Big)^n\,, \qquad \mathbb Z \ni n \ge 8\, , 
\end{align}
where $\rho_0 = \sqrt{2}\, u$ has been defined in \ref{BoundaryCondition}. The lower bound on the integer $n$ in \eqref{cky} stems from the requirement of the finiteness of the energy functional of this modified CM electroweak model. 

In \cite{Ellis:2016glu}, it has been remarked that a CM-model extension of the form \eqref{cky} is not compatible with the Higgs phenomenology at the large Hadron collider (LHC), in particular the (observed) decays of the Higgs field to two photons. Within the framework of the so-called Standard Model Effective Field Theory (SMEFT)~\cite{sanz1,sanz2}, some phenomenologically acceptable functions $h\Big(\frac{\phi^\dagger \phi}{u^2}\Big)$, which have been discussed in \cite{Ellis:2016glu} are given below:
\begin{align}\label{ffunct}
h_1\Big(\frac{\phi^\dagger \phi}{u^2}\Big) &= 5\, H^8 - 4 \, H^{10}\,, \nonumber \\
h_2\Big(\frac{\phi^\dagger \phi}{u^2}\Big) &= 6\, H^{10} - 5 \, H^{12} \,, 
\nonumber \\
h_3\Big(\frac{\phi^\dagger \phi}{u^2}\Big) &= 8\, H^{8} - 10 \, H^{10} + 3\,H^{12}\,, 
\nonumber \\
h_4\Big(\frac{\phi^\dagger \phi}{u^2}\Big) &= - 8\, H^{14}\, 
\ln(H\rho_0)
 + H^{16}\,, \, {\it etc.}\,.
\end{align}
%%%%%%%%%%%%%%%
Energy momentum tensor is given by:
\begin{equation}\label{Energy Momentum Tensor nmcoupling}
   T^{\mu\nu} =-g^{\mu\nu}\mathcal{L}+2D^{\mu}\phi^{\dagger}D^{\nu}\phi-\vec{F}^{\mu \sigma}\cdot \vec{F}^{\nu}_{\; \; \sigma}-h_i \left(
   \frac{\phi^{\dagger}\phi}{u^2}
   \right) G^{\mu\sigma}G^{\nu}_{\; \; \sigma}
\end{equation}
Energy functional is modified then as:
\begin{equation}
    E_i =\frac{4\pi }{g^2} M_W  \int^{\infty}_0dx \frac{g^2}{g^{'2}x^2}h_i(H)+\frac{4\pi}{g^2}\int^{\infty}_0 dx[(f')^2+4x^2(H')^2]+\frac{(f^2-1)^2}{2x^2}+2f^2H^2 +2\gamma x^2(H^2-1)^2]
\end{equation}
Then equations of motion, provide us with:
\begin{equation}
    f''=\frac{f(f^2-1)}{x^2}+2f H^2 
\end{equation}
\begin{equation}
    H'' =-\frac{2}{x}H'+\frac{f^2H}{2x^2}+\gamma H(H^2-1)+\frac{g^2}{16 g^{'2}x^4}\frac{dh_i}{dH}
\end{equation}
In the next subsection, we are going to briefly review string inspired extension of the electroweak model~\cite{Arunasalam:2017eyu,Mavromatos:2018kcd}, which promises a well defined energy functional for the monopole configuration.
\par Asymptotically at $x=0$ profile functions behave as:
\begin{equation}
    H(x) \xrightarrow{ x\rightarrow 0} \frac{c_1}{\sqrt{x\sqrt{\gamma}}} J_{\frac{\sqrt{3}}{2}}(x\sqrt{\gamma}) \qquad f(x)\xrightarrow{x\rightarrow 0} 1- c_2x^2
\end{equation}
And at $x\rightarrow \infty$ behave as:
\begin{equation}
    H(x)\xrightarrow{x\rightarrow \infty}1-c_4 \frac{\exp(-\sqrt{2\gamma}x)}{x} \qquad f(x)\xrightarrow{x\rightarrow \infty}c_3\exp(-\sqrt{2}x)
\end{equation}
In table \ref{AsympyFreeParameterChoMod} we showcase the values of free parameters $c_1$, $c_2$, $c_3$ and $c_4$.
\begin{table}[ht]
\caption{Monopole Mass for various model of different dielectric functions}
\centering
  \begin{tabular}{||c|c|c|c|c||}
  \hline
    $h_i (H)$  & $c_1$ & $c_2$ & $c_3$ & $c_4$ \\
    \hline

    $h_2(H)=6H^{10}-5H^{12}$                     & 1.17 & 0.85 & 3.02 & 2.30 \\
    \hline
    $h_3 (H)=8H^8-10H^{10}+3H^{12}$                    &  0.82 & 0.77 & 3.12 & 1.80 \\
    \hline
  
  \end{tabular}
  \label{AsympyFreeParameterChoMod}
\end{table}
\subsection{Finite-energy string-inspired Hypercharge-Born-Infeld extension of the CM monopole}\label{BI}

Historically, the Born-Infeld model~\cite{BI} was introduced to remove the divergences of the electron's self-energy in classical electrodynamics, resumming in a compact form to all orders in derivatives previous non-linear electrodynamics models by Euler and Heisenberg~\cite{EH1}, and Euler and Kockel~\cite{Euler:1935zz}. 
In the modern context of string theories, the BI model arises in the low-energy limit of open string theories~\cite{Fradkin:1985qd} (for a pertinent review see \cite{Zwiebach:2004tj}).
Moreover, gauge fields on the world-volumes
of D-branes are governed by (non-Abelian) Born–Infeld theory~\cite{Tseytlin:1997csa}. 

In our context such a non-linear gauge theory extension of the CM monopole leads to finite energy configurations for the monopole. Indeed, as we have seen in the previous section \ref{sec:CMclassicalmodel}, the energy divergence of the initial CM monopole configuration~\cite{Cho:1996qd} occurs due to the $1/x^2$ singular behavior at the origin $x \to 0^+$. This is the same singular behavior that one obtains when they calculate the classical self energy of a point-like charge in the electromagnetic $U_{em}(1)$ theory, and which is regulated by the Born-Infeld Electrodynamics~\cite{BI}.  

\par This motivation lead the authors of \cite{Arunasalam:2017eyu,Mavromatos:2018kcd} to considering simplified extensions of the electroweak Lagrangian \eqref{lagew} in which 
the hyopercharge kinetic terms are replaced by non-linear Born-Infeld type terms:
\begin{align}\label{BIlagew}
    \mathcal{L}^{\rm YBI} = -\frac{1}{4}\vec{F}_{\mu\nu}\cdot\vec{F}^{\mu\nu}+\beta^2\Big(1-\sqrt{1+\frac{1}{2\beta^2}G_{\mu\nu}G^{\mu\nu}-\frac{1}{16\beta^4}(G_{\mu\nu}\Tilde{G}^{\mu\nu})^2}\,\Big) +D_{\mu}\phi^{\dagger}D^{\mu}\phi-\frac{\lambda}{2}(\phi^{\dagger}\phi-u^2)^2\,,
\end{align}
where $\beta$ is the Born-Infeld parameter, which has dimensions of mass squared. In the context of string theories~\cite{Fradkin:1985qd,Tseytlin:1997csa,Zwiebach:2004tj} this parameter is proportional to the appropriate string tension, and hence the square of the string scale $M_s^2$. In the approach of \cite{Arunasalam:2017eyu,Mavromatos:2018kcd}, $\beta$ is treated as a phenomenological parameter, beyond string theory, which can be constrained by collider (current (LHC) and future) data, in particular
light-by-light scattering~\cite{Ellis:2017edi}, which has been observed by Experiment. The electroweak model \eqref{lagew} is obtained from \eqref{BIlagew} in the limit $\beta\rightarrow \infty$.

In the case of magnetic monopole configurations the dual of the hypercharge ``Maxwell'' tensor vanishes~\cite{Arunasalam:2017eyu,Mavromatos:2018kcd}, $\Tilde{G}_{\mu\nu}=0$. In the context of the model \eqref{BIlagew}, we consider the CM solution \eqref{Non Abelian Configuration}, \eqref{Abelian Configuration}, \eqref{Scalar Configuration} in the radial gauge \eqref{RadialGauge}, with boundary conditions \eqref{BoundaryCondition}. The corresponding EMT \eqref{EMT_General} in this case is given by:
\begin{equation}\label{Energy momentum tensor-Cho-String}
    T^{\mu \nu} =-g^{\mu\nu}\mathcal{L}-\frac{G^{\mu b}G^{\nu}_{\; \; b}}{\sqrt{1+\frac{1}{2\beta^2}G_{\mu\nu}G^{\mu\nu}}} -\vec{F}^{\mu\sigma} \cdot\vec{F}^{\nu}_{\; \; \; \sigma }+2(D^\mu\phi)^{\dagger}D^{\nu}\phi
\end{equation}
 As in the previous cases, by substituting the field configurations $(\ref{Non Abelian Configuration})$, $(\ref{Abelian Configuration})$, $(\ref{Scalar Configuration})$ to the purely temporal ($tt$) components of $(\ref{Energy momentum tensor-Cho-String})$, we obtain the hamiltonian density of the configuration:
\begin{align}\label{HdensCMmod}
\mathcal{H} =T^{tt}=-\mathcal{L}&=\beta^2\left(\sqrt{1+\frac{1}{(g')^2\beta^2 r^4}}-1\right) +\frac{(f')^2}{g^2r^2}\nonumber \\
&+\frac{(\rho')^2}{2} +\frac{(f^2-1)^2}{2g^2r^4}+\frac{f^2\rho^2}{4r^2}+\frac{\lambda}{2}\left(\frac{\rho^2}{2}-u^2\right)^2\,.
 \end{align}
The configuration has a well-defined energy functional, free of singularities:
\begin{align}\label{EnergyFunctionalHyperExten} 
 E &= 15.53 \sqrt{\frac{\beta_0}{(g')^3g}}\times M_W +\frac{4\pi}{g^2}M_W\int^{\infty}_0 dx [(f')^2+4x^2(H')^2+\frac{(f^2-1)^2}{2x^2}
 \nonumber \\
 &+ 2f^2H^2+2\gamma x^2(H^2-1)^2]   \,,
 \end{align}
where $\gamma = 4\lambda/g^2$, $x=M_W r$ and $\beta_0 =g \beta/M^2_W$. Once again, since we have a static configuration, we can apply the Euler-Lagrange equations on the energy functional,  and obtain the differential equations \eqref{EquationScalarProfile} and \eqref{EquationGaugeProfile}. This is to be expected since $U_Y(1)$ sector modification does not affect the $SU_L(2)$ or scalar sectors. 

In the following subsections  we proceed to determine the 
properties 
of all the CM monopole variants, the initial ones of \cite{Cho:1996qd} and those with finite energy~\cite{Cho:2013vba,Arunasalam:2017eyu,Mavromatos:2018kcd}.  We commence our study by first constructing, in subsection \ref{sec:topmagcharge}, the 
topological magnetic charge 
of the configurations. Then, in subsection \ref{NumericalResults}, we arrive at some numerical results on the properties of the solutions, which are relevant for our subsequent study of the mechanical properties of the CM solutions. This takes place in subsections \ref{EWpressureAnalysis}, where we study the pressure of the configurations, and 
\ref{sec:intforce}, where we compute the profiles of the internal force fields, and discuss the (non)satisfaction of the mechanical stability criteria. 

\subsection{Topological Magnetic Charge of the CM monopole configurations}\label{sec:topmagcharge}

When we consider the CM monopole configuration(s), we have fixed the gauge to the radial one, \eqref{Scalar Configuration}, \eqref{RadialGauge}. This raises the question as to whether the magnetic monopole solution is an artefact of the specific (radial) gauge fixing. In this subsection we shall argue (rather than proving rigorously) that this is not the case, by demonstrating the consistency of the magnetic charge value in both the radial and unitary gauges. However, we shall see that the topological arguments for the magnetic charge quantization have different origins in the two gauges. Specifically, in the radial gauge, discussed in the next section, the magnetic charge receives contributions from both the $SU_L(2)$ and hypercharge U$_{\rm Y}$(1) sectors, whilst in the unitary gauge the magnetic charge receives contributions from a U(1) subgroup of $SU_L(2)$ and the hypercharge 
U$_{\rm Y}$(1) subgroup.

\subsubsection{Radial Gauge}
In the radial gauge $(\ref{Scalar Configuration})$, we can demonstrate that the magnetic charge has the value $4\pi/e$ and is a topological quantity, which supports our argumentation on the gauge invariance of the CM monopole (and its finite-energy) variants. 
To this end, we first remark that in the radial gauge we can no longer express the physical $Z$-boson and photon fields  via $(\ref{Z-field Unitary Gauge})$ and $(\ref{EMpotentialUnitaryGauge})$, respectively, because these
configurations do not correspond to the correct mass terms in the Lagrangian density. 
 For this reason, in order to define the magnetic charge of the monopole, 
  we must first determine a {\it gauge invariant expression} for the electromagnetic field tensor. 
 To this end we use 
 Nambu's definition \cite{Nambu:1977ag} of the gauge invariant tensor is given by:
\begin{equation}\label{Nambu Definition of EM tensor}
\mathcal{F}_{\mu\nu}:= -\sin(\theta_W)(\vec{F}_{\mu\nu}\cdot \hat{\phi}) + \cos(\theta_W) G_{\mu\nu}\,,
\end{equation}
with the Higgs doublet in the radial gauge being expressed as:\footnote{It can be easily checked that in the unitary gauge the expression \eqref{Nambu Definition of EM tensor} yields the familiar expression of the electromagnetic field $(\ref{EMpotentialUnitaryGauge})$.}
\begin{equation}
    \phi=i\frac{\rho}{\sqrt{2}}\begin{pmatrix}
    \sin(\theta/2)e^{-i\psi} \\
    -\cos(\theta/2)
    \end{pmatrix}\,.
\end{equation}
Where the fundamental unit of the electric charge is defined as~\cite{Srednicki:2007qs} $e=g \, \sin(\theta_W)=g'\, \cos(\theta_W)$, with $\theta_W$ the Weinberg angle.

From \eqref{Nambu Definition of EM tensor} the magnetic charge reads:
\begin{align}\label{mgchradialgauge}
q_{m} = \oint d\vec{S}\cdot \vec{B}^{em} = \oint dS_i (-\frac{1}{2}\epsilon^{ijk}\mathcal{F}_{jk}) 
= 
   -\sin(\theta_W) \oint dS_i (-\frac{1}{2}\epsilon^{ijk} \vec{F}_{jk}\cdot \hat{\phi})+ \cos(\theta_W) \oint dS^i(-\frac{1}{2}\epsilon_{ijk}G^{jk})\,.
\end{align}
The first term on the right-hand side comes from the non Abelian tensor:
\begin{equation}
    \oint dS_i(-\frac{1}{2}\epsilon^{ijk} \vec{F}_{jk}\cdot \hat{\phi}) =\frac{1}{2g}\oint  dS^i[(f(r))^2-1]\epsilon_{ijk}\hat{\phi}\cdot (\partial^j \hat{\phi}\times \partial^k \hat{\phi})\,
\end{equation}
which is a surface integral over the spatial boundary ($r\rightarrow \infty$). Thus, we obtain:
\begin{equation}
   -\frac{1}{g}\int  dS^i[
\frac{1}{2}\epsilon_{ijk}\hat{\phi}\cdot (\partial^j \hat{\phi}\times \partial^k \hat{\phi}) ]=\frac{4\pi}{g}
\end{equation}
The integral $n=\frac{1}{4\pi}\int  dS^i\frac{1}{2}\epsilon_{ijk}\hat{\phi}^i \cdot (\partial^j \hat{\phi}\times \partial^k \hat{\phi})$ is the Brouwer degree of the map $\hat{\phi}:S^2 \rightarrow S^2$ \cite{Shnir:2005vvi}. Thus the non-Abelian contribution to the magnetic charge is purely topological.
\par On the other hand the Abelian contribution is given by:   
\begin{align}   
\int dS^i(-\frac{1}{2}\epsilon_{ijk}G^{jk}) =\int dS \cdot (\nabla \times B) = \oint d\vec{l} \cdot \vec{B}^N -\oint d\vec{l} \cdot \vec{B}^S =-\frac{i}{g'} \oint d\vec{l} \cdot (g_{Y}^{-1} \vec{\nabla}g_Y )= \frac{2}{g'} \int^{2\pi}_0 d\psi = \frac{4\pi}{g'}\,,
\end{align}
where $B^i$ is the Abelian field $(\ref{Abelian Configuration})$ and $g_Y = e^{2i\psi}$ is a $U_Y(1)$ element. Once again we have used the Wu-Yang description of the Abelian monopole \cite{Wu:1975es}. The fact that the Abelian contribution is proportional to $\oint d\vec{l} \cdot (g_{Y}^{-1} \vec{\nabla}g_Y )$ implies that this term is also a  topological one. 
\par We conclude therefore  that in the radial gauge the magnetic charge consistes of  two topological contributions:  from the Brouwer degree of the map $\hat{\phi}$ and from the winding number associated with the hypercharge group $U_Y(1)$. Adding together these two contributions we finally obtain:
\begin{equation}
    q_m = 4\pi\, \Big(\frac{\sin(\theta_W)}{g}+\frac{\cos(\theta_W)}{g'}\Big)=\frac{4\pi}{e}\,.
\end{equation}

\subsubsection{Unitary Gauge}\label{sec:gaugeinv}
\par

 The gauge transformation $U(\theta,\psi)$ \cite{Cho:2013vba}, which transforms the gauge and scalar fields from the radial gauge to the unitary gauge reads:
\begin{equation}\label{gtr}
    U(\theta,\psi)=i \begin{pmatrix}
        \cos(\theta/2) & \sin(\theta/2) e^{-i\psi} \\
        -\sin(\theta/2)e^{i\psi} & \cos(\theta/2)
    \end{pmatrix}\,, 
\end{equation}
\begin{equation}
\phi'=\frac{\rho(r)}{\sqrt{2}} \begin{pmatrix}
    0 \\
    1
\end{pmatrix}\,, 
\qquad \qquad
    \vec{A}^{'}_{\mu} = \frac{1}{g} \begin{pmatrix}
    -f(r)(\sin\psi\partial_{\mu}\theta +\sin\theta\cos\psi \partial_{\mu}\psi )\\
    f(r)(\cos\psi\partial_{\mu}\theta -\sin\theta\sin\psi \partial_{\mu}\psi )\\ 
    -(1-\cos\theta)\partial_{\mu}\psi
\end{pmatrix}\,.
\end{equation}
Notice that ({\it cf.} \eqref{BoundaryCondition}) at the spatial boundary $\phi \xrightarrow{r \rightarrow \infty} \begin{pmatrix}
    0 \\
    u
\end{pmatrix}$\,.

At this stage, we make the important remark that, on account of the gauge invariance of the pertinent actions, both radial and unitary gauges provide us with the same energy-momentum tensor components $T^{\mu\nu}=diag(\mathcal{H},\mathcal{P}_R,\mathcal{P}_\Theta,\mathcal{P}_{\Psi})$ for each model. 
This provides a further indication for the gauge independent nature of the physical observables associated with the CM electroweak magnetic monopole and its extensions. 

In this gauge, after SSB, we define, as standard~\cite{Srednicki:2007qs}, the massive $Z$-boson $Z_{\mu}$ and the massless photon fields $A_{\mu}$ via:
\begin{equation}\label{Z-field Unitary Gauge}
    Z_{\mu} = \cos(\theta_W)A^3_{\mu}- \sin(\theta_W)B_{\mu}\,
\end{equation}
\begin{equation}\label{EMpotentialUnitaryGauge}
    A^{em}_{\mu}= \sin(\theta_W)A^3_{\mu}+ \cos(\theta_W)B_{\mu}\,,
\end{equation}
respectively, where the fundamental unit of the electric charge is defined as~\cite{Srednicki:2007qs} $e=g \, \sin(\theta_W)=g'\, \cos(\theta_W)$, with $\theta_W$ the Weinberg angle. In the unitary gauge, the third component of $\vec{A}_{\mu}$, $A^3_{\mu}$, which in the radial gauge has the form $(\ref{Non Abelian Configuration})$, becomes  asymptotically (under the action of the aforementioned non-Abelian gauge transformation): 
\begin{equation}
    \vec{A}_{\mu} \xrightarrow{r \rightarrow\infty} \begin{pmatrix}
    0\\
    0 \\
    -\frac{1}{g}\Big(1-\cos(\theta) \Big)\,\partial_{\mu}\psi 
    \end{pmatrix}
\end{equation}
implying that, in the unitary gauge, $A_\mu^3(x)$ has the form of a Dirac potential \cite{Dirac:1948um}.  Together with $(\ref{Abelian Configuration})$, then, this yields an electromagnetic potential of the form:
\begin{equation}
    A^{em}_{\mu}=\sin(\theta_W)A^3_{\mu}+\cos(\theta_W)B_{\mu}=-(\frac{\sin(\theta_W)}{g}+\frac{\cos(\theta_W)}{g'})\Big(1-\cos(\theta)\Big)\,\partial_{\mu}\psi\,.
\end{equation}
Thus, in the unitary gauge we obtain a Dirac-Like monopole configuration, which has a magnetic charge given by:
\begin{align}\label{diracmch}
    q_{m} &= \int d\vec{S} \cdot \vec{B}^{em}= \int d\vec{S} \cdot (\nabla \times A^{em}) = \oint d\vec{l} \cdot \vec{A}^{emN} -\oint d\vec{l} \cdot \vec{A}^{emS}
\nonumber \\  
    &=-\frac{i}{e} \oint d\vec{l} \cdot (g^{-1}_{em} \vec{\nabla}g_{em} )= \frac{2}{e} \int^{2\pi}_0 d\psi = \frac{4\pi}{e}\,,
\end{align}
where $g_{em}= e^{2i\psi}$ denotes an element of the Abelian gauge transformations of the electromagnetic $U_{em}(1)$. It should be noticed that above, we have used once again the Wu-Yang description of the Dirac Monopole~\cite{Wu:1975es}. Moreover, the integral $\oint d\vec{l} \cdot (g^{-1}_{em} \vec{\nabla}g_{em} )$ is actually the winding number associated with the electromagnetic gauge group $U_{em}(1)$. Therefore, in the unitary gauge, the magnetic charge is $4\pi/e$ and it is indeed a topological charge, since it is associated with a (topological) winding number.

\par Note that the same calculation can be performed by using the gauge invariant definition of the electromagnetic tensor $(\ref{Nambu Definition of EM tensor})$:
\begin{equation}
    q_m =-\sin(\theta_W)\frac{i}{g}\oint d\vec{l} \cdot( g^{-1}_{SU_L(2)} \vec{\nabla}g_{SU_L(2)})-\cos(\theta_W)\frac{i}{g'} \oint d\vec{l} \cdot( g^{-1}_{Y} \vec{\nabla}g_{Y})=\frac{4\pi}{e}
\end{equation}
With $g_{SU_L(2)}=e^{2i\psi}$ and $g_Y = e^{2i\psi}$ are both Abelian elements associated with the Abelian subgroup element of $SU_L(2)$ and $U_Y(1)$. Notice that in the unitary gauge the $SU_L(2)$ contribution now becomes a winding number associated with the Abelian subgroup of $SU_L(2)$. This a very different behavior compared to the radial gauge, where such contribution is given by the Brouwer degree of the map $\hat{\phi}: S^2 \rightarrow S^2$. 

\par Thus, in both the radial and unitary gauges, the magnetic charge has a topological origin and equals $4\pi /e$. 
We consider this as an indication of the fact that the CM magnetic monopole solution (and its finite-energy variants) are not  gauge artefacts, but constitute gauge invariant solutions, with the finite-energy variants being proper solitons. Nonetheless,  the different topological origins of the magnetic charge quantization condition between the radial and unitary gauge solutions, as shown above, presents a puzzle, which may  affect the $CP^1$ topological stability arguments of the CM monopole. Indeed, as we shall show below, our mechanical criterion for stability fails for this configuration, which might be linked to above issue.   

Before closing this subsection, we remark that the anti-monopole solution is given by the following configuration~\cite{Cho:2013vba}, in the radial gauge:
\begin{align}\label{antinomR}
\widetilde \xi_{\rm rad} = i
\begin{pmatrix}
 \cos \frac{\theta}{2} \\ 
 \sin \frac{\theta}{2}\, e^{i\psi} \end{pmatrix}\,,    
\end{align} 
and by 
\begin{align}\label{antinomU}
\widetilde \xi_{\rm uni} = 
\begin{pmatrix}
 1 \\ 
 0\end{pmatrix}\,,    
\end{align} 
in the Unitary gauge. The corresponding magnetic charge turns out to be opposite in sign and equal in magnitude with the monopole, provided one changes the sign of the Higgs hypercharge quantum number accordingly.

We proceed now to discuss the internal-force profiles in the CM solutions, and examine the satisfaction of the mechanical stability criteria from this angle.

We next proceed to study numerically several properties of the CM solutions, which we shall use in sections \ref{EWpressureAnalysis}
and \ref{sec:intforce} in order to study the mechanical stability criteria of the solutions.
\subsection{Numerical Results on the CM Monopole Solutions and its variants }\label{NumericalResults}

In this section we are going to discuss some numerical results for the system of differential equations $(\ref{EquationScalarProfile})$ and $(\ref{EquationGaugeProfile})$.  
The system is solved for the standard model values~\cite{ParticleDataGroup:2024cfk} $g=0.652$, $g'=0.357$ and $\lambda=0.129$. Moreover, we provide a solution at the BPS limit, where $(\lambda,\mu) \rightarrow (0,0)$ and $M_W = 80.377$ $GeV$ fixed.
\par
\begin{figure}[ht]
    \centering
\includegraphics[width=0.5\textwidth]{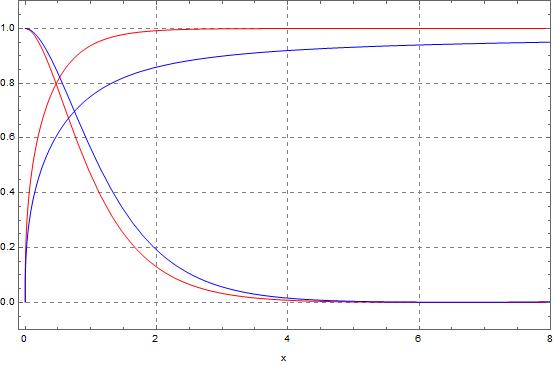}
    \caption{Solutions of $(\ref{EquationScalarProfile})$ and $(\ref{EquationGaugeProfile})$ in the CM electroweak magnetic-monopole model. The 
    $f(x)$ solutions
approach zero asymptotically, whilst the $H(x)$ solutions approach asymptotically the horizontal line $H(\infty)=1$. The $\gamma=0$ (BPS limit) solutions are represented with a blue line and the $\gamma = 1.21382$ ones with a magenta-coloured line.}
    \label{Cho-Maison-Sol}
\end{figure}

\begin{figure}[ht]
    \centering
\includegraphics[width=0.5\textwidth]{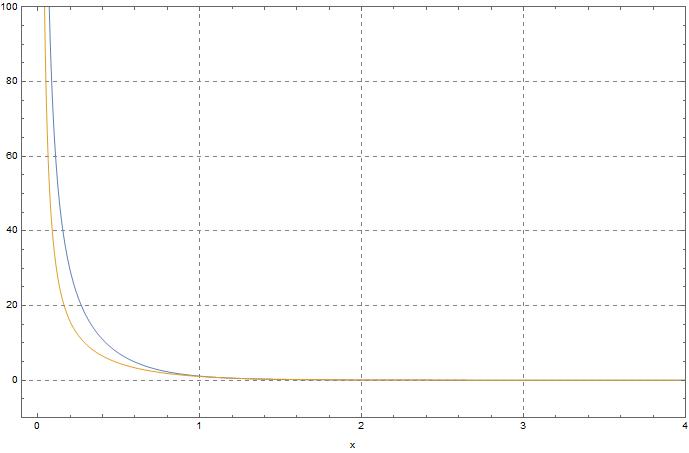}
    \caption{$SU_L(2)$ Hamiltonian density $\mathcal{H}_{SU_L(2)}$ contribution for two values of $\gamma$ in the CM electroweak magnetic-monopole model: the $\gamma= 0 $ (BPS) is represented by an orange line, whilst the $\gamma =1.2138$ (standard model) value is represented by a blue line.  }
\label{HamiltonianSU(2)constr}
\end{figure}

 In figure $\ref{Cho-Maison-Sol}$ we showcase profile functions for such values. 
From \eqref{EnergyFuncChoMaison} and \eqref{EnergyFunctionalHyperExten}, we observe that
the $SU_L(2)$ contribution to the total energy functional are common in 
the initial CM Electroweak model and its Born-Infeld extension:
{\small \begin{align}\label{SU(2) contribution to the Hamiltonian}
\int d^3x \,\mathcal{H}_{SU_L(2)} = \frac{4\pi}{g^2} M_W \times C(\gamma)
=\frac{4\pi}{g^2}M_W\int^{\infty}_0 dx \, x^2 \, \Big[(\frac{f'}{x})^2+4(H')^2+\frac{(f^2-1)^2}{2x^4}+ \frac{2f^2H^2}{x^2}+2\gamma (H^2-1)^2 \Big]\,.
\end{align}}
In figure \ref{HamiltonianSU(2)constr} we showcase the behavior of the quantity inside the square brackets in the integrant of the right-hand side of \eqref{SU(2) contribution to the Hamiltonian} ({\it i.e.}
the $SU_L(2)$ contribution $\mathcal{H}_{SU_L(2)}(x)$ to the Hamiltonian density) for two indicative values of $\gamma$, the BPS limit and the standard model value~\cite{ParticleDataGroup:2024cfk}.
Notice should be taken of the fact that, for both values of $\gamma$, such a  contribution is positive definite for all values of $ x \ge 0$, but 
diverges at $x=0$, which is to be contrasted to the case of the HP monopole Hamiltonian density $(\ref{Hamiltonian Hooft})$, which has a well defined behaviour at the origin $r=0$.
Moreover, from figure \ref{HamiltonianSU(2)constr}, the reader observes that, as the BPS limit $(\lambda,\mu)\rightarrow (0,0)$ is approached from the standard model value $\gamma = 1.21382$, the rate of approach of the $SU_L(2)$ Hamiltonian density contribution  to its singular behavior at $x=0$ becomes slower.
\par
Numerical integration then gives:
{\small \begin{equation}
  \frac{4\pi}{g^2}M_W \times C(1.21382) = 2376~{\rm GeV} \times 2.10539 = 5.00241 ~{\rm TeV}\,, \quad  
    \frac{4\pi}{g^2} M_W \times C(0) = 2376~{\rm  GeV} \times 1.82246= 3.33016~{\rm TeV}\,.
\end{equation}}
As already mentioned, in initial CM electroweak-monopole case~\cite{Cho:1996qd},  the total energy functional $(\ref{EnergyFuncChoMaison})$ is singular,  in contrast to the Born-Infeld CM extension  $(\ref{EnergyFunctionalHyperExten})$, in which case it takes on the value (the reader is invited to compare with the semi-analytic approximate estimate made in \cite{Mavromatos:2018kcd}):
\begin{equation}
    E_{SM} = (7.24731 \sqrt{\beta_0}+5.00241)~{\rm TeV}\,,
\end{equation}
whilst in the BPS limit, one obtains:
\begin{equation}
    E_{BPS}=(7.24731\sqrt{\beta_0}+3.33016)~{\rm TeV}\,.
\end{equation}
Above, we have expressed the Born-Infeld parameter $\beta$ as a dimensionless phenomenological variable, $\beta_0$. 
\par
\begin{figure}[ht]
    \centering
    % Left figure
    \begin{subfigure}{0.4\textwidth}
        \centering
        \includegraphics[width=\linewidth]{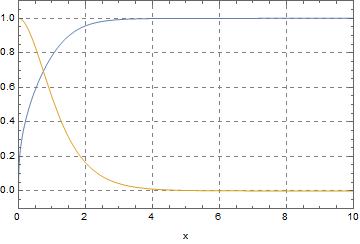} 
        \caption{Profile Functions for $h_2(H)=6H^{10}-5H^{12}$. Scalar profile function $H(x)$ is represented by blue line and gauge profile function $f(x)$ by orange line.  }
        \label{f2Profiles}
    \end{subfigure}
    \hspace{0.1\textwidth} 
    \begin{subfigure}{0.4\textwidth}
        \centering
        \includegraphics[width=\linewidth]{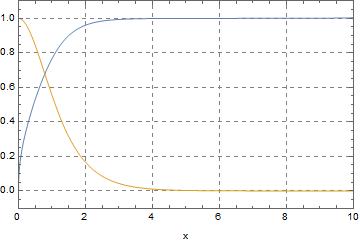} % replace with your figure file
        \caption {Profile Functions for $h_3(H)=8H^{8}-10H^{10}+3H^{12}$. Scalar profile function $H(x)$ is represented by blue line and gauge profile function $f(x)$ by orange line. }
        \label{f3Profiles}
    \end{subfigure}
    \caption{Profile functions for dielectric functions $h_2 (H)$ and $h_3 (H)$ in the extension of the electroweak model with non-minimal Higgs-field dependent dielectric-function couplings in the hypercharge sector..}
\end{figure}

 In figures \ref{f2Profiles} \ref{f3Profiles} we showcase profile functions for two characteristic examples of dielectric functions $h_i(H)$ for the standard model value $\gamma=1.21382$.

Monopole mass for various model of different dielectric functions are given in table \ref{MonopoleMassDielecticFun}.
\begin{table}[ht]
\caption{Monopole Mass for various models of different dielectric functions}
\centering
  \begin{tabular}{||c|c||}
  \hline
    $h_i (H)$  & $M$ $[\rm TeV]$ \\
    \hline

    $h_2(H)=6H^{10}-5H^{12}$                     & 12.056 \\
    \hline
    $h_3 (H)=8H^8-10H^{10}+3H^{12}$                    &  13.323 \\
    \hline
  
  \end{tabular}
  \label{MonopoleMassDielecticFun}
\end{table}

\par We are now ready to proceed with the study of the mechanical properties of the CM monopole configuration and its finite-energy variants. 

\subsection{Internal Pressure Analysis}\label{EWpressureAnalysis}
\subsubsection{CM Electroweak Model}

We calculate first the spatial diagonal components of $(\ref{Energy-Momentum-Tensor-Cho})$ in spherical polar coordinates, which correspond to the pressure of the monopole configuration. The radial pressure is given by:
\begin{equation}\label{EWRadialPressure}
    \frac{g^2}{M^4_W}\mathcal{P}_R (x) =\frac{g^2}{M^4_W}T^{r r} =-\frac{g^2}{2(g')^2 x^4} +(\frac{f'}{x})^2+4(H')^2-  \frac{2f^2H^2}{x^2}-\frac{(1-f^2)^2}{2x^4}-2\gamma (H^2-1)^2\,.
\end{equation}
We are going to show explicitly the asymptotic behavior near $x \rightarrow 0$ only for the standard model value $\gamma = 1.21382$, since for a general $\gamma$ the result is quite cumbersome, and not essential for our purposes:
\begin{equation}\label{prat0}
    \frac{g^2}{M^4_W} \mathcal{P}_{R}(x) \xrightarrow{x \rightarrow 0} -\frac{1.66774}{x^4}\,.
\end{equation}
Similarly,  for $x \rightarrow \infty$, one has: 
\begin{equation}
    \frac{g^2}{M^4_W} \mathcal{P}_{R}(x) \xrightarrow{x \rightarrow \infty} -\frac{1.66774}{x^4}
\end{equation}
The polar component of the pressure is given by:
\begin{equation}\label{EWPolarPressure}
    \frac{g^2}{M^6_W}\mathcal{P}_{\Theta} (x)=\frac{g^2}{M^6_W}T^{\theta \theta} = \frac{g^2}{2(g')^2 x^6}-4\frac{(H')^2}{x^2} +\frac{(1-f^2)^2}{2x^6}-\frac{2\gamma}{x^2} (H^2-1)^2
\end{equation}

\begin{figure}[ht]
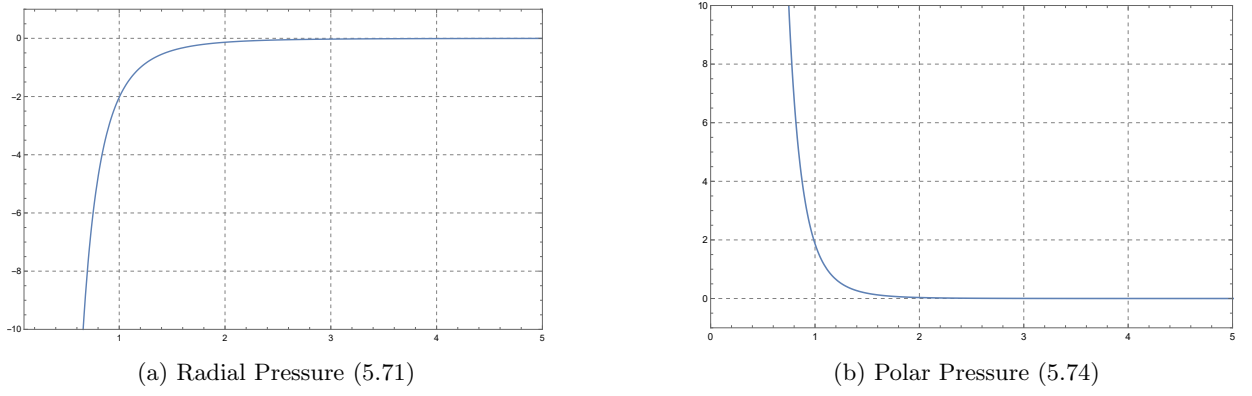

    \centering
    % Left figure
    \begin{subfigure}{0.4\textwidth}
        \centering
        \includegraphics[width=\linewidth]{RadialPressureEW.jpg} 
        \caption{Radial Pressure $(\ref{EWRadialPressure})$ }
    \end{subfigure}
    \hspace{0.1\textwidth} 
    \begin{subfigure}{0.4\textwidth}
        \centering
        \includegraphics[width=\linewidth]{PolarPressureEW.jpg} 
        \caption {Polar Pressure $(\ref{EWPolarPressure})$}
        \label{Polar PressureEW}
    \end{subfigure}
    \caption{Radial and Polar Pressure for  $\gamma =1.21382$ in the CM electroweak-magnetic monopole model.}\label{Polar and Radial Pressure EW}
\end{figure}
Once again, for standard model values of the parameters~\cite{ParticleDataGroup:2024cfk} we have:
\begin{equation}
    \frac{g^2}{M^6_W}\mathcal{P}_{\Theta}(x)\xrightarrow{ x \rightarrow 0} \frac{1.6677}{x^6}\,, \qquad 
     \frac{g^2}{M^6_W}\mathcal{P}_{\Theta}(x)\xrightarrow{ x \rightarrow \infty} \frac{1.6677}{x^6}\,.
\end{equation}
Thus, we observe that both the radial and polar components of the pressure diverge in the $x \to 0$ limit. 
In figure \ref{Polar and Radial Pressure EW} we show the radial and polar pressure for the CM initial monopole configuration. It is important to notice that the polar pressure becomes negative infinite at $x=0$, whilst the radial pressure is positive infinite at this limit. It should also be noted that the radial and polar pressure components at BPS limit are almost identical with those for $\gamma = 1.21382$. This suggest that 
the internal pressure is not sensitive to the 
Higgs self-interaction coupling 
$\lambda$. This is a very different behaviour form the $SU(2)$ HP case, where at the BPS limit the internal pressure components  vanish.
\subsubsection{Finite-energy CM monopole with non-minimal couplings between the Higgs and hypercharge sectors}\label{fpressureAnalysis}
\begin{figure}[ht]
    \centering
    % Left figure
    \begin{subfigure}{0.38\textwidth}
        \centering
        \includegraphics[width=\linewidth]{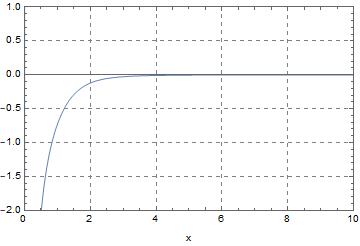} 
        \caption{Radial Pressure for $h_2(H)=6H^{10}-5H^{12}$.  }
        \label{f2RadialPressure}
    \end{subfigure}
    \hspace{0.1\textwidth} 
    \begin{subfigure}{0.38\textwidth}
        \centering
        \includegraphics[width=\linewidth]{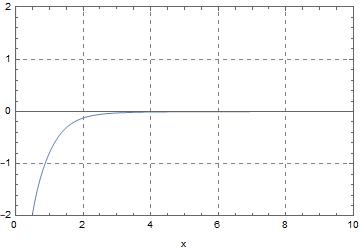} % replace with your figure file
        \caption {Radial Pressure for $h_3(H)=8H^{8}-10H^{10}+3H^{12}$.  }
        \label{f3RadialPressure}
    \end{subfigure}
    \caption{Radial pressure for dielectric functions $h_2 (H)$ and $h_3 (H)$, in the extension of the electroweak model with non-minimal Higgs-field dependent dielectric-function couplings in the hypercharge sector. }
\end{figure}

\begin{figure}[ht]
    \centering
    % Left figure
    \begin{subfigure}{0.38\textwidth}
        \centering
        \includegraphics[width=\linewidth]{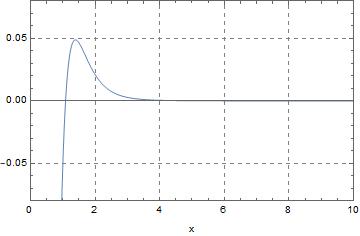} 
        \caption{Polar Pressure for $h_2(H)=6H^{10}-5H^{12} $.}
        \label{f2PolarPressure}
    \end{subfigure}
    \hspace{0.1\textwidth} 
    \begin{subfigure}{0.38\textwidth}
        \centering
        \includegraphics[width=\linewidth]{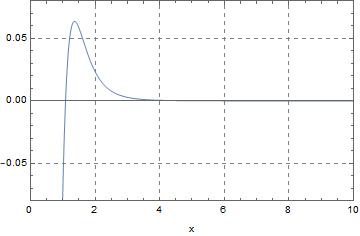} % replace with your figure file
        \caption {Polar Pressure for $h_3(H)=8H^{8}-10H^{10}+3H^{12}$.  }
        \label{f3PolarPressure}
    \end{subfigure}
    \caption{Polar pressure for dielectric functions $h_2 (H)$ and $h_3 (H)$, in the extension of the electroweak model with non-minimal Higgs-field dependent dielectric-function couplings in the hypercharge sector.}
\end{figure}

We calculate the diagonal spatial elements of $(\ref{Energy Momentum Tensor nmcoupling})$, which correspond to the pressure of the monopole configuration. The radial pressure is given by:
\begin{equation}\label{EWRadialPressurenmc}
     \frac{g^2}{M^4_W}\mathcal{P}^i_R (x) =\frac{g^2}{M^4_W}T^{r r} =-\frac{g^2}{2(g')^2 x^4}h_i(H) +(\frac{f'}{x})^2+4(H')^2-  \frac{2f^2H^2}{x^2}-\frac{(1-f^2)^2}{2x^4}-2\gamma (H^2-1)^2\,.
\end{equation}
The polar pressure is given by:
\begin{equation}\label{EWPolarPressurenmc}
    \frac{g^2}{M^6_W}\mathcal{P}^i_{\Theta} (x)=\frac{g^2}{M^6_W}T^{\theta \theta} = \frac{g^2}{2(g')^2 x^6}h_i(H)-4\frac{(H')^2}{x^2} +\frac{(1-f^2)^2}{2x^6}-\frac{2\gamma}{x^2} (H^2-1)^2
\end{equation}
In figures \ref{f2RadialPressure} \ref{f3RadialPressure}  \ref{f2PolarPressure} \ref{f3PolarPressure} we showcase radial and polar pressure for various models.

Near $x\rightarrow0$ we obtain the following behaviors for the radial pressure:
\begin{equation}
    \mathcal{P}^2_R (x) \xrightarrow{x\rightarrow 0}-0.717 x^{\sqrt{3}-3}-0.281 x^{5\sqrt{3}-9}\xrightarrow{x\rightarrow0}-\infty
\end{equation}
\begin{equation}
\mathcal{P}^3_R(x) \xrightarrow{x\rightarrow 0} -0.352 x^{\sqrt{3}-3}-0.045 x^{4\sqrt{3}-8}+0.013x^{5\sqrt{3}}\xrightarrow{ x\rightarrow 0} -\infty
\end{equation}
As for polar pressure we get near $x\rightarrow 0$:
\begin{equation}
    \mathcal{P}^2_\Theta (x) \xrightarrow{x\rightarrow 0}-0.115 x^{6 \sqrt{3}-12}+0.281 x^{5 \sqrt{3}-11}+2.93 x^{\sqrt{3}-3}-0.582 x^{\sqrt{3}-4}-0.262 x^{\sqrt{3}-5}-\frac{0.983}{x^2}\xrightarrow{x\rightarrow 0}-\infty
\end{equation}
$$
    \mathcal{P}^3_\Theta (x) \xrightarrow{x\rightarrow 0}0.001 x^{6 \sqrt{3}-12}+0.022 x^{5 \sqrt{3}-9}-0.013 x^{5 \sqrt{3}-11}-0.058 x^{4 \sqrt{3}-8}
$$
\begin{equation}
    +0.045 x^{4 \sqrt{3}-10}-0.14 x^{2 \sqrt{3}-4}+1.439 x^{\sqrt{3}-3}-0.129
   x^{\sqrt{3}-5}-\frac{1.242}{x^2}\xrightarrow{x\rightarrow0}-\infty
\end{equation}
\subsubsection{String-Inspired Born-Infeld Extension of the Electroweak Model}
We calculate the diagonal spatial elements of $(\ref{Energy momentum tensor-Cho-String})$, which correspond to the pressure of the monopole configuration. The radial pressure is given by:
\begin{align}\label{StringPressureRadial}  
\frac{g^2}{M^4_W}\mathcal{P}_R (x) = \frac{g^2}{M^4_W} T^{r r} &= -\beta^2_0(\sqrt{1+\frac{g^2}{(g')^2\beta^2_0 x^4}}-1)+(\frac{f'}{x})^2+4(H')^2-\frac{2f^2H^2}{x^2} \nonumber \\
   & -\frac{(1-f^2)^2}{2x^4}-2\gamma (H^2-1)^2\,.
\end{align}
As $x \rightarrow 0$ the radial pressure for $\gamma=1.21382$ behaves as:
\begin{equation}
    \frac{g^2}{M^4_W}\mathcal{P}_R (x) \xrightarrow{x \rightarrow 0}-\beta_0 \frac{1.8263}{x^2  }\,,
\end{equation}
whilst in the limit $x \rightarrow \infty$ it behaves as:
\begin{equation}
    \frac{g^2}{M^4_W}\mathcal{P}_R (x) \xrightarrow{x \rightarrow \infty} -\frac{1.3387}{x^4}\,.
\end{equation}
\par The polar component of the pressure is given by:
\begin{align}\label{StringPressurePolar} 
\frac{g^2}{M^6_W}\mathcal{P}_{\Theta} (x)= \frac{g^2}{M^6_W}T^{\theta \theta} &=-\frac{\beta^2_0}{x^2}(\sqrt{1+\frac{g^2}{\beta^2_0(g')^2 x^4}}-1)+\frac{g^2}{(g')^2x^6\sqrt{1+\frac{g^2}{\beta^2_0 (g')^2 x^4}}}-\frac{4(H')^2}{x^2}
\nonumber \\ &+\frac{(1-f^2)^2}{2x^6}-\frac{2\gamma}{x^2} (H^2-1)^2   \,,
\end{align}
while the azimuthal component of the pressure reads:
\begin{equation}
\frac{g^2}{M^6_W}\mathcal{P}_{\Psi}(x)=\frac{g^2}{M^6_W}T^{\psi\psi} = \frac{g^2}{M^6_W}\frac{1}{sin^2(\theta)}\mathcal{P}_{\Theta}(x) \,.
\end{equation}
As $x\rightarrow 0$, the polar pressure behaves as:
\begin{equation}
    \frac{g^2}{M^6_W}\mathcal{P}_{\Theta}(x)\xrightarrow{ x \rightarrow 0} \frac{\beta^2_0}{x^2}-\frac{\beta^2_0}{x^2} \sqrt{\frac{3.336}{\beta^2_0 x^4}+1}+\frac{3.336}{x^6 \sqrt{\frac{3.336}{\beta^2_0 x^4}+1}}-0.66
   x^{\sqrt{3}-5}\,,
\end{equation}
whilst in the $x\rightarrow \infty$ limit it asymptotes to:
\begin{equation}
    \frac{g^2}{M^6_W}\mathcal{P}_{\Theta}(x) \xrightarrow{ x\rightarrow \infty} \frac{1.6677}{x^6}\,.
\end{equation}
\begin{figure}[ht]
    \centering
    % Left figure
    \begin{subfigure}{0.4\textwidth}
        \centering
        \includegraphics[width=\linewidth]{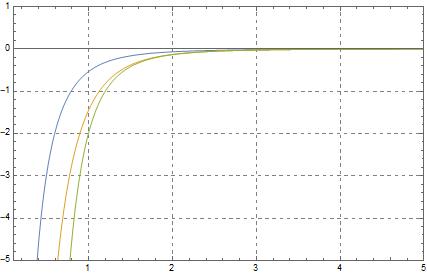} 
        \caption{Radial Pressure $(\ref{StringPressureRadial})$ }
    \end{subfigure}
    \hspace{0.1\textwidth} 
    \begin{subfigure}{0.4\textwidth}
        \centering
        \includegraphics[width=\linewidth]{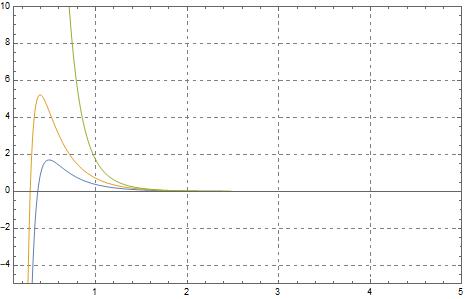} 
        \caption {Polar Pressure $(\ref{StringPressurePolar})$}
        \label{Polar PressureString}
    \end{subfigure}
    \caption{Radial and Polar Pressure for  $\gamma =1.21382$ ( Standard Model Value) in the string-inspired Born-Infeld CM-monopole model. $\beta_0$ values $0.1,1,5$ are represented with blue, orange and green lines respectively.}\label{Polar and Radial Pressure String}
\end{figure}
In figure $\ref{Polar and Radial Pressure String}$ we show the radial and polar pressure components for the Born-Infeld extension of the electroweak model.
We observe that the  increase of the parameter $\beta_0$ causes the radial pressure $(\ref{StringPressureRadial})$ to approach its negative singular behavior faster. 
For the polar pressure $(\ref{StringPressurePolar})$, we observe a different pattern. In the pure electroweak model, the polar pressure $(\ref{EWPolarPressure})$ is positively singular near $x=0$, but in the modified string-inspired Born-Infeld extension of the CM model, such a positive behaviour becomes negative, while creating a positive pressure region. Such a region becomes larger and larger as $\beta_0$ decreases.   
Just as with the $SU_L(2)$ Hamiltonian density $(\ref{SU(2) contribution to the Hamiltonian})$, the radial and polar pressure components in the CM electroweak model and its finite-energy variants are singular at $x=0$.

\subsection{Internal Force Field}\label{sec:intforce}
\subsubsection{CM Electroweak Model}
We commence our discussion with the study of the polar and azimuthal forces, defined in section \ref{InternalForceGeneralStructure}, which are expressed in terms of the radial component of the pressure $\mathcal{P}_R (x)$.
We observe that the singular asymptotic behavior of the latter near the origin $x=0$ ({\it cf.} 
\eqref{prat0}), 
$\mathcal{P}_R(x)\thicksim1/x^4$, causes both polar and azimuthal force components to be singular at $x=0$, both diverging as $1/x^2$:
\begin{equation}
    \frac{g^2}{M^2_W}\mathcal{F}_{\theta}(\theta)=-\pi\sin(\theta)\lim_{r \to 0}[x^2 \frac{g^2}{M^2_W}\mathcal{P}_R (x)] \, \sim \frac{1}{r^2} \, \rightarrow \infty\,, 
\qquad 
    \mathcal{F}_{\Psi}= -\frac{\pi}{2}\lim_{r \to 0}[r^2 \mathcal{P}_R (r)] \, \sim \frac{1}{r^2} \, \rightarrow \infty
\end{equation}
Such a singular behavior will cause rotational instability of the monopole configuration, which seems to be forced to rotate absurdly fast near its centre. 
\par
\begin{figure}[ht]
    \centering
\includegraphics[width=0.4\textwidth]{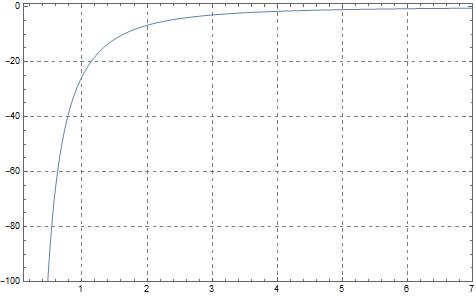}
    \caption{Radial force $(\ref{RForceEW})$ for the standard model value $\gamma=1.21382$ in the CM electroweak magnetic-monopole model. }
\label{RadialForceEW}
\end{figure}

 The total Radial force component:
\begin{equation}\label{RForceEW}
  \frac{g^2}{M^2_W} \mathcal{F}_R (x) = 4\pi x^2 \frac{g^2}{M^2_W}\mathcal{P}_R (x) \,,
\end{equation}
is plotted in figure \ref{RadialForceEW} for the standard model value $\gamma=1.21382$. 
The radial force also diverges near the origin $r \to 0$ of the monopole configuration, and is negative, since the radial pressure approaches negative infinity in that region \eqref{prat0}.
Thus, the configuration will collapse at its centre. 

From the above discussion therefore, 
one concludes that, since all the components of the total force diverge near the origin, and the radial force is directed towards the centre in that region, 
the infinite-energy CM monopole configuration is mechanically unstable by our criteria.  At any rate, as already mentioned, its infinite energy, has rendered this configuration unphysical, so the additional drawback of instability should not come as a surprise. 

We proceed next to examine the mechanical stability of the string-inspired Born-Infeld finite-energy CM extension. Due to their finite energy feature, such solutions behave as proper solitons, and have physical significance, as they are in principle produced through particle collisions, and also in relevant processes in the early Universe. 

\subsubsection{CM Electroweak Model Energy Momentum Tensor Decomposition into Short and Long-range parts}
\par 
In what follows, we shall apply the analysis of section \ref{slEMTparts}
to this case, in order to subtract the long-range electromagnetic part from the total EMT of the model, so as to isolate the short-range part, and examine the mechanical stability criteria on this part.
We remind the reader that it is this short-range part that satisfies the mechanical stability for the HP monopole.

To this end, we need first to evaluate the Electromagnetic  tensor of the model,  defined in \eqref{Nambu Definition of EM tensor}.
In the radial gauge where $\hat{\phi}=-\hat{r}$, the components of this tensor are:
\begin{equation}
    \mathcal{F}_{r\theta}=\mathcal{F}_{r \psi}=0\,, \qquad
    g\mathcal{F}_{\theta \psi} = -\sin(\theta)Q(x)
\end{equation}
where,
\begin{equation}
    gQ(x)= -\sin(\theta_W)\Big(1-f^2(x)\Big)+\frac{g}{g'}\, \cos(\theta_W)\,,
\end{equation}
which is associated with the magnetic charge across the magnetic monopole configuration. Indeed,  with the spatial components of the magnetic field given by $\mathcal{B}^i=-\frac{1}{2}\epsilon^{ijk}\mathcal{F}_{jk}$ one obtains the standard expression  for the magnetic charge $q_m$:
\begin{equation}\label{qmcm}
q_m = \oint dS_r B^r =\frac{4\pi}{e}
\end{equation}
Moreover, it is useful for our analysis below to note that the magnetic charge density $ \vec{\nabla} \cdot \vec{\mathcal{B}}=4\pi \rho_M (r) $ is given by:
\begin{equation}\label{mcdcm}
     \frac{g}{M^3_W} \rho_{M} (x)=\frac{1}{4\pi x^2}\frac{d Q}{dx}\,.
\end{equation}

The electromagnetic EMT is given in \eqref{ememt}.
Its radial and polar components for the  electroweak CM monopole are given by:
\begin{equation}\label{ememtCM}
    \frac{g^2}{M^4_W}\mathcal{P}^{LR}_R(x)= \frac{g^2}{M^4_W}\mathcal{T}^{rr}=-\frac{Q^2(x)}{2x^4} \,, \qquad 
    \frac{g^2}{M^6_W}\mathcal{P}^{LR}_{\Theta}(x)=\frac{g^2}{M^6_W}\mathcal{T}^{\theta\theta}=\frac{Q^2(x)}{2x^6}\,.
\end{equation}
Then the (radial and polar) components of the short-range pressure, defined via
\eqref{LongRangeRadialPressureHooft} and 
\eqref{LongRangePolarHooft} respectively,  are given by:
\begin{align}\label{sremtcm}
    \frac{g^2}{M^4_W}\mathcal{P}^{SR}_R (x) &= -\frac{g^2}{2(g')^2 x^4} +(\frac{f'}{x})^2+4(H')^2-  \frac{2f^2H^2}{x^2}-\frac{(1-f^2)^2}{2x^4}-2\gamma (H^2-1)^2+\frac{Q^2(x)}{2x^4}\,,
    \nonumber \\
    \frac{g^2}{M^6_W}\mathcal{P}^{SR}_{\Theta}(x) & =\frac{g^2}{2(g')^2 x^6}-4\frac{(H')^2}{x^2} +\frac{(1-f^2)^2}{2x^6}-\frac{2\gamma}{x^2} (H^2-1)^2-\frac{Q^2(x)}{2x^6}\,.
\end{align}
The short-range force components have been defined  
in \eqref{srforce}, and imply that 
the short-range radial force component in this case is given by:
\begin{equation}\label{srradialforceCM}
    \frac{g^2}{M^2_W}\mathcal{F}^{SR}_R(x)= 4\pi x^2 \frac{g^2}{M^2_W}\mathcal{P}^{SR}_R(x)
\end{equation}
Separation of the EMT of the model into short and long range parts suggests that the equilibrium conditions \eqref{SRequilibruimEqHooft} must be modified in the following way:
\begin{equation}\label{SRequilibruimEqEW}
  \frac{d \mathcal{P}^{SR}_R (x)}{dx}+\frac{2}{x}( \mathcal{P}^{SR}_R (x)-x^2  \mathcal{P}^{SR}_{\Theta}(x))= \frac{\mathcal{P}_{ext}(x)}{x} \,,
\end{equation}
and
\begin{equation}\label{LRequilibruimEqEW}
   \frac{d \mathcal{P}^{LR}_R (x)}{dx}+\frac{2}{x}( \mathcal{P}^{LR}_R (x)-x^2  \mathcal{P}^{LR}_{\Theta} (x))= -\frac{\mathcal{P}_{ext}(x)}{x} \,,
\end{equation}
where we have introduced the external pressure:
\begin{equation}
     \frac{g^2}{M^4_W}\mathcal{P}_{ext}(x) =\frac{4\pi \rho_M (x) Q(x)}{x}\,.
\end{equation}
\par Integrating out Eqs.~\eqref{SRequilibruimEqEW}), \eqref{LRequilibruimEqEW}, we obtain:
\begin{equation}
     \mathcal{P}^{SR}_R (x)+P_{ext}(x) = \Sigma^{SR}(x)\,, \qquad
 \mathcal{P}^{LR}_R (x)-P_{ext}(x) = \Sigma^{LR}(x)\,,
\end{equation}
where, 
{\small \begin{equation}
P_{ext}(x) =\frac{1}{x^2} \int^{\infty}_x dx'x'\mathcal{P}_{ext}(x') \,, \quad
    \Sigma^{SR}(x)=-\frac{2}{x^2}\int^{\infty}_x  dx'x^{'3}\mathcal{P}^{SR}_{\Theta}(x')\,, \quad 
    \Sigma^{LR}(x)=-\frac{2}{x^2}\int^{\infty}_x  dx'x^{'3}\mathcal{P}^{LR}_{\Theta}(x')\,.
\end{equation}}
\par Such a pressure gives rise to a Coulomb force associated with the interaction of magnetically charged sphere $Q(r)$ acting on the magnetic charge density $\rho_M (r)$. The short-range equation $(\ref{SRequilibruimEqEW})$ describes the balance between the ``short-range stress”, which  tends to pull the monopole inwards, towards the centre, and the
repulsive magnetic ``Coulomb force”, which pushes the monopole outwards. On the other hand, the long-range equation $(\ref{LRequilibruimEqEW})$ describes the magnetostatic equilibrium between the ``Coulomb stress”, which is responsible for pushing the monopole outwards, and the magnetic ``Coulomb force”, which pulls the monopole inwards, towards its center.
Therefore, the total radial force is given by:
\begin{equation}\label{Total Radial Force EW}
    \vec{\mathcal{F}}^{SR}_{Rtotal}(x)=\vec{\mathcal{F}}^{SR}_R(x)+\vec{F}_{ext}(x)=4\pi x^2( \mathcal{P}^{SR}_R (x)+P_{ext}(x))\hat{r}\,,
\end{equation}
where we have defined the external force as 
$\vec{F}_{ext}(x) =4\pi x^2P_{ext}(x)\hat{r}$.
\begin{figure}[ht]
    \centering
\includegraphics[width=0.5\textwidth]{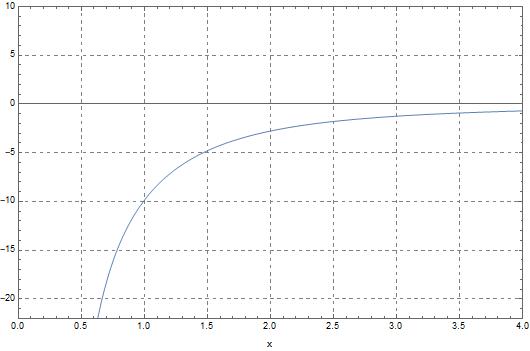}
    \caption{Total short range radial force  $(\ref{Total Radial Force EW})$ for the standard model value $\gamma=1.21382$ in the CM electroweak-magnetic-mononopole model. }
    \label{TotalSRForceEW}
\end{figure}
In figure $\ref{TotalSRForceEW}$ we depict the total short-range force as a function of $x$, for the standard model value $\gamma=1.21382$ \cite{ParticleDataGroup:2024cfk}. The reader should notice that such a force {\it violates} the local stability condition $\mathcal{F}^{SR}_{Rtotal}(x)\geq 0$ for the entirety of the radial force domain. Taking into account also the infinite energy of the configuration, this suggests that further regularization of the model might be required. 

\par Furthermore, we can calculate the short-range polar and azimuthal forces via:
\begin{equation}
    \mathcal{F}^{SR}_{\Theta} (\theta)=2\pi \sin(\theta) \int^{\infty}_0 dx' x^{'3}\mathcal{P}^{SR}_{\Theta}(x') \rightarrow -\infty
\end{equation}
\begin{equation}
    \mathcal{F}^{SR}_{\Psi} =\pi  \int^{\infty}_0 dx' x^{'3}\mathcal{P}^{SR}_{\Theta}(x') \rightarrow -\infty
\end{equation}
Such forces are singular, since near $x=0$ we have $x^3\mathcal{P}^{SR}_{\Theta}(x)\thicksim -1/x^3$ ({\it cf.} \eqref{sremtcm}). Therefore, even if we subtract the long-range contribution, the polar and azimuthal forces will cause an angular instability for the CM monopole solution. \\ \\
In the next section, we shall discuss the mechanical stability criteria for the Born-Infeld regularised version of the CM monopole, which has finite-energy density~\cite{Arunasalam:2017eyu,Mavromatos:2018kcd}.
\subsubsection{Finite-energy CM monopole with non-minimal couplings between the Higgs and hypercharge sectors}\label{sec:nmcoupl}
\begin{figure}[ht]
    \centering
    % Left figure
    \begin{subfigure}{0.35\textwidth}
        \centering
        \includegraphics[width=\linewidth]{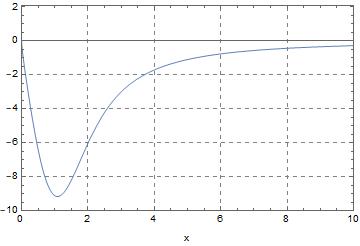} 
        \caption{Radial force for $h_2(H)=6H^{10}-5H^{12}$.  }
        \label{f2RadialForce}
    \end{subfigure}
    \hspace{0.1\textwidth} 
    \begin{subfigure}{0.35\textwidth}
        \centering
        \includegraphics[width=\linewidth]{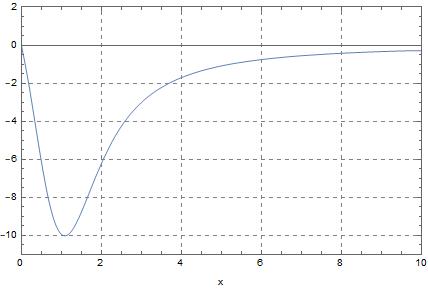} % replace with your figure file
        \caption {Radial force for $h_3(H)=8H^{8}-10H^{10}+3H^{12}$.  }
        \label{f3RadialForce}
    \end{subfigure}
    \caption{Radial force for dielectric functions $h_2 (H)$ and $h_3 (H)$  in the modified electroweak model with non-minimally coupled dielectric function in the hypercharge sector .}
\end{figure}

By using results from \ref{fpressureAnalysis} we can calculate polar and azimuthal forces. We obtain:
\begin{equation}
     \frac{g^2}{M^2_W}\mathcal{F}^2_{\Theta}(\theta)= -\pi\sin(\theta)\lim_{x \to 0 }[x^2\frac{g^2}{M^4_W}\mathcal{P}^2_R(x)]=0 \qquad \frac{g^2}{M^2_W}\mathcal{F}^2_{\psi}=-\frac{\pi}{2}\lim_{x \to 0  }[x^2 \frac{g^2}{M^4_W}\mathcal{P}^2_R(x)]=0
\end{equation}
\begin{equation}
     \frac{g^2}{M^2_W}\mathcal{F}^3_{\Theta}(\theta)= \pi\sin(\theta)\lim_{x \to 0 }[x^2\frac{g^2}{M^4_W}\mathcal{P}^3_R(x)]=0 \qquad \frac{g^2}{M^2_W}\mathcal{F}^3_{\psi}=-\frac{\pi}{2}\lim_{x \to 0  }[x^2 \frac{g^2}{M^4_W}\mathcal{P}^3_R(x)]=0
\end{equation}
This means that these particular models describe monopole configurations with a purely radial internal force field. 
As for the radial force we obtain:
\begin{equation}
    \frac{g^2}{M^2_W}\mathcal{F}^i_R(x)=4\pi x^2\frac{g^2}{M^2_W}\mathcal{P}^i_R (x)
\end{equation}
In figures $\ref{f2RadialForce}$ $\ref{f3RadialForce}$ we showcase radial force for various models. 

\subsubsection{Finite-energy CM monopole with non-minimal couplings between the Higgs and hypercharge sectors: Energy Momentum Tensor Decomposition into Short and Long-range parts}

In what follows, we shall apply the analysis of section \ref{slEMTparts}
to this case, in order to subtract the long-range electromagnetic part from the total EMT of the model, so as to isolate the short-range part, and examine the mechanical stability criteria on this part.
We remind the reader that it is this short-range part that satisfies the mechanical stability for the HP monopole.

To this end, we need first to evaluate the Electromagnetic  tensor of the model,  defined in \eqref{Nambu Definition of EM tensor}. For various models electromagnetic energy momentum tensor is given by:
\begin{equation}
    \mathcal{T}^i_{\mu\nu}=\{1+[h_i(H)-1]\cos^2(\theta_W)\}(\mathcal{F}_{\mu a}\mathcal{F}^a_{\; \; \nu}+\frac{g_{\mu\nu}}{4}\mathcal{F}_{ab}\mathcal{F}^{ab})
\end{equation}
In the radial gauge where $\hat{\phi}=-\hat{r}$, the components of this tensor are:
\begin{equation}
    \mathcal{F}_{r\theta}=\mathcal{F}_{r \psi}=0\,, \qquad
    g\mathcal{F}_{\theta \psi} = -\sin(\theta)Q(x)
\end{equation}
where,
\begin{equation}
    Q(x)= -\sin(\theta_W)\Big(1-f^2(x)\Big)+\frac{g}{g'}\, \cos(\theta_W)\,,
\end{equation}
which is associated with the magnetic charge across the magnetic monopole configuration. Indeed,  with the spatial components of the magnetic field given by $\mathcal{B}^i=-\frac{1}{2}\epsilon^{ijk}\mathcal{F}_{jk}$ one obtains the standard expression  for the magnetic charge $q_m$:
\begin{equation}\label{fqmcm}
q_m = \oint dS_r B^r =\frac{4\pi}{e}
\end{equation}
Moreover, it is useful for our analysis below to note that the magnetic charge density $ \vec{\nabla} \cdot \vec{\mathcal{B}}=4\pi \rho_M (r) $ is given by:
\begin{equation}
     \frac{g}{M^3_W} \rho_{M} (x)=\frac{1}{4\pi x^2}\frac{d Q}{dx}\,.
\end{equation}
Long range radial and polar components are given by:
\begin{equation}
    \frac{g^2}{M^4_W}\mathcal{P}^{iLR}_R(x)= \frac{g^2}{M^4_W}\mathcal{T}^{irr}=-\{1+[h_i(H)-1]\cos^2(\theta_W)\}\frac{Q^2(x)}{2x^4}
\end{equation}    
\begin{equation}
    \frac{g^2}{M^6_W}\mathcal{P}^{iLR}_{\Theta}(x)=\frac{g^2}{M^6_W}\mathcal{T}^{i\theta\theta}=\{1+[h_i(H)-1]\cos^2(\theta_W)\}\frac{Q^2(x)}{2x^6}
\end{equation}
Then the (radial and polar) components of the short-range pressure, are given by:
\begin{equation}
    \frac{g^2}{M^4_W}\mathcal{P}^{iSR}_R (x) = -\frac{g^2}{2(g')^2 x^4}h_i(H) +(\frac{f'}{x})^2+4(H')^2-  \frac{2f^2H^2}{x^2}-\frac{(1-f^2)^2}{2x^4}-2\gamma (H^2-1)^2+\{1+[h_i(H)-1]\cos^2(\theta_W)\}\frac{Q^2(x)}{2x^4}
\end{equation}
\begin{equation}
    \frac{g^2}{M^6_W}\mathcal{P}^{iSR}_{\Theta}(x)  =\frac{g^2}{2(g')^2 x^6}h_i(H)-4\frac{(H')^2}{x^2} +\frac{(1-f^2)^2}{2x^6}-\frac{2\gamma}{x^2} (H^2-1)^2-\{1+[h_i(H)-1]\cos^2(\theta_W)\}\frac{Q^2(x)}{2x^6}
\end{equation}
The short-range force components have been defined  
in \eqref{srforce}, and imply that 
the short-range radial force component in this case is given by:
\begin{equation}\label{SRradialforceEWMod}
    \frac{g^2}{M^2_W}\mathcal{F}^{iSR}_R(x)= 4\pi x^2 \frac{g^2}{M^2_W}\mathcal{P}^{iSR}_R(x)
\end{equation}
Separation of the EMT of the model into short and long range parts suggests that the equilibrium conditions \eqref{SRequilibruimEqHooft} must be modified in the following way:
\begin{equation}\label{SRequilibruimEqEWMod}
  \frac{d \mathcal{P}^{iSR}_R (x)}{dx}+\frac{2}{x}( \mathcal{P}^{iSR}_R (x)-x^2  \mathcal{P}^{iSR}_{\Theta}(x))= \frac{\mathcal{P}^i_{ext}(x)}{x} \,,
\end{equation}
and
\begin{equation}\label{LRequilibruimEqEWMod}
   \frac{d \mathcal{P}^{iLR}_R (x)}{dx}+\frac{2}{x}( \mathcal{P}^{iLR}_R (x)-x^2  \mathcal{P}^{iLR}_{\Theta} (x))= -\frac{\mathcal{P}^i_{ext}(x)}{x} \,,
\end{equation}
where we have introduced the external pressure:
\begin{equation}
     \frac{g^2}{M^4_W}\mathcal{P}^i_{ext}(x) =\{1+[h_i(H)-1]\cos^2(\theta_W)\}\frac{4\pi \rho_M (x) Q(x)}{x}\,.
\end{equation}
\par Such a pressure gives rise to a Coulomb force associated with the interaction of magnetically charged sphere $Q(x)$ acting on the magnetic charge density $\rho_M (x)$. The short-range equation $(\ref{SRequilibruimEqEWMod})$ describes the balance between the ``short-range stress”, which  tends to pull the monopole inwards, towards the center, and the
repulsive magnetic ``Coulomb force”, which pushes the monopole outwards. On the other hand, the long-range equation $(\ref{LRequilibruimEqEWMod})$ describes the magnetostatic equilibrium between the ``Coulomb stress”, which is responsible for pushing the monopole outwards, and the magnetic ``Coulomb force”, which pulls the monopole inwards, towards its center.
Therefore, the total radial force is given by:
\begin{equation}\label{Total Radial Force EW Mod}
    \vec{\mathcal{F}}^{SR}_{Rtotal}(x)=\vec{\mathcal{F}}^{SR}_R(x)+\vec{F}_{ext}(x)=4\pi x^2( \mathcal{P}^{SR}_R (x)+P_{ext}(x))\hat{r}\,,
\end{equation}

\begin{figure}[ht]
    \centering
    % Left figure
    \begin{subfigure}{0.4\textwidth}
        \centering
        \includegraphics[width=\linewidth]{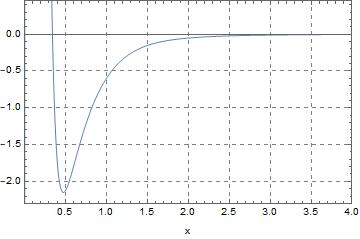} 
        \caption{Total short range pressure $\mathcal{P}^{SR}_R(x)+P_{ext}(x)$ for $h_2(H)=6H^{10}-5H^{12}$.  }
        \label{f2totalSRPressure}
    \end{subfigure}
    \hspace{0.1\textwidth} 
    \begin{subfigure}{0.4\textwidth}
        \centering
        \includegraphics[width=\linewidth]{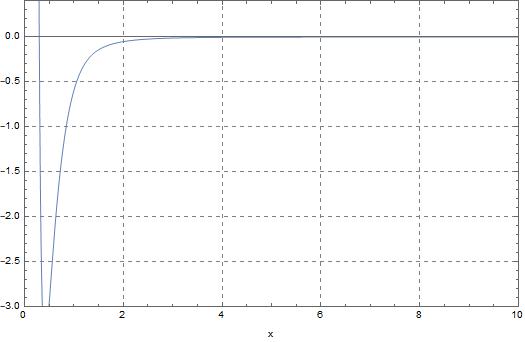} % replace with your figure file
        \caption {Total short range pressure $\mathcal{P}^{SR}_R(x)+P_{ext}(x)$ for $h_3(H)=8H^{8}-10H^{10}+3H^{12}$.  }
        \label{f3totalSRPressures}
    \end{subfigure}
    \caption{Total short-range pressure for dielectric functions $h_2 (H)$ and $h_3 (H)$ in the modified electroweak model with non-minimally coupled dielectric function in the hypercharge sector .}
\end{figure}
In figures $\ref{f2totalSRPressure}$ $\ref{f3totalSRPressures}$ we depict the total short-range pressure as a function of $x$, for the standard model value $\gamma=1.21382$ \cite{ParticleDataGroup:2024cfk}. The reader should notice that such a force (whose behaviour is similar to that of the pressure, by construction, see \eqref{Total Radial Force EW Mod}) {\it violates} the local stability condition $\mathcal{F}^{SR}_{Rtotal}(x)\geq 0$ for the majority of the radial force domain. On the other hand, near the origin $x=0$, the short-range radial force approaches positive infinity. Therefore, this instability might 
suggest that further regularization of the model might be in order. 

\par Moreover we calculate the short-range polar and azimuthal forces via:
\begin{equation}
    \mathcal{F}^{SR}_{\Theta} (\theta)=2\pi \sin(\theta) \int^{\infty}_0 dx' x^{'3}\mathcal{P}^{SR}_{\Theta}(x') 
\end{equation}
\begin{equation}
     \mathcal{F}^{SR}_{\Psi} =\pi  \int^{\infty}_0 dx' x^{'3}\mathcal{P}^{SR}_{\Theta}(x')
\end{equation}
And we find that:
\begin{equation}
    \mathcal{F}^{2SR}_{\Theta} (\theta)\rightarrow -\infty \qquad \mathcal{F}^{3SR}_{\Theta} (\theta)\rightarrow-\infty
\end{equation}
\begin{equation}
    \mathcal{F}^{2SR}_{\Psi} \rightarrow -\infty\qquad \mathcal{F}^{3SR}_{\Psi} \rightarrow -\infty
\end{equation}
This is due the fact that $\int^{\infty}_0 dx \mathcal{P}^{iSR}_{\Theta}(x)x^3 \rightarrow -\infty$, since $\mathcal{P}^{iSR   }_{\Theta}(x)x^3 \xrightarrow{x\rightarrow 0}-\frac{g^2}{2g'^2x^3}\sin^2(\theta_W)\cos^2(\theta_W)$.
\subsubsection{String-Inspired Born-Infeld Extension of the CM Model: Internal Force field}

This model has been discussed in section \ref{BI}.
We first examine the polar and azimuthal forces, defined in \ref{InternalForceGeneralStructure}, which in this case are given by:
\begin{equation}
    \frac{g^2}{M^2_W}\mathcal{F}_{\theta}(\theta)=-\pi\sin(\theta)\lim_{r \to 0}[x^2 \frac{g^2}{M^2_W}\mathcal{P}_R (x)]=1.8263\pi \beta_0\times\sin(\theta)\,, \qquad 
    \mathcal{F}_{\Psi}= -\frac{\pi}{2}\lim_{r \to 0}[r^2 \mathcal{P}_R (r)]=0.91315\pi\beta_0 \,,
\end{equation}
where we have used the asymptotic behavior mentioned in section \ref{Electroweak Monopole Analysis}. We observe that the Born-Infeld extension regularizes the polar and azimuthal forces. Thus, the Born-Infeld extended monopole configuration is now free from angular instabilities. 
\par
\begin{figure}[ht]
        \centering
        \includegraphics[width=0.5\textwidth]{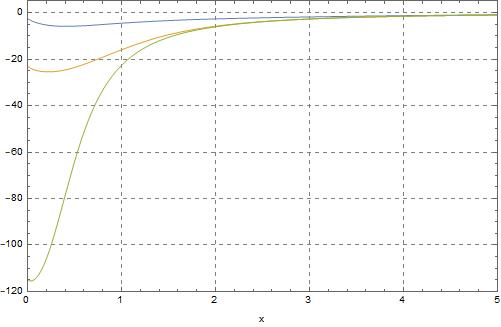} 
          \caption{Radial  force component for  $\gamma =1.21382$,  Standard Model Value $(SM)$  in the string-inspired Born-Infeld electroweak model. Such components are given for $\beta_0$ values $0.1,1,5$, which are associated with blue, yellow and green line respectively.}\label{Radial Force String}
\end{figure}

  The radial force component of the monopole configuration is given by:
\begin{equation}\label{RadialForceString} 
   \frac{g^2}{M^2_W}\mathcal{F}_R(r)=  4\pi x^2\frac{g^2}{M^2_W}\mathcal{P}_R (x)
\end{equation}
In figure $\ref{Radial Force String}$ we show the radial  component of the force for various values of the dimensionless Born-Infeld parameter, $\beta_0$, and the standard-model value $\gamma =1.21382$.
We observe that the radial force is also free of singularities, negative and becomes smaller and smaller as we decrease $\beta_0$.  

\subsubsection{String-Inspired Born-Infeld Model: Energy Momentum Tensor Decomposition into Short and Long-range parts}
\par The EM tensor \eqref{ememt} for this model, 
in the radial gauge where $\hat{\phi}=-\hat{r}$, 
has the following components:
\begin{equation}
    \mathcal{F}_{r\theta}=\mathcal{F}_{r \psi}=0\,, \qquad 
    g\mathcal{F}_{\theta \psi} = -\sin(\theta)Q(x)\,,
\end{equation}
with
\begin{equation}
    gQ(x)= -sin(\theta_W)(1-f^2(x))+\frac{g}{g'}\, \cos(\theta_W)\,,
\end{equation}
which, as in the previously studied case of the initial CM monopole, is associated with the magnetic charge density \eqref{mcdcm}, and thus ultimately with the monopole magnetic charge, which remains the same as in the initial CM case, \eqref{qmcm}.

\par In order to obtain the long-range component of the EMT in this non-linear model, one considers the gauge sector of the string-inspired model \eqref{BIlagew}:
\begin{equation}
    S_{gauge} = \int d^4x \sqrt{-g}[-\frac{1}{4}\vec{F}_{\mu\nu}\cdot \vec{F}^{\mu\nu}+\beta^2(1-\sqrt{1+\frac{1}{2\beta^2}G_{\mu\nu}G^{\mu\nu}})]
\end{equation}
Since we are interested in a (static) magnetic monopole configuration, we consider $\Tilde{G}_{\mu\nu}=0$, Then, we perform a gauge-invariant transformation in the field space to obtain the physical fields:
\begin{equation}
    \vec{F}_{\mu\nu}\cdot \hat{\phi}= -\sin(\theta_W)\mathcal{F}_{\mu\nu} -\cos(\theta_W)Z_{\mu\nu}\,, \qquad 
    G_{\mu\nu} = \cos(\theta_W) \mathcal{F}_{\mu\nu}-sin(\theta_W)Z_{\mu\nu}\,,.
\end{equation} 
It suffices for our purposes to determine the long-range contributions to the EMT to ignore the 
$Z_{\mu\nu}$ contributions, thus obtaining the following non-linear action for the electromagnetic field:
\begin{equation}\label{SBI}
    S_{EM}=\int d^4x \sqrt{-g}[-\frac{1}{4}\mathcal{F}_{\mu\nu}\mathcal{F}^{\mu\nu}+\frac{\cos^2(\theta_W)}{4}\mathcal{F}_{\mu\nu}\mathcal{F}^{\mu\nu}+\beta^2(1-\mathcal{R}) ]\,,
\end{equation}
where $\mathcal{R}$ is defined as:
\begin{equation}
    \mathcal{R}=\sqrt{1+\frac{\cos^2(\theta_W)}{2\beta^2}\mathcal{F}_{ab}\mathcal{F}^{ab}}\,.
\end{equation}
The electromagnetic EMT is given by: 
\begin{align}\label{BIemt}
    \mathcal{T}^{\mu\nu} &= \frac{2}{\sqrt{-g}}\frac{\delta S_{EM}}{\delta g_{\mu\nu}}|_{g_{\mu\nu}=\eta_{\mu\nu}} =
     -\mathcal{F}^{\; \; \mu}_{\sigma}\mathcal{F}^{\sigma \nu}+\cos^2(\theta_W)\mathcal{F}^{\; \; \mu}_{\sigma}F^{\sigma \nu}(1-\frac{1}{\mathcal{R}}) \nonumber \\
&+ g^{\mu\nu}[\frac{1}{4}\mathcal{F}_{ab}\mathcal{F}^{ab}-\frac{\cos^2(\theta_W)}{4}\mathcal{F}_{ab}\mathcal{F}^{ab}-\beta^2(1-\mathcal{R})]\,,    
\end{align}
which is non-linear, as exptected from the non-linear nature of the Born-Infeld electrodynamics \eqref{SBI}.

The radial and polar components of \eqref{BIemt} are given by:
\begin{equation}
    \frac{g^2}{M^4_W}\mathcal{T}^{rr}=-\frac{Q^2(x)}{2x^4}[1-\cos^2(\theta_W)-\frac{2x^4 \beta^2_0}{Q^2(x)}(1-\mathcal{R}(x))]\,,
\end{equation}
\begin{equation}
     \frac{g^2}{M^4_W}\mathcal{T}^{\theta \theta} =\frac{Q^2(x)}{2x^6}[1+\frac{cos^2(\theta_W)}{\mathcal{R}(x)}+\frac{2x^6\beta^2_0}{Q^2(x)}(1-\mathcal{R}(x))]\,,
\end{equation}
respectively.

For completeness we note at this point that 
the long-range components of the pressure  are the same as the ones in the initial electroweak model \eqref{ememtCM}. We proceed then with the short-range internal pressure defined via \eqref{sremt}. The short-range radial pressure is given by:    
\begin{align}    
    \frac{g^2}{M^4_W}\mathcal{P}^{SR}_R (x) &= -\beta^2_0(\sqrt{1+\frac{g^2}{(g')^2\beta^2_0 x^4}}-1)+(\frac{f'}{x})^2+4(H')^2-\frac{2f^2H^2}{x^2}-\frac{(1-f^2)^2}{2x^4}-2\gamma(H^2-1)^2
\nonumber \\
    &+\frac{Q^2(x)}{2x^4}[1-\cos^2(\theta_W)-\frac{2x^4 \beta^2_0}{Q^2(x)}(1-\mathcal{R}(x))]
\end{align}
On the other hand, the short-range polar pressure is:
\begin{align}
\frac{g^2}{M^6_W}\mathcal{P}^{SR}_{\Theta} (x) &= -\frac{\beta^2_0}{x^2}(\sqrt{1+\frac{g^2}{\beta^2_0(g')^2 x^4}}-1)+\frac{g^2}{(g')^2x^6\sqrt{1+\frac{g^2}{\beta^2_0 (g')^2 x^4}}}-\frac{4(H')^2}{x^2} +\frac{(1-f^2)^2}{2x^6}-\frac{2\gamma}{x^2}(H^2-1)^2
\nonumber \\
    &-\frac{Q^2(x)}{2x^6}[1+\frac{\cos^2(\theta_W)}{\mathcal{R}(x)}+\frac{2x^6\beta^2_0}{Q^2(x)}(1-\mathcal{R}(x))]
\end{align}
We proceed with the short-range internal force field components. The short-range radial force component is given by:
\begin{equation}\label{SRradialForceString}
   \frac{g^2}{M^2_W}\vec{\mathcal{F}}^{SR}_R(r)= 4\pi x^2 \frac{g^2}{M^2_W} \mathcal{P}^{SR}_R(x)\hat{r}
\end{equation}
Separation of the EMT into short- (SR) and long-range (LR) parts suggests the following equilibrium condition in this case:
\begin{equation}\label{SRequilibruimEqString}
  \frac{d \mathcal{P}^{SR}_R (x)}{dx}+\frac{2}{x}( \mathcal{P}^{SR}_R (x)-x^2  \mathcal{P}^{SR}_{\Theta}(x))= \frac{\mathcal{P}_{ext}(x)}{x} \,,
\end{equation}
\begin{equation}\label{LRequilibruimEqString}
   \frac{d \mathcal{P}^{LR}_R (x)}{dx}+\frac{2}{x}( \mathcal{P}^{LR}_R (x)-x^2  \mathcal{P}^{LR}_{\Theta}(x))= \frac{\mathcal{P}_{ext}(x)}{x} \,,
\end{equation}
\par Upon integration, Eqs.~\eqref{SRequilibruimEqString}, \eqref{LRequilibruimEqString}) yield:
\begin{equation}
     \mathcal{P}^{SR}_R (x)+P_{ext}(x) = \Sigma^{SR}(x)
\,, \qquad 
 \mathcal{P}^{LR}_R (x)-P_{ext}(x) = \Sigma^{LR}(x)
\end{equation}
where, 
\begin{equation}
P_{ext}(x) = \frac{1}{x^2}\int^{\infty}_x dx'x'\mathcal{P}_{ext}(x') \,.
\end{equation}
\par As in the initial CM monopole configuration, we obtain an induced  Coulomb force associated with the interaction of a magnetically charged sphere $Q(r)$ acting on the magnetic charge density $\rho_M (r)$. The short-Range equation \eqref{SRequilibruimEqString} describes the balance between the attractive ``short-range stress”, pulling the monopole towards its center, and the repulsive magnetic ``Coulomb force”,  pushing the monopole outwards. On the other hand, the long-range equation \eqref{LRequilibruimEqString} describes a magnetostatic equilibrium between the ``Coulomb stress”, which pushes the monopole
outwards, and the magnetic ``Coulomb force”, which pulls the monopole inwards, towards its center. The total radial force in the Born-Infeld finite-energy extension of the CM is given by:
\begin{equation}\label{Total Radial Force String}
    \vec{\mathcal{F}}^{SR}_{Rtotal}(x)=\vec{\mathcal{F}}^{SR}_R(x)+\vec{F}_{ext}(x)=4\pi x^2( \mathcal{P}^{SR}_R (x)+P_{ext}(x))\hat{r}
\end{equation}
\begin{figure}[ht]
    \centering
\includegraphics[width=0.5\textwidth]{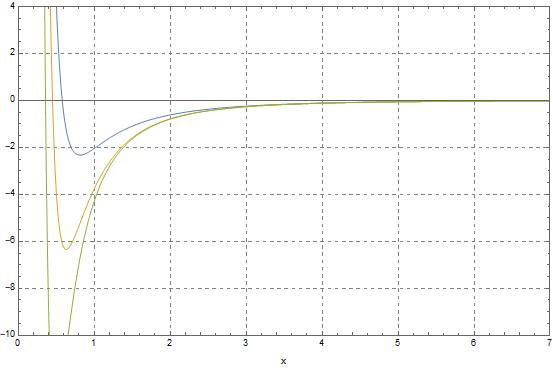}
    \caption{Total short-range pressure in $(\ref{Total Radial Force String})$ for the standard model value $\gamma=1.21382$, in the string-inspired Born-Infeld electroweak model. Green, orange and blue represent $\beta_0$ values $5,$ $1$ and $0.1$, respectively. }\label{TotalSRForceString}
\end{figure}
In figure \ref{TotalSRForceString} we depict the total short range-pressure for the standard model value $\gamma=1.21382$. The behavior is similar to the 
CM monopole with non-minimal couplings between the Higgs and hypercharge sectors, discussed previously. 
Indeed, such a force (whose behaviour is similar to that of the pressure, by
construction) violates the local stability condition $\mathcal{F}^{SR}_{Rtotal}(x)\geq 0$ for the majority of the radial force domain, but near the origin $x=0$ the short-range radial force approaches positive infinity. Therefore, at this particular region the local stability criterion is satisfied, but it remains singular, despite the finite-energy of the Born-Infeld CM extension. 
\par The short range polar and azimuthal forces, on the other hand, can be calculated via:
\begin{equation}
    \mathcal{F}^{SR}_{\Theta} (\theta)=2\pi \sin(\theta) \int^{\infty}_0 dx' x^{'3}\mathcal{P}^{SR}_{\Theta}(x') \rightarrow -\infty
\end{equation}
\begin{equation}
    \mathcal{F}^{SR}_{\Psi} =\pi  \int^{\infty}_0 dx' x^{'3}\mathcal{P}^{SR}_{\Theta}(x') \rightarrow -\infty
\end{equation}
Such an integral is singular, since:
\begin{equation}
    x^3\mathcal{P}^{SR}_{\Theta}(x)\xrightarrow{ x\rightarrow 0}-\beta_0\frac{ 1.8264}{ x}-0.66\, x^{\sqrt{3}-2}-\frac{1.28307}{x^3}+\frac{3.336}{\sqrt{x^6+\frac{3.336}{\beta^2_0}x^2}}
\end{equation}
Therefore, we observe that the short range polar and azimuthal forces are, like the case of the initial CM monopole, singular, indicating angular instability.  

\subsubsection{Further Non-linear-Hypercharge Extensions  of the Electroweak model }\label{sec:extnBI}

In this subsection, for completeness, we shall extend briefly our discussion on stability criteria to recent non-linear extensions of the hypercharge sector of the electroweak monopole, generalizing the Born-Infeld model, discussed above, which are given in \cite{DeFabritiis:2021eah}. That work considered two kinds of 
non-linear hypercharge sector 
extensions of the electroweak monopole: 
\begin{equation}
    \mathcal{L}_i=-\frac{1}{4}\vec{F}^{\mu\nu} \cdot \vec{F}_{\mu\nu}+\mathcal{L}_{Y_i}+D_{\mu}\phi^{\dagger}D^{\mu}\phi-\frac{\lambda}{2}\left(\phi^{\dagger}\phi -\frac{\mu^2}{\lambda}\right)^2\,, \quad i=\rm log\,, \rm exp \,,
\end{equation}
where $\mathcal{L}_{Y_i}, \, i=\rm log, \rm exp$, 
correspond to the following hypercharge extensions:
\begin{equation}
    \mathcal{L}_{Y_{\rm log}} =-\beta^2 \log(1+\frac{G_{\mu\nu}G^{\mu\nu}}{4\beta^2})= -\beta^2 \log (1+\frac{1}{2(g')^2\beta^2 r^4})\,,
\end{equation}
and 
\begin{equation}
     \mathcal{L}_{Y_{\rm exp}} =\beta^2[-1+\exp(-\frac{G_{\mu\nu}G^{\mu\nu}}{4\beta^2})]=\beta^2[-1+\exp(-\frac{1}{2(g')^2\beta^2 r^4})]\,.
\end{equation}
The corresponding EMT is given by:
\begin{equation}
    T^{\mu\nu}=-g^{\mu\nu}\mathcal{L}+t^{\mu\nu}_{Y_i}  -\vec{F}^{\mu\sigma} \cdot\vec{F}^{\nu}_{\; \; \; \sigma }+2(D^\mu\phi)^{\dagger}D^{\nu}\phi\,, \quad i=\rm log\,, \rm exp \,,
\end{equation}
where, 
\begin{equation}
    t^{\mu\nu}_{Y_{\rm log}}= -\frac{G^{\mu b}G^{\nu }_{\;  b}}{1+\frac{G_{ab}G^{ab}}{4\beta^2}}\,, \qquad 
    t^{\mu\nu}_{Y_{\rm exp}} = -G^{\mu b}G^{\nu}_{\;  b}\exp[-\frac{G_{ab}G^{ab}}{4\beta^2}]\,,
\end{equation}
for the logarithmic and exponential hypercharge extensions, respectively. Following the analysis in \cite{DeFabritiis:2021eah}, the field configurations are the same as in the CM case.
For concreteness, 
we restrict ourselves to the radial gauge in what follows. 

For the {\it Logarithmic Model}, the  expressions for the Radial $\mathcal{P}_R(x)$ 
and polar components $\mathcal{P}_{\Theta}(x)$  of the pressure 
are given by:
\begin{equation}
    \frac{g^2}{M^4_W}\mathcal{P}_R(x) = (\frac{f'}{x})^2+4(H')^2-\frac{(1-f^2)^2}{2x^4}-\frac{2f^2H^2}{x^2}-2\gamma (H^2-1)^2-\beta^2_0 \log (1+\frac{g^2}{2(g')^2 \beta^2_0 x^4})\,,
\end{equation}
and
\begin{equation}
    \frac{g^2}{M^6_W}\mathcal{P}_{\Theta}(x)=\frac{g^2}{g'^2x^2}\frac{1}{1+\frac{g^2}{2g'^2\beta^2_0 x^4}}-\frac{\beta^2_0}{x^2}\log(1+\frac{g^2}{2g'^2\beta^2_0 x^4})-4(\frac{H'}{x})^2+\frac{(1-f^2)^2}{2x^6}-\frac{2\gamma}{x^2}(H^2-1)^2\,,
\end{equation}
respectively. 

On the other hand, the corresponding components for the {\it Exponential Model}, are:
\begin{equation}
    \frac{g^2}{M^4_W}\mathcal{P}_R (x)=-\beta^2_0[1-\exp(-\frac{g^2}{2(g')^2 \beta^2_0 x^4})]+(\frac{f'}{x})^2+4(H')^2-\frac{(1-f^2)^2}{2x^4}-\frac{2f^2H^2}{x^2}-2\gamma (H^2-1)^2\,,
\end{equation}
and
\begin{equation}
    \frac{g^2}{M^6_W}\mathcal{P}_{\Theta}(x)=-\frac{\beta^2_0}{x^2}(1-\exp(-\frac{g^2}{2(g')^2\beta^2_0 x^4}))+\frac{g^2}{(g')^2x^6}\exp(-\frac{g^2}{2(g')^2 \beta^2_0 x^4})-4(\frac{H'}{x})^2+\frac{(1-f^2)^2}{2x^6}-\frac{2\gamma}{x^2}(H^2-1)^2\,.
\end{equation}

As in previous cases, our stability criterion will be examined with reference to the short-range contributions
to the 
corresponding Force components in the models, after subtraction of the (long-range) electromagnetic (EM) contribution from the EMT, as done in section \ref{slEMTparts}, see eq.~\eqref{sremt}. The EM field tensor is given by:
\begin{equation}
    \mathcal{F}_{\mu\nu}=-\sin(\theta_W)\vec{F}_{\mu\nu}\cdot \hat{\phi} +\cos(\theta_W)G_{\mu\nu}\,,
\end{equation}
with components:
\begin{equation}
    \mathcal{F}_{r\psi}=\mathcal{F}_{r\theta}=0 \quad \quad g\mathcal{F}_{\theta \psi } =-\sin(\theta)Q(x)
\end{equation}
where $Q(x)$ is given by:
\begin{equation}
    Q(x)=-\sin(\theta_W)(1-f^2(x))+\frac{g}{g'}\cos(\theta_W)
\end{equation}
Then, by calculating the appropriate EMT $\mathcal{T}^{\mu\nu}$ for each model we can calculate the short-range contribution $\mathcal{T}^{SR\mu\nu}$, as in our previous discussion, \eqref{sremt}:
\begin{equation}\label{sremtextBI}
    \mathcal{T}^{SR\mu\nu}=T^{\mu\nu}-\mathcal{T}^{\mu\nu}
\end{equation}

Separation of the EMT into short- (SR) and long-range (LR) parts suggests the following equilibrium condition in this case:
\begin{equation}\label{SRequilibruimEq}
  \frac{d \mathcal{P}^{SR}_R (x)}{dx}+\frac{2}{x}( \mathcal{P}^{SR}_R (x)-x^2  \mathcal{P}^{SR}_{\Theta}(x))= \frac{\mathcal{P}_{ext}(x)}{x} \,,
\end{equation}
\begin{equation}\label{LRequilibruimEq}
   \frac{d \mathcal{P}^{LR}_R (x)}{dx}+\frac{2}{x}( \mathcal{P}^{LR}_R (x)-x^2  \mathcal{P}^{LR}_{\Theta}(x))= -\frac{\mathcal{P}_{ext}(x)}{x} \,,
\end{equation}
\par Upon integration, Eqs.~\eqref{SRequilibruimEq}, \eqref{LRequilibruimEq}) yield:
\begin{equation}
     \mathcal{P}^{SR}_R (x)+P_{ext}(x) = \Sigma^{SR}(x)
\,, \qquad 
 \mathcal{P}^{LR}_R (x)-P_{ext}(x) = \Sigma^{LR}(x)
\end{equation}
where, 
\begin{equation}
P_{ext}(x) = \frac{1}{x^2}\int^{\infty}_x dx'x'\mathcal{P}_{ext}(x') \,.
\end{equation}

For both models, the
Polar and Azimuthal forces are zero, whilst the radial component of the force is non vanishing.  

\begin{figure}[ht]
    \centering
\includegraphics[width=0.35\textwidth]{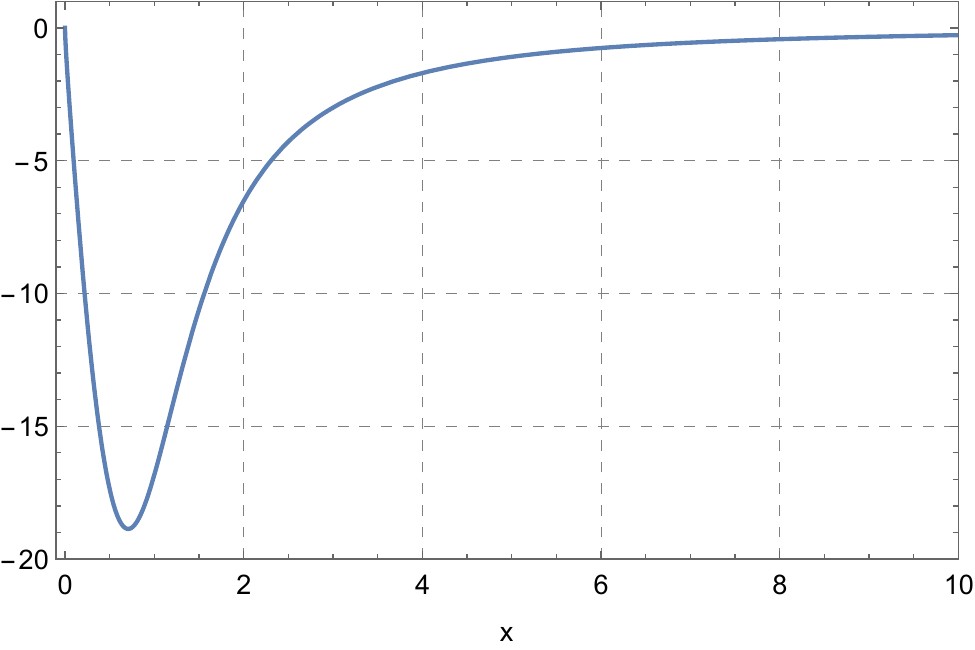}
    \caption{$\mathcal{F}_R(x)$ for $\beta_0=1$ for the logarithmic model.}\label{Radial Force Log}
\end{figure}

For the logarithmic model, the radial force 
is given by:
\begin{equation}
    \frac{g^2}{M^2_W}\mathcal{F}_R(x)=4 \pi x^2 \frac{g^2}{M^2_W}\mathcal{P}_R(x)\,,
\end{equation}
and is depicted in figure \ref{Radial Force Log}. 

In order to calculate the EM EMT we consider:
\begin{equation}
    S_{gauge}=\int d^4x (-\frac{1}{4}\vec{F}_{\mu\nu}\cdot \vec{F}^{\mu\nu}-\beta^2 \log (1+\frac{1}{4\beta^2}G_{\mu\nu}G^{\mu\nu}))\,,
\end{equation}
and the EM action can be obtained from this by appropriate projections:
\begin{equation}
    S_{EM} =\int d^4x\sqrt{-g}[-\beta^2\log(1+\frac{\cos^2(\theta_W)}{4\beta^2}\mathcal{F}_{ab}\mathcal{F}^{ab})-\frac{1}{4}\mathcal{F}_{ab}\mathcal{F}^{ab}+\frac{1}{4}\cos^2(\theta_W)\mathcal{F}_{ab}\mathcal{F}^{ab}]\,.
\end{equation}
The relevant EM EMT, then, reads:
\begin{align}\label{ETEMT}
    \mathcal{T}^{\mu\nu}&=\frac{2}{\sqrt{-g}}\frac{\delta S_{EM}}{\delta g_{\mu\nu}}=-g^{\mu\nu}(-\beta^2\log(1+\frac{\cos^2(\theta_W)}{4\beta^2}\mathcal{F}_{ab}\mathcal{F}^{ab})-\frac{1}{4}\mathcal{F}_{ab}\mathcal{F}^{ab}+\frac{1}{4}\cos^2(\theta_W)\mathcal{F}_{ab}\mathcal{F}^{ab}) \nonumber \\
    & -\cos^2(\theta_W)\frac{\mathcal{F^{\;  \mu}_{\sigma}}\mathcal{F}^{\sigma \nu}}{1+\frac{1}{4\beta^2}\mathcal{F}^{ab}\mathcal{F}^{ab}}-\mathcal{F}^{\; \mu}_{\sigma}\mathcal{F}^{\sigma \nu}+\cos^2(\theta_W)\mathcal{F}^{\; \mu}_{\sigma} \mathcal{F}^{\sigma \nu}\,.
\end{align}
The long-range pressure components are given by:
\begin{equation}
    \frac{g^2}{M^4_W}\mathcal{P}^{LR}_R(x)=-\beta^2_0 \log (1+\frac{\cos^2(\theta_W)}{2\beta^2_0}\frac{Q^2(x)}{x^4})-\frac{Q^2(x)}{2 x^4}+\frac{Q^2(x)}{2 x^4}\cos^2(\theta_W)
\end{equation}
\begin{equation}
    \frac{g^2}{M^6_W}\mathcal{P}^{LR}_{\Theta}(x)=\frac{g^2}{M^6_W}\mathcal{T}^{\theta \theta} = \frac{Q^2(x)}{x^6}(-\cos^2(\theta_W)+1 +\frac{\cos^2(\theta_W )}{1+\frac{Q^2(x)}{2\beta^2_0 x^4}})-\frac{\beta^2_0}{x^2} \log (1+\frac{\cos^2(\theta_W)}{2\beta^2_0}\frac{Q^2(x)}{x^4})-\frac{Q^2(x)}{2 x^6}+\frac{Q^2(x)}{2 x^6}\cos^2(\theta_W)\,.
\end{equation}
On subtracting such components, using \eqref{sremtextBI}, we obtain the corresponding short-range (SR) prerssure and internal force components. 

\begin{figure}[ht]
    \centering
\includegraphics[width=0.35\textwidth]{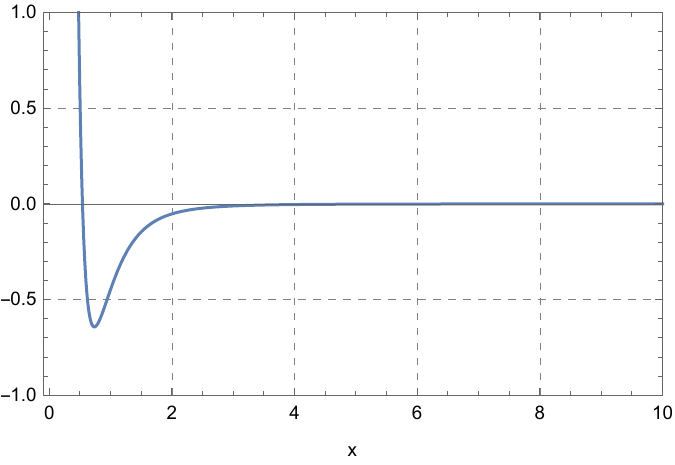}
    \caption{$\mathcal{P}^{SR}_R(x)+P_{ext}(x)$ for $\beta_0=1$}\label{Stability Cond β=1}
\end{figure}

Then, the Laue's local-stability condition is expressed as:
\begin{equation}
    \mathcal{P}^{SR}_R(x)+P_{ext}(x)\geq 0
\end{equation}
In figure \ref{Stability Cond β=1} we demonstrate the {\it violation} of this condition in the case of the logarithmic model.

\begin{figure}[ht]
    \centering
\includegraphics[width=0.35\textwidth]{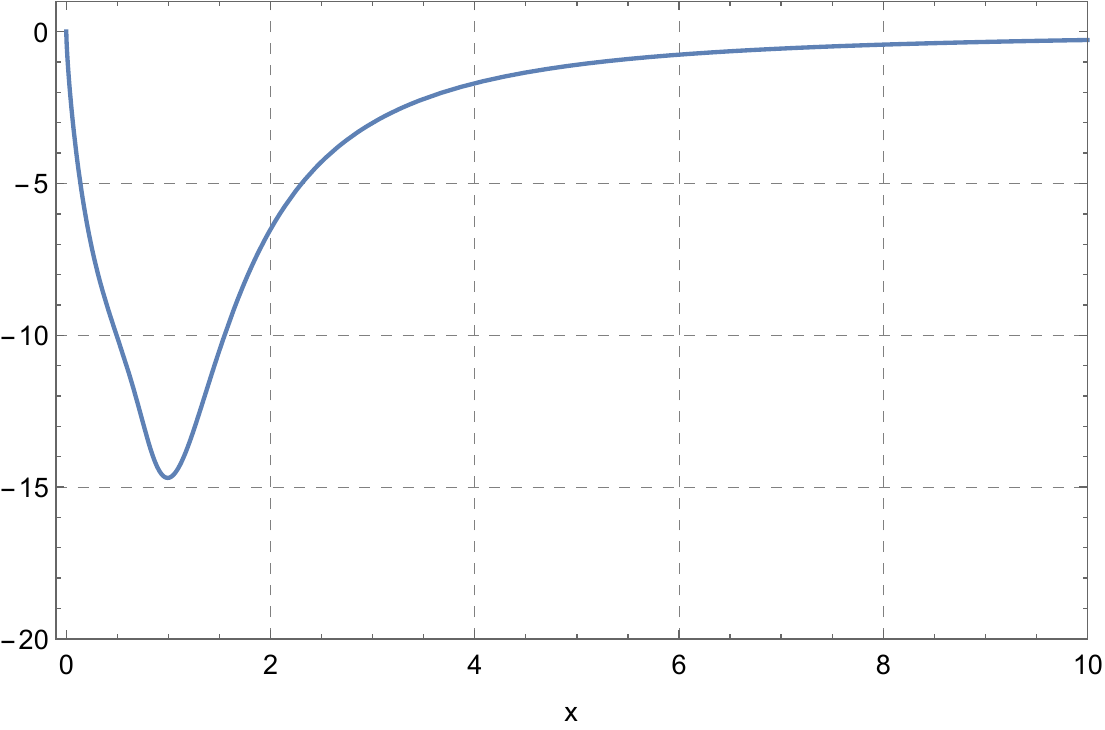}
    \caption{$\mathcal{F}_R(x)$ for $\beta_0=1$}\label{Radial Force Exp}
\end{figure}

In the case of the {\it Exponential Model}, the radial force is given by:
\begin{equation}
    \frac{g^2}{M^2_W}\mathcal{F}_R(x)=4 \pi x^2 \frac{g^2}{M^2_W}\mathcal{P}_R(x)\,,
\end{equation}
and is depicted in figure \ref{Radial Force Exp}. 

\begin{figure}[ht]
    \centering
\includegraphics[width=0.35\textwidth]{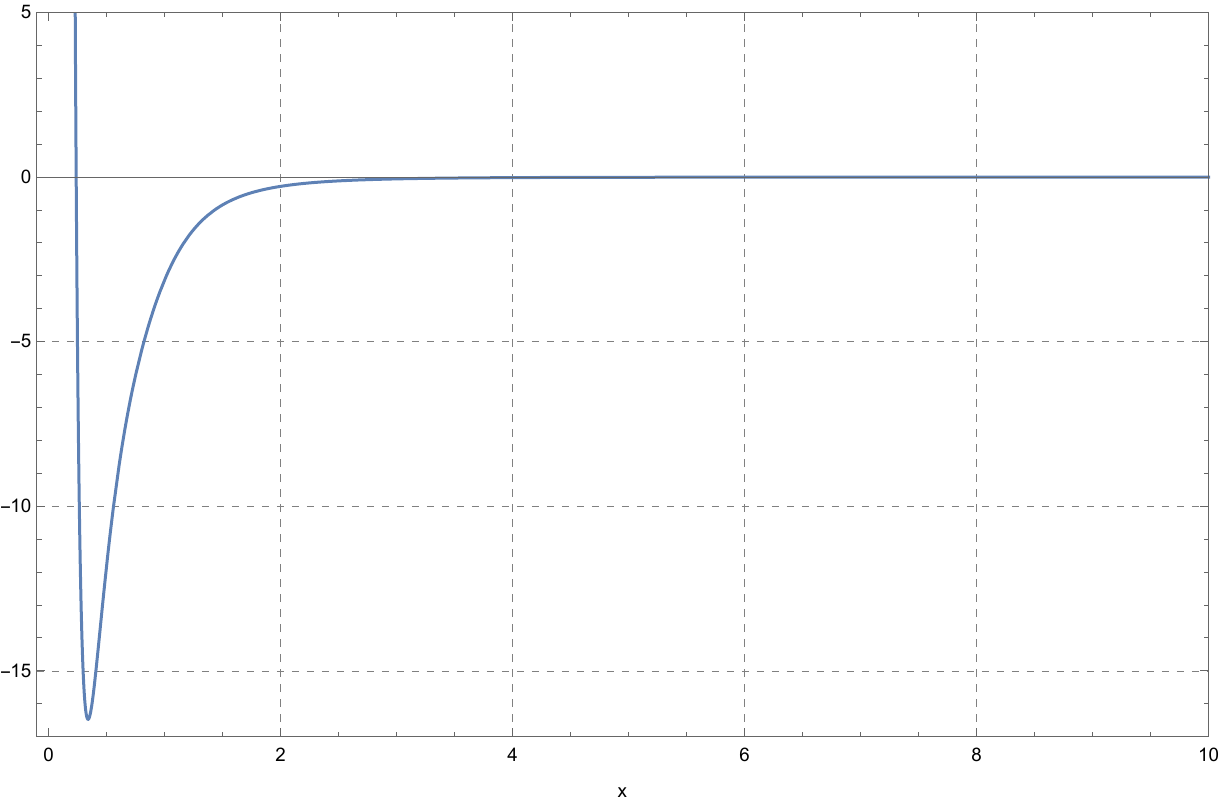}
    \caption{$\mathcal{P}^{SR}_R(x)+P_{ext}(x)$ for $\beta_0=1$}\label{Stability Cond β=1 Exp}
\end{figure}

As in the logarithmic case, in order to calculate the EM EMT we consider:
\begin{equation}
    S_{gauge}=\int d^4x [-\frac{1}{4}\vec{F}_{\mu\nu}\cdot \vec{F}^{\mu\nu}-\beta^2(1- \exp (-\frac{1}{4\beta^2}G_{\mu\nu}G^{\mu\nu}) )]\,.
\end{equation}
The EM action in this case reads:
\begin{equation}
    S_{EM} =\int d^4x\sqrt{-g}[-\frac{1}{4}\mathcal{F}_{ab}\mathcal{F}^{ab}+\frac{1}{4}\cos^2(\theta_W)\mathcal{F}_{ab}\mathcal{F}^{ab} +\beta^2(-1+\exp(-\frac{\cos^2(\theta_W)}{4\beta^2}\mathcal{F}_{ab}\mathcal{F}^{ab}))]\,,
\end{equation}
and, therefore, the EM EMT is given by:
\begin{align}\label{emEMTexp}
    \mathcal{T}^{\mu\nu}&=\frac{2}{\sqrt{-g}}\frac{\delta S}{\delta g_{\mu\nu}}=-g^{\mu\nu}[-\frac{1}{4}\mathcal{F}_{ab}\mathcal{F}^{ab}+\frac{1}{4}\cos^2(\theta_W)\mathcal{F}_{ab}\mathcal{F}^{ab} +\beta^2(-1+\exp(-\frac{\cos^2(\theta_W)}{4\beta^2}\mathcal{F}_{ab}\mathcal{F}^{ab}))] \nonumber \\
&    -\mathcal{F}^{\; \mu}_{\sigma}\mathcal{F}^{\sigma \nu}+\cos^2(\theta_W)\mathcal{F}^{\; \mu}_{\sigma} \mathcal{F}^{\sigma \nu}+\cos^2(\theta_W)\mathcal{F}^{\; \mu}_{\sigma}\mathcal{F}^{\sigma \nu}\exp(-\frac{\cos^2(\theta_W)}{4\beta^2}\mathcal{F}_{ab}\mathcal{F}^{ab})\,,
\end{align}
whose components are:
\begin{equation}
    \frac{g^2}{M^4_W}\mathcal{P}^{LR}_R (x) = -\beta^2_0[1-\exp(-\frac{\cos^2(\theta_W)}{2\beta^2_0}\frac{Q^2(x)}{x^4})]-\frac{Q^2(x)}{2x^4}[1-\cos^2(\theta_W)]\,,
\end{equation}
\begin{align}
    \frac{g^2}{M^6_W}\mathcal{P}^{LR}_\Theta (x) &=-\frac{\beta^2_0}{x^2}[1-\exp(-\frac{\cos^2(\theta_W)}{2\beta^2_0}\frac{Q^2(x)}{x^4})]-\frac{Q^2(x)}{2x^6}[1-\cos^2(\theta_W)] \nonumber \\
&    +\frac{Q^2(x)}{x^6}(-\cos^2(\theta_W)+1 -\cos^2(\theta_W )\exp(-\frac{\cos^2(\theta_W)}{2\beta^2_0}\frac{Q^2(x)}{x^4}))\,.
\end{align}
Then, local stability is guaranteed if the following (Laue's) criterion is valid:
\begin{equation}\label{radSRexp}
    \mathcal{P}^{SR}_R(x)+P_{ext}(x)\geq 0
\end{equation}
In figure \ref{Stability Cond β=1 Exp} we demonstrate the {\it violation} of the condition \eqref{radSRexp}.

This concludes our discussion on the mechanical stability of these non-linear extensions of the hypercharge sector of the electroweak model. 
From the above discussion it seems that the situation resembles that of the non-minimal Higgs-hypercharge-sector coupling, discussed in section \ref{sec:nmcoupl}.

\subsection{Energy Conditions}\label{encm}
In section \ref{Energy Conditions} we have introduced weak, strong and dominant energy conditions. These are given by:
\begin{equation}
    WEC=
    \begin{cases}
  \mathcal{H}\geq 0 \,, \quad  {\rm  timelike-vectors} \\
  \mathcal{H}+\mathcal{P}_R \geq 0 \,, \quad {\rm null-vectors}
\end{cases}
\end{equation}
\begin{equation}
    SEC=
    \begin{cases}
  \mathcal{H}+\mathcal{P}_R+2r^2\mathcal{P}_{\Theta}\geq 0 \,, \quad  {\rm timelike-vectors} \\
  \mathcal{H}+\mathcal{P}_R \geq 0 \,, \quad {\rm null-vectors}
\end{cases}
\end{equation}
\begin{equation}
    DEC=
    \begin{cases}
  (\mathcal{H})^2\geq 0 \,, \quad    {\rm timelike-vectors} \\
  (\mathcal{H})^2-(\mathcal{P}_R)^2 \geq 0 \,, \quad {\rm null-vectors}
\end{cases}
\end{equation}
By using numerical results from this section, we find that electroweak monopole and the modified electroweak monopole with dielectric function satisfy all energy conditions. On the other hand, the string-inspired Born-Infeld, logarithmic and exponential (hypercharge sector)  also satisfies all energy conditions, but violates strong energy condition for timelike vectors.

\section{Discussion: Interpretation of the results in terms of Stability}
\label{concl}

In this concluding section we are going to give a review of the main results of our analysis. We commence the discussion with the well established 't Hooft-Polyakov monopole in $SU(2)$ Georgi-Glashow model. 
 From the results of section \ref{HPReview} we understand that
such a monopole configuration has a well defined energy momentum tensor throughout its domain, resulting to a finite monopole mass and finite radial internal force field. Therefore, from this section we draw a picture for the mechanical properties of a well-established magnetic monopole configuration. Moreover, at the BPS limit of such model the spatial energy momentum tensor components vanish and the monopole configuration behaves like isotropic matter. In addition to this by extracting the long-range contribution from the energy momentum tensor, Laue local stability criterion $\mathcal{F}^{SR}_{Rtotal}(r)>0$ $\forall r$ is satisfied, just as it was showcased in \cite{Panteleeva:2023aiz}.

\par Moving on to the electroweak monopole we have studied its mechanical properties at section \ref{Electroweak Monopole Analysis}. It is well known that electroweak monopole is ill-defined, at least classically, in Minkowski space-time, in the sense that it has singular energy due to the $U_Y (1)$ contribution to the energy functional. In addition to this, the topological argument $\pi_2 (CP^1)=\mathbb{Z}$ based on the $CP^1$ structure of the Higgs doublet in the radial gauge is not beyond doubt, since we have demonstrated explicitly that the magnetic charge quantization conditions between the radial and
unitary gauge solutions have different topological origins. This might be related with the instabilities we find in the current work.

We have shown that the spatial energy-momentum tensor elements, as well as the associated components of the internal force field, in the CM monopole, are also singular. The fact that the internal force field is singular provides us with a non-perturbative criterion for the configuration's (in)stability, given that such a configuration cannot be stable, since the (diverging) internal force field will crash it immediately. By extracting the long-range contribution from the energy momentum tensor, the local stability criterion of Laue, $\mathcal{F}^{SR}_{Rtotal}(r)>0$, is violated. Such an unstable behavior could arise from the fact that the magnetic charge quantization conditions have different topological origins between the different gauges. Furthermore, the internal force field of the monopole configuration has non-radial behavior, which constitutes a very different behaviour compared to that of the well established 't Hooft-Polyakov monopole. Also, the BPS limit of the electroweak monopole solution has very different mechanical properties. In particular, the spatial energy momentum tensor elements are non-vanishing at the BPS limit and, thus, the electroweak monopole does not behave like isotropic matter. 

 \par On the other hand, we have also analysed the mechanical properties of the magnetic monopole solution in the 
 modified electroweak model with Born-Infeld and dielectric function extensions (non-minimal couplings) in the $U_Y(1)$ sector. In both models, monopoles have a well defined energy and the energy momentum tensor components are all singular at $x=0$, and so are their associated  internal total force components, but not their energy functionals which are well defined and finite. The internal force field of the Born-Infeld model is not radial, since the polar and azimuthal internal force field components are finite and non-zero. This is an indication that the monopole configuration could be subject to rotation under a particular perturbation associated with either the monopole profile itself (quantum effects) or the space-time background  ({\it e.g.} gravitational wave). As for the non-minimally coupled dielectric-function models, the internal force field is purely radial, since polar and azimuthal force vanish. 
A similar situation characterises the further extensions of the hypercharge sector discussed in \cite{DeFabritiis:2021eah} with logarithmic and  exponential extensions. 
 Last, but not least, after subtracting the long range contribution of the energy momentum tensor in these last three types of models, we observe a violation of the Laue local stability criterion, except at a region near the origin $r\rightarrow0$,  
where  
 $\mathcal{F}^{SR}_{Rtotal}(r)>0$. A summary of 
 the results discussed above is given in tables \ref{Outlook1}, \ref{Outlook2}, for the convenience of the reader. 
\begin{table}[ht]
\caption{}
\centering
  \begin{tabular}{|c|c|}
  \hline
    \textbf{Model}  & \textbf{Laue} \textbf{Local} \textbf{Stability} \textbf{Criterion} \\
  \hline

    $SU(2)$ $Georgi$-$Glashow$ $Model$ & \CheckmarkBold $(\mathcal{F}^{SR}_{Rtotal}(r)
    >0$ $\forall$ $r$)  \\
    \hline
     $Electroweak$ $Model$ $(EM)$ & \XSolidBrush  \\
    \hline
    $Dielectric$ $Function$ $Modification$ & \XSolidBrush ~(except at $x\rightarrow 0$, where  $\mathcal{F}^{SR}_{Rtotal}(x\to 0)> 0$ ) \\
    \hline
 $String$-$Inspired$ $Hypercharge$ $Modification$ & \XSolidBrush ~(except at $x\rightarrow 0$, where  $\mathcal{F}^{SR}_{Rtotal}(x\to 0)> 0$)  \\
    \hline
 $Logarithmic$ $Hypercharge$ $Modification$ & \XSolidBrush ~(except at $x\rightarrow 0$, where  $\mathcal{F}^{SR}_{Rtotal}(x\to 0)> 0$) \\
    \hline
  $Exponetial$ $Hypercharge$ $Modification$ & \XSolidBrush ~(except at $x\rightarrow 0$, where  $\mathcal{F}^{SR}_{Rtotal}(x\to 0)> 0$) \\
    \hline
  \end{tabular}
  \label{Outlook1}
\end{table}

\begin{table}[ht]
\caption{}
\centering
  \begin{tabular}{|c|c|}
  \hline
    \textbf{Model}   & \textbf{Internal} \textbf{Force} \textbf{Remarks} \\
  \hline

    $SU(2)$ $Georgi$-$Glashow$ $Model$ &  Radial Force Field \\
    \hline
     $Electroweak$ $Model$ $(EM)$ & Singular Polar and Azimuthal Forces \\
    \hline
    $Dielectric$ $Function$ $Modification$ &  Radial Force Field \\
    \hline
 $String$-$Inspired$ $Hypercharge$ $Modification$ & Well Behaved Polar and Azimuthal Forces  \\
    \hline
 $Logarithmic$ $Hypercharge$ $Modification$ & Radial Force Field \\
    \hline
  $Exponetial$ $Hypercharge$ $Modification$ &  Radial Force Field \\
    \hline
  \end{tabular}
  \label{Outlook2}
\end{table}
 \par It is important to stress that, as discussed in subsection \ref{sec:gaugeinv}, the internal force field and pressure are the same between the radial and unitary gauges in the CM electroweak magnetic monopole and its finite-energy extensions, thereby supporting the physical consistency of the pertinent configurations. This is an important remark, because an explicit proof of gauge invariance of the CM model is not yet available, and as we have seen, there is a puzzling behavior as to the topological origin of the magnetic charge quantization condition, which is different between the two gauges.

 In regards to the energy conditions, we find that both electroweak monopole and the modified finite-energy electroweak monopole solutions with non-trivial, non-minimally coupled dielectric function, satisfy them. 
  On the other hand, the string-inspired Born-Infeld electroweak monopole, as well as the monopole in the non-linear (logarithmic or exponential) 
 extensions of the hypercharge sector of the electroweak model, violate only the strong energy condition for timelike vectors, respecting all others. However, we remark that violation of the strong energy condition is not uncommon among general relativistic conditions, and hence we do not ascribe to it any particular significance. 

 \color{black} In a nutshell, what we have found by this analysis is the following: 
\par
(i) For the HP monopole, the fact that the radial short range force points outwards ($\mathcal{F}^{SR}_{Rtotal}(r)>0$) suggests that such configuration is {\it stable}. After all, short range force is responsible for the structure of the inner core and since such force points outwards in this case, monopole configuration tends to be {\it mechanically} stable. This is the essence of Laue local stability criterion for radial short range forces.
We stress that the use of the word 'tends' here is made because, as we explained previously, complete quantitative results on stability are usually  associated with perturbations of space-time background or a monopole profile (quantum effects), beyond those of linear stability.

\par 
(ii) On the other hand, the CM configuration, which at any rate is unphysical due to its infinite energy, unless properly regularized,  
is associated with a short range radial force which points inwards ($\mathcal{F}^{SR}_{Rtotal}(r)<0$) and is singular at the center of the configuration ($\mathcal{F}^{SR}_{Rtotal}(r\rightarrow 0)\rightarrow -\infty$). Following the reasoning stated for the well-established HP monopole, the inner structure of such a configuration will tend to collapse, hence the mechanical instability, according to Laue's criterion. Moreover, the configuration is characterized by angular instabilities, which are also singular at the centree of CM monopole, This suggests that the configuration could be
subject to rotation under a particular perturbation, which in principle renders it unstable. 

\par 
(iii) Modified electroweak models, namely dielectric, logarithmic and exponential modifications, which unlike the CM configuration, are characterized by finite energy,  suffer the same fate. Although all of them are characterized by the absence of angular forces, nonetheless they are associated with the existence of a short range radial force pointing inwards. Towards the center of the configuration, this short-range radial force becomes singular and points outwards. This suggests that further modifications could save these monopole configurations from collapsing under the action of perturbations, since their center does not appear to collapse. 
\par     
(iv) Last but not least, the finite-energy electroweak model with String-Inspired non-linear Hypercharge modifications (of Born-Infeld-like terms) differs from the previous models, insofar as mechanical stability is concerned, since it is associated with well-behaved angular forces. Although its radial short-range force behaves in a similar manner as in the previous monopole cases, nonetheless the angular forces appear to be finite. This suggests that under perturbations such monopole configurations could be subject to rotation, but the finite nature of angular forces indicates that the configuration may not be destroyed.   
\color{black}

\par  In case of instabilities of finite-energy solutions, we remark that, as the monopoles are composites, they will eventually decay in their constituent particles from the corresponding SU(2) sectors of the model, 
namely $W^\pm$ (and Higgs $h^\pm$ bosons, but the latter will play the r\^ole of the 
Goldstone modes in the spontaneously broken phase, so they will decouple from the physical spectrum). Given that in collisions, the magnetic monopoles are produced in monopole-antimnopole pairs, a similar decay process will characterize the unstable anti-monopoles. Such an excess of standard model charged gauge bosons might be detectable in the MAPP1 detector of the MoEDAL-MAPP LHC Experiment~\cite{MoEDAL-MAPP:2022kyr}. Unfortunately, such decaying unstable monopoles are not highly ionizing, so they are not suitable for detection by the MoEDAL-LHC experiment~\cite{MoEDAL:2014ttp}, unless the latter is equipped with a sufficiently large number of time-pix subdetectors. Nonetheless, such decaying unstable monopoles are in principle detectable (through the $W^\pm$ excess produced by their decays) by ATLAS- and CMS- LHC experiments, provided they are of sufficiently low mass. Another issue, which is still to be calculated regarding the unstable monopoles is their life time. This is a complicated issue, which needs to be studied separately, given the non-perturbative nature of the composite magnetic monopole configuration. One needs to determine the precise decay modes, which is a topic that is left for a future work. 

Moreover, these instabilities we found may provide a way out of the puzzle for the cosmic relic abundance of the stable `t Hooft-Polyakov monopoles, 
calculated in \cite{Zeldovich:1978wj} for the range of masses of order $5-10$~TeV, which lies 
well above the experimental upper limits for stable magnetic-monopole relics, as per the relevant cosmic searches for such objects~\cite{Mavromatos:2020gwk}. Indeed, if the magnetic electroweak monopoles that exist in nature are of the type of the finite-energy unstable ones, discussed here, then upon cosmic production, they will quickly decay into charged gauge bosons (after the electroweak cosmic era), leaving  no stable relics behind. 
These are issues that deserve further study.

\par As an outlook, 
we should 
mention our plans to study electroweak 
magnetic monopole solutions of the field equations  of systems with Born-Infeld extensions in the whole $SU_L(2)\times U_Y(1)$ gauge sector. In such a case there is some preliminary evidence that magnetic monopoles exist only above some critical value of the Born-Infeld parameter~\cite{Grandi:1999rv}. We hope to address this issue, together with the pertinent criteria for stability, in a future work.

\section*{Acknowledgments}

 We thank colleagues from the MoDEAL-MAPP collaboration for their interest in this work, and discussions.
The work of NEM is supported in part by the UK Science and Technology Facilities research Council (STFC) 
under the research grant  ST/X000753/1.
 NEM also acknowledges participation in the COST Association Actions CA21136 “Addressing observational
tensions in cosmology with systematics and fundamental physics (CosmoVerse)” and CA23130 ”Bridging high and
low energies in search of quantum gravity (BridgeQG)”.

\appendix
\section{Numerical Techniques}
In this short Appendix we discuss the numerical techniques used in the main text. In all of the models examined in this work, we need, at some point,  to solve a system of differential equations. For example, in the electroweak model of CM~\cite{Cho:1996qd} we have the system: 
\begin{equation}
    f'' =\frac{f(f^2-1)}{x^2}+2fH^2
\end{equation}
\begin{equation}
    H''=-\frac{2}{x}H'+\frac{f^2H}{2x^2}+\gamma H(H^2-1)
\end{equation}
\begin{equation}
    H(0) =0 \quad H(\infty)=1 \qquad \qquad f(0)=1 \quad f(\infty)=0
\end{equation}
For a solution of this problem, with the given boundary conditions, we have used the shooting method together with the stiffness switching method. The shooting method provides   numerical results at boundaries, while the stiffness switching method is used for a numerical solution of the system throughout the domain of the definition of the profile functions $H(x)$ and $f(x)$. This particular system is stiff, therefore the stiffness switching is a suitable method for solution, given that, at each stiff point of the domain, the method chooses the appropriate solver.

\bibliography{Refs_updated}

\end{document}